\documentclass[11pt]{article}
\pdfoutput=1
\usepackage{jheppub}
\usepackage{tabu}
\usepackage[vcentermath]{youngtab}
\usepackage[usenames,dvipsnames,table]{xcolor}
\usepackage{graphicx,subfig}
\usepackage{amsmath,amssymb,amsthm,amsfonts,multirow,array,bm,bbm,esint}
\usepackage[mathscr]{eucal}
\usepackage[bbgreekl]{mathbbol}
\usepackage{epsf,grffile}
\usepackage{slashed}
\usepackage[numbers,sort&compress]{natbib}
\usepackage{array,tikz-cd}

\usepackage{subfig}
\usepackage{float}
\usetikzlibrary{snakes}
\usetikzlibrary{shapes.misc}

\definecolor{rust}{rgb}{0.8,0.2,0.2}

 \def\shadeB{\cellcolor{blue!5}}
\def\shadeR{\cellcolor{red!5}}

\def\bR {\mathbb{R}}
\def\bZ {\mathbb{Z}}

\def\bC {\mathbb{C}}

\def\be{\begin{equation}}
\def\ee{\end{equation}}
\def\bea{\begin{eqnarray}}
\def\eea{\end{eqnarray}}

\makeatletter\@addtoreset{equation}{section}\makeatother

\newcommand{\tr}{{\rm tr\,}}

\def\be{\begin{equation}}
\def\ee{\end{equation}}
\def\bea{\begin{eqnarray}}
\def\eea{\end{eqnarray}}
\def\ie{\begin{equation}\begin{aligned}}
\def\fe{\end{aligned}\end{equation}}

\newcommand{\A}{{\alpha}}

\newcommand{\Tr}{{\rm Tr\,}}
\newcommand{\la}{\langle}
\newcommand{\ra}{\rangle}

\makeatletter
\newcommand{\subalign}[1]{%
  \vcenter{%
    \Let@ \restore@math@cr \default@tag
    \baselineskip\fontdimen10 \scriptfont\tw@
    \advance\baselineskip\fontdimen12 \scriptfont\tw@
    \lineskip\thr@@\fontdimen8 \scriptfont\thr@@
    \lineskiplimit\lineskip
    \ialign{\hfil$\m@th\scriptstyle##$&$\m@th\scriptstyle{}##$\crcr
      #1\crcr
    }%
  }
}
\makeatother

\title{Supersymmetric Landau-Ginzburg Tensor Models}

\author{Chi-Ming Chang, Sean Colin-Ellerin,  Mukund Rangamani}

\affiliation[]{
Center for Quantum Mathematics and Physics (QMAP),  \\
Department of Physics, University of California, Davis, CA 95616 USA.}

\emailAdd{wychang@ucdavis.edu, scolinellerin@ucdavis.edu, mukund@physics.ucdavis.edu}

\abstract{ 
We study two dimensional $\mathcal{N} = (2, 2)$ Landau-Ginzburg models with tensor valued superfields with the aim of constructing large central charge superconformal field theories which are solvable using large $N$ techniques. We demonstrate the viability of such constructions and motivate the study of anisotropic tensor models. Such theories are a  novel deformation of tensor models where we break the continuous symmetries while preserving the large $N$ solvability. Specifically, we examine theories with superpotentials involving tensor contractions chosen to pick out melonic diagrams. The anisotropy is introduced by further biasing individual terms by  different coefficients, all of the same order, to retain large $N$ scaling. We carry out a detailed analysis of the resulting low energy fixed point and comment on potential applications to holography. Along the way we also examine gauged versions of the models (with partial anisotropy) and find  generically that such theories have a non-compact Higgs branch of vacua.
}
\begin{document}
\maketitle

%%%%%%%%%%%%%%%%%%%%%%%%%%%%%%%%%%%%%%%%%%%%%%%%%%%

%~~~~~~~~~~~~~~~~~~~~~~~~~~~~~~~~~~~~~~~~~~~~~~~
\section{Introduction}
\label{sec:intro}
%~~~~~~~~~~~~~~~~~~~~~~~~~~~~~~~~~~~~~~~~~~~~~~
  
As is well known, large $N$ field theories have played a key role in the holographic gauge/gravity duality. The emergence of semiclassical gravitational dynamics in this context is predicated on the existence of a large number of degrees of freedom (measured by central charge $c \sim N^\alpha$ with $\alpha \geq1$), with a relatively small number of collective excitations of low energy (roughly $\rho(E)  \sim \mathcal{O}(1)$ for $E \ll c$).\footnote{ In general we require that the density of low lying states does not scale with $c$. Necessary bounds are available in two dimensions where $\rho(E) \lesssim e^{\gamma E}$ with $\gamma >0$, cf., \cite{Hartman:2014oaa}. }  Solvable large $N$ models could therefore provide valuable insight into the holographic map. However, planar matrix models \cite{tHooft:1973alw} which have been investigated for decades are intractable in general. This  can be seen for example from the interacting two-matrix model which exhibits spectral features required above, but is not solvable \cite{Aharony:2003sx}. On the other hand, large $N$ vector models whilst solvable, lead to at best weakly interacting dynamics and possessing nearly conserved higher spin currents. In recent years a new possibility has arisen: melonic models which appear to provide a happy middle ground of being technically tractable and having non-trivial dynamics. 

Motivated by these considerations, we will attempt to construct large $c$ interacting two-dimensional CFTs. One approach would be to study a family of theories with central charge $c$ as a parameter which can be dialed appropriately. For instance, the D1-D5 CFT is obtained as (deformation of) a symmetric product CFT \cite{Strominger:1996sh}, the sigma model with target $X^N/S_N$ where $X$ is either $K3$ or $T^4$ in the large $N$ limit ($c\sim N$). Another possibility is to take scaling limits of a family of WZW coset models \cite{Gaberdiel:2010pz} (or even WZW models themselves \cite{Kiritsis:2010xc}). The aforementioned examples however end up being the analog of vector-like modes; in the large $N$ limit such theories typically have higher spin symmetry. For the orbifold example with sufficient supersymmetry one has a moduli space: far out away from  the orbifold point, one expects to recover semiclassical gravity.   
Given this status quo, we seek to ask if we can conjure up interesting CFTs, exploiting lessons learnt from the analysis of theories with melonic structure.

While field theories with melonic large $N$ expansions have been studied in the literature for a while, recent interest in them owes to similar structures being relevant for the analysis of the disordered Sachdev-Ye-Kitaev  (SYK) model \cite{Sachdev:1992fk,Kitaev:2015aa, Maldacena:2016hyu,Kitaev:2017awl}, which has been argued to provide a 1d toy model for holography. Inspired by this, various groups have examined tensor models exhibiting melonic dominance, focusing on quantum mechanical models of colored  \cite{Witten:2016iux,Bonzom:2011zz}, or uncolored \cite{Klebanov:2016xxf,Carrozza:2015adg} tensor valued degrees of freedom,  where it is sufficient to write down potential terms involving tetrahedral tensor contractions  (i.e., those which guarantee only melonic diagrams contribute in the large $N$ limit). These models are investigated extensively in the literature: recent reviews include \cite{Delporte:2018iyf,Klebanov:2018fzb}.

There have also been serious attempts to find non-trivial IR fixed points in higher dimensional tensor models in the large $N$ expansion \cite{Giombi:2017dtl,Giombi:2018qgp}. Much of the analysis on this front has been carried out for bosonic models which generically suffer from an unbounded from below Hamiltonian. Even when one can stabilize the spectrum, by cleverly truncating a supersymmetric theory as in \cite{Giombi:2018qgp}, the resulting fixed points often end up having operators violating unitarity bounds. This intransigence of these models can in 
part be traced to the fact that while the Lagrangian may be engineered to only contain terms that give rise to melonic diagrams, renormalization effects induce non-tetrahedral tensor contractions (for example the so called pillow or double-sum terms).  One therefore wonders if it is even possible to find non-trivial, solvable higher dimensional tensor models, which could provide further insight into the holographic AdS/CFT correspondence. 

What the aforementioned set of investigations suggest is that we should focus on situations where we have some symmetry principle preventing contamination from the non-melonic sector. Happily, we know a context where this can be achieved, viz.,  situations where we can use supersymmetric non-renormalization theorems to avoid generation of non-tetrahedral tensor 
contractions during the renormalization group flow. This requires us to focus on examples of tensor models with at least $4$ supercharges in $d \leq 4$. Of primary interest will be theories with $\mathcal{N} = (2,2)$ in $d=2$. 

We will focus on Landau-Ginzburg (LG) models with a set of tensor valued chiral superfields, as well as gauged models. The former can be simply understood as a natural upgrade of usual LG models with tensor valued chiral superfields. The latter can be likewise realized by gauging some of the global symmetry, or independently motivated by a hybridization of matrix and vector models. For instance, one can consider a vector valued set of matrices which can be used to construct matrix-vector models, cf., \cite{Ferrari:2017ryl} who motivated such models by imagining the D0-brane matrix model with a large number of transverse dimensions. In this case, one can explicitly view the  melonic diagrams as a subset of standard planar  diagrams (by considering ribbon graphs decorated with an internal line corresponding to the vector label). 

Now $\mathcal{N}=(2,2)$  models in two dimensions are well studied in the literature as they play a central role in mirror symmetry through the LG/Calabi-Yau correspondence \cite{Martinec:1989in,Vafa:1988uu,Greene:1988ut} cf., \cite{Witten:1993yc,Hori:2003ic} for detailed reviews.  The general idea of $(2,2)$ LG models is that we have a non-trivial RG flow driven by a relevant superpotential. The IR dynamics is altogether controlled by the superpotential, so one expects to find a low energy superconformal field theory once constraints from anomaly cancellation are taken into account. As such this is a powerful statement, relying on supersymmetry to argue for a critical fixed point. Indeed many of the early checks involve the usual matching of protected quantities such as the chiral ring \cite{Lerche:1989uy} and elliptic genera \cite{Witten:1993jg}. Bringing  tensor valued fields into the game, provides us with an opportunity to use large $N$ melonic techniques to solve the theory. At the level of two-point functions we do not learn any more information than we already knew from the supersymmetric analysis, but the pay-off lies in being able to compute other observables, such as four-point functions by resummation of a class of ladder diagrams, which in turn gives us non-chiral spectral data of the low-energy fixed point. 

While it appears at face value that we have engineered a perfect blend of supersymmetry and melonic diagrammatics, the situation will turn out to be a lot more complicated. Standard tetrahedral tensor contractions which have hitherto been investigated in the literature will result in superpotentials which have flat directions, resulting in the low-energy theory having many moduli. 
We would ideally like to construct rigid models, which are moduli free, to avoid strong IR effects in two dimensions. One reason for the presence of the moduli is the fact that the tensor valued fields of interest typically have a large global symmetry (which in itself is a problem as they potentially give rise to relevant operators \cite{Murugan:2017eto} and a high degeneracy of low-lying states \cite{Bulycheva:2017ilt,Choudhury:2017tax}). Two natural possibilities come to mind: either gauge the symmetries to focus on the singlet sector, or consider explicit breaking of the symmetry whilst retaining solvability.

We will demonstrate that the most efficacious choice is to break the symmetry by considering \emph{anisotropic tensor models}. These will have tensor valued fields with index contractions inspired by the tetrahedral structure, except that we will bias individual contractions with slightly different couplings. By a judicious choice of the couplings we will show that the theory retains large $N$ solvability, which we illustrate explicitly for weak anisotropy, where we can use perturbative arguments. It will be important to note that the anisotropic couplings are specified once and for all in the microscopic Lagrangian, and so we  are dealing with a genuine quantum field theory (and not an ensemble thereof).\footnote{ One could have alternately considered disordered models where the couplings are averaged over some suitable chosen ensemble. Such models have been investigated in the literature before: \cite{Murugan:2017eto} examined theories with two supercharges, while \cite{Bulycheva:2018qcp} has analyzed disordered  $(2,2)$ LG models, and \cite{Peng:2018zap,Ahn:2018sgn} have analyzed two dimensional $(0,2)$ disordered models. We will have use for some of the results derived therein when we turn to the explicit solution of our models.} On the other hand the gauged models turn out to still possess flat directions. Once we gauge the models we of course also have to worry about the gauge sector which can typically lead to non-compact Coulomb branches. This we will be able to cure, but generically find that we will be stuck with non-compact Higgs branches in the large $N$ limit. 

The outline of the paper is as follows: we describe the broad class of models we will focus on in \S\ref{sec:lg2d}. In \S\ref{sec:rg} we review the RG flow of $\mathcal{N} = (2,2)$ models, in particular the non-renormalization of the superpotential and the problems that arise when the classical moduli space of vacua has flat directions. Furthermore, we remind readers of general properties of Landau-Ginzburg models, computing central charges, etc. In \S\ref{sec:diagrammatics} we then turn to the details of the large $N$ analysis, arguing that for the models with global symmetry one can import results obtained in \cite{Bulycheva:2018qcp} before turning to the anisotropic models. We then turn to a detailed discussion of the existence of flat directions and moduli in our models in \S\ref{sec:moduli}, arguing that one can engineer anisotropic models which are moduli free. We then turn in  \S\ref{sec:gauge} to analyze 
models where we gauge (part of) the global symmetry.  While these fail to produce moduli-free IR fixed points, there are several technical features of these models which we found to be interesting and unexplored in the literature (in particular, it is possible to construct moduli-free theories for small rank tensors). We undertake a detailed analysis of the phase structure and compute elliptic genera in \S\ref{sec:egen} to confirm some of our findings. 
We conclude in \S\ref{sec:discussion} with a discussion of open questions. Some technical results which are helpful in our analysis can be found in the  Appendices. Specifically, Appendix~\ref{sec:conventions} outlines our $\mathcal{N}=(2,2)$ conventions. In Appendix \ref{sec:unitarity} we establish that the part of the spectrum of the low energy fixed point we can access is consistent with unitarity. Finally, Appendix \ref{sec:nomoduliproof} contains the details of our proof that the anisotropic models are moduli-free which is obtained using the theory of resultants. Appendix \ref{sec:higss2} has some further results regarding the gauged models for low rank theories.

%~~~~~~~~~~~~~~~~~~~~~~~~~~~~~~~~~~~~~~~~~~~~~~~
\section{Melonic Landau-Ginzburg tensor models}
\label{sec:lg2d}
%~~~~~~~~~~~~~~~~~~~~~~~~~~~~~~~~~~~~~~~~~~~~~~

Two-dimensional $\mathcal{N}=(2,2)$ supersymmetric models  can be easily realized in superspace. We follow the conventions in \cite{Witten:1993yc} which are summarized  in Appendix \ref{sec:conventions}. The primary `matter' fields of interest are chiral and anti-chiral superfields $\mathscr{O}(z,\bar z,\theta^{\pm}, \overline{\theta}^{\pm})$, and $\overline{\mathscr{O}}(z,\bar z,\theta^{\pm},\overline{\theta}^{\pm})$, obeying 
$\overline{D}_{\dot{\alpha}} \mathscr{O}  =0$ and $D_{\alpha} \overline{\mathscr{O} } =0$ respectively. They 
admit a component expansion:
\begin{equation}
\begin{split}
\mathscr{O}({\sf Z}) &= 
	O(y)+ \, \theta^{\alpha} \, \psi_{\scriptscriptstyle{O} \alpha}(y)+\frac{1}{2}\theta^{\alpha}\theta_{\alpha}\, F_{\scriptscriptstyle{O}}(y) \,,\\ 
\overline{\mathscr{O}}({\sf Z}) &=
	 \overline{O}(\overline{y})+\, \overline{\theta}_{\dot{\alpha}}\, \overline{\psi}_{\scriptscriptstyle{O}}^{\dot{\alpha}}(\overline{y})+\frac{1}{2}
	 \overline{\theta}_{\dot{\alpha}}\overline{\theta}^{\dot{\alpha}}\,\overline{F}_{\scriptscriptstyle{O}}(\overline{y}) \,,
\end{split}
\label{eq:multiplets}
\end{equation} 
with the chiral, anti-chiral coordinates $y, \overline{y}$ defined in \eqref{eq:ydef}. We will use  ${\sf Z} = (z,\overline{z},\theta^\pm,\overline{\theta}^\pm)$ to denote the superspace coordinate, with $z, \overline{z}$ being the usual complex coordinates in two-dimensional Euclidean space. R-charge assignments are given in Table~\ref{tab:Rfsym}.

We will be interested in situations where $\mathscr{O} $ is a tensor valued field transforming under some symmetry group $G$. We will  exemplify some choices below. However, even without further specification, we can say that the theories of interest are captured by writing down a K\"ahler potential $K(\mathscr{O} ,\overline{\mathscr{O} })$ and a superpotential $W(\mathscr{O} )$  for these matter fields.
In situations where $G$ is a global symmetry, we will take the K\"ahler potential to correspond to a flat metric in field space. The only choice we will make is to engineer the superpotential  $W(\mathscr{O} )$ to ensure melonic dominance.
To wit, the simplest supersymmetric action is given by
\begin{equation}
S = \int d^2z \, d^4\theta \, \overline{\mathscr{O} }\, \mathscr{O}  - \int\, d^2 z\, d^2 \theta \, W(\mathscr{O} ) -  \int\, d^2 z\, d^2 \overline{\theta} \, \overline{W}(\overline{\mathscr{O} }).
\label{eq:susyaction}
\end{equation}	
We will first study this simple system as it will turn out to be amenable to direct large $N$ diagrammatic analysis. Later in \S\ref{sec:gauge} we will also be interested in situations where we gauge the symmetry $G$ (or some subgroup $H \subset G$ thereof). For now let us continue to a more complete specification of our theories.

%~~~~~~~~~~~~~~~~~~~~~~~~~~~~~~~~~~~~~~~~~~~~~~~
\subsection{The models}
\label{sec:models}
%~~~~~~~~~~~~~~~~~~~~~~~~~~~~~~~~~~~~~~~~~~~~~~

To ensure that we have a theory with melonic diagrams dominating, we will take $\mathscr{O} $ to be a tensor valued field\footnote{  We will use the basic superfield label to refer to the models as indicated, and $\mathscr{O} $ when we wish to make model independent statements.}  with a quartic interaction term in the superpotential that obeys the tetrahedral contraction structure, where for each pair of superfields there is exactly one index contraction. There are a-priori several choices we can make, which we can categorize into two broad classes:\footnote{  For ease of discussion we will focus below on the case the tensors interact via a quartic superpotential. It is possible to generalize this to arbitrary $q$-fold interactions, as we shall comment on during the course of our discussion (though we will often refrain from writing out explicit tensors and their contractions).}
\begin{itemize}
\item \emph{Colored tensors $\mathscr{B}$}: Following \cite{Witten:2016iux} we pick a collection of chiral superfields $\{\mathscr{B}_a\}$ transforming under $G = U(N)^6/\mathbb{Z}_{2}^2$. Each $\mathscr{B}_a$ transforms in the fundamental of some of the gauge groups and as anti-fundamental in  others. Labeling the components of $G$ as $U(N)_{ab}$, we can summarize the representation content succinctly as in Table \ref{tab:repcontent}.
	\begin{table}[h]
	\begin{center}
		\begin{tabular}{| l | l | l | l | l | l | l |}
		\hline
		& $U(N)_{01}$ & $U(N)_{02}$ & $U(N)_{03}$ & $U(N)_{12}$ & $U(N)_{13}$ & $U(N)_{23}$ \\
		\hline\hline
		$\mathscr{B}_{0}$ & $N$ & $N$ & $\overline{N}$ & $1$ & $1$ & $1$ \\
		$\mathscr{B}_{1}$ & $\overline{N}$ & $1$ & $1$ & $N$ & $N$ & $1$ \\
		$\mathscr{B}_{2}$ & $1$ & $\overline{N}$ & $1$ & $\overline{N}$ & $1$ & $N$ \\
		$\mathscr{B}_{3}$ & $1$ & $1$ & $N$ & $1$ & $\overline{N}$ & $\overline{N}$ \\
		\hline
		\end{tabular}
		\caption{Representation content of superfields, where $1$ is the trivial representation and $N,\overline{N}$ are the fundamental and anti-fundamental representations, respectively.}
		\label{tab:repcontent}
	\end{center}
	\end{table}
Choosing the superpotential (nb: index placement correlates with representation)
\begin{equation}
	W_4(\{\mathscr{B}_{a}\}) = g\,  (\mathscr{B}_{0})^{i_{_{01}}i_{_{02}}}_{i_{_{03}}} \, (\mathscr{B}_{1})_{i_{_{01}}}^{i_{_{12}} i_{_{13}}} \, 
	(\mathscr{B}_{2})^{i_{_{23}}}_{i_{_{02}} i_{_{12}}}\, (\mathscr{B}_{3})_{i_{_{23}} i_{_{13}}}^{i_{_{03}}} 
\label{eq:colorW}
\end{equation}	
suffices to ensure that the large $N$  expansion is controlled by melonic diagrams \cite{Witten:2016iux}. We can study the theory with the symmetry $G$ being either global or gauged. 
\item \emph{Uncolored tensors $\mathscr{X}$}: 	Alternately, we can consider a single tensor-valued field $\mathscr{X}^{a_1a_2a_3}$ transforming under some symmetry group which acts on the indices independently, cf.,  \cite{Klebanov:2016xxf,Carrozza:2015adg}. One simple choice is to take $\mathscr{X}^{a_1a_2a_3}$ to transform under $O(N)^3$ though clearly other choices are possible. 
The index subscripts are correlated with each of the components of the global symmetry group. This would have been more transparent if we choose to work with $G = O(N_1) \times O(N_2) \times O(N_3)$, but it will be sufficient to focus on the case $N_1 = N_2 = N_3 = N$. One can pictorially differentiate the indices with color in a triple-line notation, cf., Fig.~\ref{fig:3-loop_vacuum_supergraph}, where the colors red, green, and blue correspond to index $1$, $2$, and $3$, respectively. The superpotential can simply be taken to be the tetrahedral contraction of indices, viz., 
\begin{equation}
W_4(\mathscr{X}) = \frac{1}{4}\, g\,\mathscr{X}^{a_1a_2a_3}\,\mathscr{X}^{a_1b_2b_3}\,\mathscr{X}^{b_1a_2b_3}\,\mathscr{X}^{b_1b_2a_3}
\label{eq:uncolorW}
\end{equation}	

\item \emph{Matrix-vectors $\mathscr{Y}$:} A particular example of uncolored tensor models is obtained with fields $\mathscr{Y}^{a\,I}_{\;b}$ transforming in the adjoint of $U(N)$ (indices $a,b$) and the fundamental of $O(M)$ (index $I$). For this case we prefer to keep $M\neq N$ to retain separate information about the matrix and vector structures. The tetrahedral superpotential \eqref{eq:uncolorW} can then be simplified to a single trace potential, viz.,
\begin{equation}
W_4(\mathscr{Y}) = g\, \Tr{(\mathscr{Y}^I\, \mathscr{Y}^J\, \mathscr{Y}^I \, \mathscr{Y}^J)}  \equiv g\, \mathscr{Y}^{a\,I}_{\;b} \, \mathscr{Y}^{b\,J}_{\;c}\, \mathscr{Y}^{c\,I}_{\;d}\, \mathscr{Y}^{d\,J}_{\;a}\,.
\label{eq:mvW}
\end{equation}	
\end{itemize}
Due to the tetrahedral contraction structure, the melonic supergraphs dominate the sum over supergraphs, and the theories are exactly solvable in the large $N$ limit defined as: 
\begin{equation}
N\to\infty,\quad\text{fixing}\quad J^2 = 
\begin{cases}
& g^2\, N^{3} \,, \quad \quad\; \textrm{colored and uncolored tensors}, \\
& g^2\, N^2\, M \,, \quad \textrm{matrix-vector}.
\end{cases}
\label{eq:Jdef}
\end{equation}	

While we analyze all three models to some degree, of primary interest to us will be certain deformations of the  uncolored tensor and its cousin the matrix-vector model. It is worth noting that the matrix-vector model has some nice features in that we can identify the melonic diagrams as a subclass of planar diagrams with an internal vector `decoration'. One can view this as arising from a kind of Veneziano limit \cite{Veneziano:1976wm} where we scale the `flavor' degrees of freedom commensurately with the `color' degrees of freedom in the planar large $N$ expansion.\footnote{ The adjectives flavor and color obviously refer here to the standard QCD parlance.} It has the advantage of making certain aspects of the large $N,M$ counting more transparent.   We should note that a similar philosophy has been advocated earlier in \cite{Ferrari:2017ryl}. These constructions were inspired by the D0-brane matrix model where the adjoint valued scalars carry a spacetime index; \cite{Ferrari:2017ryl} wished to view these theories as matrix models on branes living in a spacetime with the number of dimensions taken large. 

%~~~~~~~~~~~~~~~~~~~~~~~~~~~~~~~~~~~~~~~~~~~~~~~
\subsection{Anisotropic deformation}
\label{sec:anisotropic}
%~~~~~~~~~~~~~~~~~~~~~~~~~~~~~~~~~~~~~~~~~~~~~~

Our models admit {\it anisotropic} deformations of the superpotential that preserve the tetrahedral contraction structure, but break the glabal symmetry $G$. We will focus on the uncolored tensor model, where the deformed superpotential will be taken to be:\footnote{ When we write the deformed models we eschew the use of  Einstein summation convention since the sum over index contractions is no longer homogeneously weighted.}
\ie\label{eqn:deformed_superpotential}
W_4(\mathscr{X}) =\frac{1}{4}\, g \sum_{a_1,a_2,a_3,b_1,b_2,b_3=1}^N \A_{a_1b_1,a_2b_2,a_3b_3}\; \mathscr{X}^{a_1a_2a_3}\mathscr{X}^{a_1b_2b_3}\mathscr{X}^{b_1a_2b_3}\mathscr{X}^{b_1b_2a_3}\,.
\fe
The anisotropic deformation parameters $\A_{a_1b_1,a_2b_2,a_3b_3}$ satisfy the relation
\ie\label{eqn:Z2xZ2}
\A_{a_1b_1,a_2b_2,a_3b_3}  =\A_{b_1a_1,b_2a_2,a_3b_3}  =\A_{a_1b_1,b_2a_2,b_3a_3}  =\A_{b_1a_1,a_2b_2,b_3a_3},
\fe
and are defined only up to the scaling
\ie
g\sim \lambda \, g,\quad \A_{a_1b_1,a_2b_2,a_3b_3}\sim \lambda^{-1} \,\A_{a_1b_1,a_2b_2,a_3b_3}.
\label{eqn:scaling}
\fe
The ${\rm O}(N)^3$ symmetry of the isotropic model \eqref{eq:uncolorW} is broken to the discrete symmetry $\bZ_2^{3N}$. The isotropic model introduced in 
\eqref{eq:uncolorW} is obtained for the choice $\A_{a_1b_1,a_2b_2,a_3b_3} =1$ for all choices of $\{a_1, b_1, a_2, b_2, a_3, b_3\}$.

We argue that the melonic diagrams still dominate in the large $N$ limit as long as the anisotropic deformation parameters are chosen such that the coupling constants $
g\, \A_{a_1b_1,a_2b_2,a_3b_3}$ are all of order $N^{-\frac{3}{2}}$, i.e.
\ie\label{eqn:melonic_cond}
N\to\infty,\quad{\rm fixing}\quad g\,\A_{a_1b_1,a_2b_2,a_3b_3}  \;N^{\frac{3}{2}} \quad\forall \;\A_{a_1b_1,a_2b_2,a_3b_3}.
\fe
By partially fixing the scaling ambiguity \eqref{eqn:scaling}, we can rewrite the above condition as\footnote{ The remaining scaling ambiguity is \eqref{eqn:scaling} with the $\lambda$ of order one in the large $N$ limit.}
\ie
g= {\cal O}(N^{-\frac{3}{2}})\quad{\rm and}\quad \A_{a_1b_1,a_2b_2,a_3b_3}= {\cal O}(N^0).
\fe

\begin{figure}[H]
\centering
\subfloat{
\begin{tikzpicture}

\draw[red, thick] (0,0) circle (2cm);

\draw[red, thick] 
(-1.5,0) to[out=30,in=150] (1.5,0) 
(-1.5,0) to[out=-30,in=-150] (1.5,0);

\draw [green, thick,domain=20:160] (0,0) plot ({1.5*cos(\x)}, {1.5*sin(\x)});
\draw[green, thick] 
(-1.40954,0.51303) to[out=30,in=150] (1.40954,0.51303);
 
\draw [green,thick,domain=-20:-160] (0,0) plot ({1.5*cos(\x)}, {1.5*sin(\x)});
\draw[green,thick,shift={(0,0)}] 
(-1.40954,-0.51303) to[out=-30,in=-150] (1.40954,-0.51303);

\draw [blue,thick,domain=0:180] (0,0) plot ({1.75*cos(\x)}, {1.75*sin(\x)});
\draw[blue,thick,shift={(0,0)}] 
(-1.75,0) to[out=-45,in=-135] (1.75,0);

\draw [blue,thick,domain=190:350] (0,0) plot ({1.75*cos(\x)}, {1.75*sin(\x)});
\draw[blue,thick,shift={(0,0)}] 
(-1.5,0.25) to[out=30,in=150] (1.5,0.25);

\end{tikzpicture}
}
\caption{A three-loop vacuum supergraph in the triple line notation. Indices with subscript $1$, $2$, and $3$ are colored red,  green, and blue, respectively. The outer (inner) red loop is associated to $a_1$ ($b_1$), while the top (respectively, bottom) green and blue loops correspond to $a_2$ and $b_2$ (respectively, $a_3$ and $b_3$) in Eq.~\eqref{eq:vac_nL6_nv2}. 
 }
\label{fig:3-loop_vacuum_supergraph}
\end{figure}
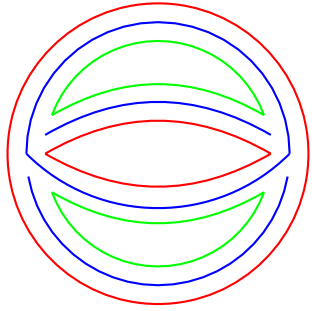

For simplicity, let us first focus on vacuum supergraphs. Consider a vacuum supergraph with $n_V$ vertices and $n_L$ index loops. Each vertex contributes a factor of the deformation parameter $\A_{a_1b_1,a_2b_2,a_3b_3}$ along with a factor of $g$. The supergraph is proportional to 
\ie\label{eqn:comb_alpha}
\sum_{a_1,a_2,\cdots, a_{n_L}=1}^N(\A^{n_V})_{a_1a_2\cdots a_{n_L}},
\fe
after summing over index loops. Here $(\A^{n_V})_{a_1a_2\cdots a_{n_L}}$ denotes the collection of $n_V$ factors of the deformation parameters from the $n_V$ vertices of the supergraph
and the summation is over the legs that participate in the loops. For example, the three-loop vacuum supergraph shown in Fig.~\ref{fig:3-loop_vacuum_supergraph} has $n_L=6$, $n_V=2$, and is proportional to
\ie\label{eq:vac_nL6_nv2}
\sum_{a_1,a_2,a_3,b_1,b_2,b_3=1}^N|\A_{a_1b_1,a_2b_2,a_3b_3}|^2.
\fe
The indices $a_1$, $b_1$ correspond to the red lines, $a_2$, $b_2$ correspond to the green lines, and $a_3$, $b_3$ correspond to the blue lines.
In the large $N$ limit, \eqref{eqn:comb_alpha} scales as $N^{n_L}$ by the condition \eqref{eqn:melonic_cond}. An equivalent way to see this is to note that we are summing a set of $\mathcal{O}(N^0)$ numbers when we are considering the deformed model, while the isotropic model is the same sum with unit weight for each summand. Thus, it should come as no surprise that the large $N$ scaling remains the same as in the isotropic model. Hence, the melonic dominance of the anisotropic models directly follows from the melonic dominance of the isotropic models. For supergraphs with external legs one can easily extend the above argument as the reader can verify.

There is a special class of anisotropic deformation that breaks the ${\rm O}(N)^3$ symmetry down to $(\bZ_2^N\rtimes S_N)^3$,
\ie\label{eqn:Z2xSN_model}
\A_{a_1b_1,a_2,b_2,a_3,b_3}&=\A_1+ \A_2\,\delta_{a_1b_1}+ \A_3\,\delta_{a_2b_2}+ \A_4\,\delta_{a_3b_3}
\\
&\quad+\A_5\,\delta_{a_1b_1}\,\delta_{a_2b_2}+\A_6\,\delta_{a_1b_1}\,\delta_{a_3b_3}+\A_7\,\delta_{a_2b_2}\,\delta_{a_3b_3} +\A_8\,\delta_{a_1b_1}\,\delta_{a_2b_2}\,\delta_{a_3b_3}\,.
\fe

%~~~~~~~~~~~~~~~~~~~~~~~~~~~~~~~~~~~~~~~~~~~~~~~
\section{Renormalization group flow}{}
\label{sec:rg}
%~~~~~~~~~~~~~~~~~~~~~~~~~~~~~~~~~~~~~~~~~~~~~~

Having introduced the models we can now turn to analyzing their low energy dynamics. From Eq.~\eqref{eq:susyaction} it is clear that the superpotential is relevant and will drive the theory away from the free field limit. We now review some of the standard arguments which allow us to control this flow, using the non-renormalization theorems to argue that other tensor contractions which would spoil melonic dominance in the large $N$ limit are not induced, and  further comment on the properties of the IR fixed point.

%~~~~~~~~~~~~~~~~~~~~~~~~~~~~~~~~~~~~~~~~~~~~~~~
\subsection{Non-renormalization of the superpotential}{}
\label{sec:nrt}
%~~~~~~~~~~~~~~~~~~~~~~~~~~~~~~~~~~~~~~~~~~~~~~

Our focus on $\mathcal{N}=(2,2)$ supersymmetry owes to the non-renormalization of the superpotential. The standard argument using holomorphy says that the any quantum correction should be a holomorphic function of the chiral superfields and the uplift of coupling constants to  (background) chiral superfields. The theories are also characterized by left and right moving $R$-symmetries which we combine into a vector $R$-symmetry ${\rm U}(1)_V$ and an axial $R$-symmetry ${\rm U}(1)_A$.  The former controls terms we can write down in the superpotential  (the latter will be temporarily irrelevant). We may in addition have other non-anomalous global flavour symmetries in the problem (in addition to the symmetry group $G$). The charge assignments relevant for our models are summarized in 
Table~\ref{tab:Rfsym}.

\begin{table}[h]
\centering
\begin{tabular}{| c | c | c | c | c | c |}
\hline
& ${\rm U}(1)_{V}$ & ${\rm U}(1)_{A}$ & ${\rm U}(1)_{L}$ & ${\rm U}(1)_{R}$ & ${\rm U}(1)_f$ 
\\
\hline
$\theta^{+}$ & $1$ & $ 1$ & 0 & 2 & $0$ 
\\
$\theta^{-}$ & $1$ & $-1$ & 2 & 0 & $0$ 
\\
$\overline{\theta}^{+}$ & $-1$ & $-1$ & 0 & $-2$ & $0$ 
\\
$\overline{\theta}^{-}$ & $-1$ & $1$ & $-2$ & 0 & $0$ 
\\
\hline
$\mathscr{O} = \{\mathscr{B}, \mathscr{X}, \mathscr{Y}\}$ & $\frac{1}{2}$ & $0$ & ${1\over 4}$ & ${1\over 4}$ & $1$ 
\\	 
$g$ & 0 & 0 & 0 & 0 &  $-4$
\\
\hline
\end{tabular}
\caption{Charge assignments under the $R$-symmetries and global symmetries relevant for the non-renormalization argument.}
\label{tab:Rfsym}
\end{table}

For this discussion we will not need to specifically distinguish the isotropic and anisotropic models. It will therefore suffice for us to talk about a single quartic coupling $g$ for the most part of our discussion. Under renormalization group flow, the IR effective superpotential must be a holomorphic function in $g$ and $\mathscr{O} $ and should  have ${\rm U}(1)_V$ R-charge $2$ and be neutral under the 
${\rm U}(1)_f$ flavor symmetry. We can then immediately write down an ansatz for the effective superpotential at energy scale $\mu$ :
\begin{equation}
W_\text{eff} = \; g \, f(\mathscr{O} ),
\label{eq:Weff}
\end{equation}	
where $f(\mathscr{O} )$ is a homogeneous holomorphic function of homogeneity degree 4. 
For example, some terms in the effective superpotential are
\ie
f(\mathscr{O} ) = a_{4}(\mu)\left(\mathscr{O} ^4\right)_\text{melonic} +  b_{4}(\mu)\left(\mathscr{O} ^4\right)_\text{other contractions} +   \cdots
\fe
where by $(\mathscr{O} ^4)_\text{melonic}$ we refer to the index contractions presented in \eqref{eq:colorW}-\eqref{eq:mvW}, as well as the one appearing in anisotropic superpotential \eqref{eqn:deformed_superpotential}, depending on the specific model in question.  All other contractions between the fields are lumped into $\left(\mathscr{O} ^4\right)_\text{other contractions}$ and a few of them are summarized in Table \ref{tab:4pids}.

\begin{table}[h]
\centering
	\begin{tabular}{| c | c || c |}
	\hline
		& Uncolored Tensor & Matrix-vector model \\
	\hline
Pillow 
	&  $\mathscr{X}^{a_1a_2a_3}\mathscr{X}^{b_1a_2a_3}\mathscr{X}^{a_1b_2b_3}\mathscr{X}^{b_1b_2b_3}$
	& $\Tr{(\mathscr{Y}^I\, \mathscr{Y}^I \,\mathscr{Y}^J \,\mathscr{Y}^J)}$
	\\
Double sum/trace 
	& $\mathscr{X}^{a_1a_2a_3}\mathscr{X}^{a_1a_2a_3}\mathscr{X}^{b_1b_2b_3}\mathscr{X}^{b_1b_2b_3}$
	&
	$\left(\Tr{(\mathscr{Y}^I \mathscr{Y}^I)}\right)^2 $\\
	\hline
	\end{tabular}
\caption{Quartic monomials that can be constructed from our tensor-valued fields.}
\label{tab:4pids}
\end{table}

For the colored tensor, the non-renormalization theorem is immediate: there are no holomorphic quartic terms that are possible given the representation content. Only the tetrahedral term of \eqref{eq:colorW} is admissible and matching the result with the UV superpotential we conclude that the coupling is not renormalized.

The effective superpotential in the weak coupling limit $g\to 0$ must match with the UV bare superpotential. In particular, $a_{4}(\mu) =1$ and the other coefficients are zero. This establishes that the quartic tetrahedral superpotential is unrenormalized both for the isotropic and the anisotropic models. 

%~~~~~~~~~~~~~~~~~~~~~~~~~~~~~~~~~~~~~~~~~~~~~~~
\subsection{Renormalization of the K\"ahler potential}{}
\label{sec:Rkahler}
%~~~~~~~~~~~~~~~~~~~~~~~~~~~~~~~~~~~~~~~~~~~~~~

It will be important that the K\"ahler potential does get renormalized. The higher order corrections to the K\"ahler potential cannot involve more derivatives than the canonical kinetic term $\overline{\mathscr{O}}\mathscr{O}$. All of these would have positive mass dimensions, so naively one would expect such corrections to be irrelevant in the IR. Therefore, the IR effective K\"ahler potential admits a schematic expansion as
\ie
K_{\rm eff}(\mathscr{O},\overline{\mathscr{O}})=Z_2\, \overline{\mathscr{O}}\,\mathscr{O}+Z_4\, (\overline{\mathscr{O}}\,\mathscr{O})^2 +\cdots.
\label{eqn:kahler_expand}
\fe
If the bosonic potential generated by the superpotential $W_4(\mathscr{O})$ has a unique minimum so that the classical moduli space is trivial, then the RG flow would generate finite positive anomalous dimension for the superfield $\mathscr{O}$. Hence, the higher order terms in the expansion \eqref{eqn:kahler_expand} are more irrelevant than the leading term, and can be ignored in the IR. One therefore broadly expects the theory to have only wavefunction renormalization, and the IR dynamics be dominated by the superpotential.

Let us focus on the uncolored tensor model as the matrix-vector models can be understood as a special case of this argument. The IR effective K\"ahler potential takes the general form 
\ie
K_{\rm eff}(\mathscr{X},\overline{\mathscr{X}})= \sum_{a_1,a_2,a_3,b_1,b_2,b_3=1}^N\, Z_{a_1a_2a_3,b_1b_2b_3}\,\overline{\mathscr{X}}^{a_1a_2a_3}\, \mathscr{X}^{b_1b_2b_3}.
\label{eqn:effective_kahler}
\fe
The $\bZ^{3N}_2$ symmetry of the anisotropic model constrains the $Z_{a_1a_2a_3,b_1b_2b_3}$ to be
\ie
Z_{a_1a_2a_3,b_1b_2b_3}& =
Z_{a_1a_2a_3} \; \delta_{a_1b_1}\delta_{a_2b_2}\delta_{a_3b_3}.
\label{eqn:Z3N_constraint}
\fe
For the isotropic model, the ${\rm O}(N)^3$ symmetry further constrains $Z_{a_1a_2a_3} = Z$. To normalize the kinetic term, we define the renormalized fields
\ie\label{eqn:renor_superfield}
\widetilde{\mathscr{X}}^{a_1a_2a_3}=\sqrt{Z_{a_1a_2a_3}}\; \mathscr{X}^{a_1a_2a_3}.
\fe
The superpotential \eqref{eqn:deformed_superpotential} can be rewritten in terms of the renormalized fields as 
\ie
W_4(\mathscr{X}) = \frac{1}{4}\, \sum_{a_1,a_2,a_3,b_1,b_2,b_3=1}^N \widetilde\A_{a_1b_1,a_2b_2,a_3b_3}\;  \widetilde{\mathscr{X}}^{a_1a_2a_3}\,
\widetilde{\mathscr{X}}^{a_1b_2b_3}\,\widetilde{\mathscr{X}}^{b_1a_2b_3}\,\widetilde{\mathscr{X}}^{b_1b_2a_3},
\fe
where the `physical coupling' $\widetilde \A_{a_1b_1,a_2b_2,a_3b_3} $ is given by
\ie
\widetilde \A_{a_1b_1,a_2b_2,a_3b_3}=g\, \A_{a_1b_1,a_2b_2,a_3b_3} \left(Z_{a_1a_2a_3}\,Z_{a_1b_2b_3}\, Z_{b_1a_2b_3}\,
Z_{b_1b_2a_3}\right)^{-\frac{1}{2}}.
\fe
We note in particular that the wavefunction renormalization preserves the tetrahedral contraction structure in the superpotential. This ensures the  melonic dominance we seek in the large $N$ limit. In \S\ref{sec:aam}, we will show that in the IR conformal limit the physical coupling $\widetilde \A_{a_1b_1,a_2b_2,a_3b_3}$ is independent of the bare coupling $g$ which sets the overall interaction strength. Therefore, the anisotropic deformations induce exactly marginal deformations in the IR fixed point. In particular, we have an IR conformal manifold which is a projective space parametrized by the (projective) coordinates $\A_{a_1b_1,a_2b_2,a_3b_3}$.

There is a potential subtlety with this argument. The tetrahedral superpotentials we have written down could have a non-trivial moduli space of classical vacua. This is obviously the case for the colored tensor model, where we have a quartic monomial obtained from contracting four different fields. It is less obvious for the uncolored models, but one can explicitly demonstrate their existence for the isotropic potentials \eqref{eq:uncolorW} and \eqref{eq:mvW} (as we do in \S\ref{sec:moduli}). Moreover, given the homogeneity of our superpotential, it is clear that the classical moduli space is non-compact.

In two spacetime dimensions we should be integrating over this moduli space. The flat directions comprise physical degrees of freedom and, being gapless, dominate the IR dynamics. The details of what happens when we do so, depends on the moduli space geometry, with a potential danger of 
destabilizing the fixed point that we naively inferred above (eg., by developing a dynamical mass gap). This will indubitably happen unless we land upon a moduli space which admits a Ricci flat metric. Should this be the case, we would end up with an IR fixed point, which may nevertheless have a continuous spectrum from the non-compact directions.

Ideally, therefore, we would  like to construct models with an isolated classical vacuum. We will discuss this in more detail in \S\ref{sec:moduli}, and argue that the flat directions can be lifted by turning on generic enough anisotropic deformation in the superpotential \eqref{eqn:deformed_superpotential}. For now, we will carry out the naive analysis at large $N$, before turning to the question about removing all flat directions. The reader is urged to bear these caveats in mind as we undertake our first pass at solving these models.

%~~~~~~~~~~~~~~~~~~~~~~~~~~~~~~~~~~~~~~~~~~~~~~~
\subsection{IR fixed point}
\label{sec:IRfix}
%~~~~~~~~~~~~~~~~~~~~~~~~~~~~~~~~~~~~~~~~~~~~~~

A nice feature of the $\mathcal{N}=(2,2)$ Landau-Ginzburg models is that they are expected to flow to a superconformal fixed point given the above non-renormalization arguments \cite{Witten:1993jg}. Typically, the argument for the IR fixed point is made by appealing to supersymmetry protected quantities, such as the chiral ring \cite{Vafa:1988uu}, or the elliptic genus \cite{Witten:1993jg}. This is usually the case where we have a strong coupling fixed point which lies outside the purview of perturbation theory. For the melonic theories however, we will have the happy advantage of being able to carry out a large $N$ analysis and examine the spectrum of the fixed point explicitly. To set the stage for this discussion, let us note some salient and well-known facts about Landau-Ginzburg models. 

For the theory to attain an IR fixed point, the superpotential must transform quasi-homogeneously $W \to \Lambda^{-1} \, W$ under a scaling of the fields and couplings. Requiring the coupling be marginal fixes the scaling dimensions of all the fields. For the quartic superpotential we immediately conclude that 
\begin{equation}
\Delta (\mathscr{O} ) = \frac{1}{4}.
\label{eq:DeltaX4}
\end{equation}	
One can equivalently arrive at this conclusion by noting the $R$-charge assignments in Table~\ref{tab:Rfsym} and using the emergent superconformal symmetry. 

Furthermore, general arguments from the $R$-symmetry anomaly matching and the structure of ${\cal N}=2$ superconformal algebra lead to the IR central charge  \cite{Vafa:1988uu,Witten:1993jg}. The ${\rm U}(1)_R\times{\rm U}(1)_L$ $R$-symmetry flows to the ${\rm U}(1)_k\times {\rm U}(1)_k$ current algebra of the ${\cal N}=(2,2)$ superconformal algebra. The level $k$ of the current algebra can be determined by the ${\rm U}(1)_R$ symmetry anomaly matching, where each supermultiplet contributes $(J_R-1)^2-J_R^2$. Also, by the ${\cal N}=(2,2)$ superconformal algebra, the central charge $c$ is related to the level $k$ by $c=3k$. Putting everything together, we find 
\begin{equation}
\begin{split}
&c = 3k = {3\over 2}\, \times(\text{number of chiral superfields})
\\
&\Longrightarrow \;\;
\{c (\mathscr{B}), c(\mathscr{X}), c(\mathscr{Y}) \} =  \frac{3}{2} \big\{ 4 N^3,   N^3,  N^2\, M\big\}.
\label{eq:c4}
\end{split}
\end{equation}	
We will independently verify these central charges by solving the four-point function and extracting the contribution of the stress-tensor.

%~~~~~~~~~~~~~~~~~~~~~~~~~~~~~~~~~~~~~~~~~~~~~~~
\section{Explicit analysis of low energy fixed point}
\label{sec:diagrammatics}
%~~~~~~~~~~~~~~~~~~~~~~~~~~~~~~~~~~~~~~~~~~~~~~

As noted above, we expect that the RG flow lands us on a superconformal fixed point (modulo subtleties with flat directions of the superpotential). The simplicity of the melonic models is that we can check the properties of the fixed point explicitly in large $N$ perturbation theory.  For the isotropic models, our task is made even simpler by the fact that the analysis has already been carried out in the literature in the related context of disordered SYK models in \cite{Murugan:2017eto} and especially \cite{Bulycheva:2018qcp} which analyzes $(2,2)$ models in two dimensions. We simply need to adapt the results to the case at hand. For the most part we will be brief and only note some salient points of the analysis, referring the reader to  \cite{Bulycheva:2018qcp} for further details, though we will also take the opportunity to comment on some technical issues in the computation of the four-point function.

For the anisotropic models,  correlation functions in the large $N$ limit are computed by the same set of Feynman diagrams as in the isotropic models. However, there is more structure to uncover here since the anisotropy coefficients enter non-trivially into various computations. This will be particularly important in the computation of the four-point function using ladder diagrams. 

\begin{table}[h]
\centering
	\begin{tabular}{| c | c | c | c |}
	\hline
		& $\mathscr{O} _q$ & Symmetry & $c$ \\
	\hline
		Colored & $(\mathscr{B}_{a})_{i_{0a} \ldots i_{(a-1)a}}^{i_{a(a+1)}\ldots i_{a(q-1)}} \equiv \mathscr{B}_a^{(q)} $ & $U(N)^{(q+1)(q+2)/2}/\mathbb{Z}_{2}^{(q-1)(q+2)/2}$ & $ 3\, q\, \left( 1-\frac{2}{q}\right) \, N^{q-1}$ \\
		Uncolored & $\ \mathscr{X}^{a_1\, a_2 \, \ldots a_{q-1}} \equiv \mathscr{X}^{A_q} $ & $O(N)^{q-1}$ & 
		$3 \left( 1-\frac{2}{q}\right) \, N^{q-1}$ \\
		Matrix-vector & 	$\mathscr{Y}^{a\,I_1\,I_2\,\cdots I_{q-3}}_{\;b} \equiv \mathscr{Y}^{I_q}$ & 
		$SU(N)\times O(M)^{q-3}$
		 &  
		$3 \left( 1-\frac{2}{q}\right) N^2\, M^{q-3}$ \\
	\hline
	\end{tabular}
\caption{Generalization to models with q-fold interactions.}
\label{tab:qfolddata}
\end{table}
We directly work in superspace and, since it is straightforward to consider arbitrary $q$-body interaction in the superpotential, upgrade to tensors with rank $q-1$ transforming under a symmetry $G_q$. We summarize some of the relevant data for these models in 
Table~\ref{tab:qfolddata}. We will not write out the superpotential explicitly, apart from noting that for $q>6$ there may  potentially be multiple index contraction structures (cf., \cite{Gubser:2018yec}) that guarantee melonic dominance in the suitable large $N$ limit. We will assume for the sake of simplicity that we have picked one such term in writing the superpotential (eg., the maximally single-trace term of \cite{Klebanov:2019jup}). 

%~~~~~~~~~~~~~~~~~~~~~~~~~~~~~~~~~~~~~~~~~~~~~~~
\subsection{Isotropic model}
%~~~~~~~~~~~~~~~~~~~~~~~~~~~~~~~~~~~~~~~~~~~~~~

We begin our discussion with the isotropic models. As there is no broad difference between the three classes of models we introduced, we will simply analyze them en masse. We first review the basic Schwinger-Dyson equations which gives us the information about the low energy fixed point and then turn to the computation of the four-point function of the chiral superfields.

%~~~~~~~~~~~~~~~~~~~~~~~~~~~~~~~~~~~~~~~~~~~~~~~
\subsubsection{Two-point function}
\label{sec:sd}
%~~~~~~~~~~~~~~~~~~~~~~~~~~~~~~~~~~~~~~~~~~~~~~

The starting point for understanding the low energy dynamics is the two-point function of the superfield $\mathscr{O}$. We have the superspace correlation function:
\begin{equation}
\mathcal{G}_q({\sf Z}_{12}) = \langle \overline{\mathscr{O} }_q({\sf Z}_{1})\, \mathscr{O} _q({\sf Z}_{2}) \rangle,
\label{eq:superG}
\end{equation}	
where ${\sf Z}_{i} = (z_{i},\overline{z}_{i},\theta^\pm_{i},\overline{\theta}^\pm_{i})$ is the superspace coordinate (we work in Euclidean spacetime). The contribution to this two-point function can be obtained from the leading melonic diagrams, which by the standard analysis leads to the large $N$ super-Schwinger-Dyson equation 
\begin{equation}
 D_{+1}D_{-1}\mathcal{G}_q({\sf Z}_{13})+J_q^2\, \int d^{2}z_{2}\,d^{2}\theta_{2}\; \mathcal{G}_q({\sf Z}_{12})\mathcal{G}_q({\sf Z}_{32})^{q-1} = \overline{\theta}_{13}^{+}\overline{\theta}_{13}^{-}\, \delta(\Theta_{13})\, \delta(\overline{\Theta}_{13}),
\label{eq:sd2}
\end{equation}	
where we have defined the supertranslation invariant combinations  
\begin{equation}
\Theta_{12} = z_{12}+2\overline{\theta}_{1}^{+}\theta_{2}^{+}+\theta_{1}^{+}\overline{\theta}_{1}^{+}+\theta_{2}^{+}\overline{\theta}_{2}^{+} \qquad \mathrm{and} \qquad \overline{\Theta}_{12} = \overline{z}_{12}+2\overline{\theta}_{1}^{-}\theta_{2}^{-}+\theta_{1}^{-}\overline{\theta}_{1}^{-}+\theta_{2}^{-}\overline{\theta}_{2}^{-}.
\label{eq:stinv}
\end{equation}	

The coupling constant $J$ is the melonic analog of the 't Hooft coupling at large $N$ and is given by:
\begin{equation}
J_q^2 = 
\begin{cases}
& g^2\, N^{q-1} \,, \quad \quad\; \textrm{colored and uncolored tensor} \\
& g^2\, N\, M^\frac{q-2}{2} \,, \quad \textrm{matrix-vector}
\end{cases}
\label{eq:Jdef}
\end{equation}	

It is easy to convince oneself that there is a low energy solution to the above of the scaling form, obtainable by dropping the contribution from the K\"ahler term. The superconformal ansatz 
\begin{equation}
\mathcal{G}_q({\sf Z}_{12}) = \frac{b_q}{\Theta_{12}^{\Delta_q} \; \overline{\Theta}_{12}^{\Delta_q}},
\label{eq:sconfansatz}
\end{equation}	
solves \eqref{eq:sd2} in this limit with 
\begin{equation}
\Delta_q = \frac{1}{q} \,, 
\qquad 
\mathrm{and} \qquad 
b_q = \frac{1}{(4\pi^2\, J_q^2)^{1/q}}.
\label{eq:Delbq}
\end{equation}	
This is precisely the conformal dimension expected from $R$-symmetry. The superfield $\mathscr{O}$ has ${\rm U}(1)_{L} \times {\rm U}(1)_{R}$ charge given by $(Q_{L},Q_{R}) = (\frac{1}{q},\frac{1}{q})$, which is preserved under RG flow. In the IR superconformal field theory, the superfield $\mathscr{O}$ corresponds to a chiral primary operator so $(\Delta_{q},\overline{\Delta}_{q}) = (Q_{L},Q_{R})$, in perfect agreement with the solution of the super-Schwinger-Dyson equation.

%~~~~~~~~~~~~~~~~~~~~~~~~~~~~~~~~~~~~~~~~~~~~~~
\subsubsection{Four-point function}
\label{sec:spectrum}
%~~~~~~~~~~~~~~~~~~~~~~~~~~~~~~~~~~~~~~~~~~~~~~

We can explore further properties of the model, in particular the spectral data for certain low lying conformal primaries. The essential idea is to compute the four-point function of the superfields $\langle \overline{\mathscr{O} }({\sf Z}_1)\, \mathscr{O} ({\sf Z}_2) \, 
\overline{\mathscr{O} }({\sf Z}_3)\, \mathscr{O}({\sf Z}_4)  \rangle$ and decompose this into an OPE expansion. The computation turns out to be tractable as the four-point function in the melonic theory is captured by doing a ladder resummation \cite{Maldacena:2016hyu,Kitaev:2017awl}. The result actually can be obtained from the eigenvalue of a certain conformal kernel \cite{Bulycheva:2018qcp} which ends up being equal (up to a sign) to the kernel for the bosonic superoperator channel in the two-dimensional $\mathcal{N} = (1,1)$ SYK model analyzed in \cite{Murugan:2017eto}. We will focus on the uncolored model, but the results are almost identical for all of the models with the only difference being factors of $N$.

Specifically, the four-point function of interest is
\begin{equation}
\frac{\sum_{A^{q},B^{q}} \ \langle \overline{\mathscr{X}}^{A_{q}}({\sf Z}_{1})\mathscr{X}^{A_{q}}({\sf Z}_{2})\overline{\mathscr{X}}^{B_{q}}({{\sf Z}}_{3})\mathscr{X}^{B_{q}}({\sf Z}_{4})\rangle}{ \sum_{A^{q}}\; \langle \overline{\mathscr{X}}^{A_{q}}({\sf Z}_{1})\mathscr{X}^{A_{q}}({\sf Z}_{2})\rangle\   \sum_{B^{q}}\;\langle\overline{\mathscr{X}}^{B_{q}}({\sf Z}_{3})\mathscr{X}^{B_{q}}({\sf Z}_{4})\rangle} = 1+\frac{1}{N^3}\mathcal{F}({\sf Z}_1, {\sf Z}_2,{\sf Z}_3,{\sf Z}_4)+{\cal O}(N^{-4})\label{eq:fullfourptfn}.
\end{equation}	
The first subleading term $\mathcal{F}({\sf Z}_1, {\sf Z}_2,{\sf Z}_3,{\sf Z}_4)$ can be computed by an infinite sum of ladder diagrams,
\ie\label{eqn:sum_ladders}
&{\cal F}({\sf Z}_1, {\sf Z}_2,{\sf Z}_3,{\sf Z}_4) = \sum_{n=0}^\infty{\cal F}_n({\sf Z}_1, {\sf Z}_2,{\sf Z}_3,{\sf Z}_4) ,
\\
&{\cal F}_n({\sf Z}_1, {\sf Z}_2,{\sf Z}_3,{\sf Z}_4)=\int d {\sf Z}_1' d{\sf Z}_2' \; K_q({\sf Z}_1, {\sf Z}_2;{\sf Z}_1', {\sf Z}_2')\; {\cal F}_{n-1}({\sf Z}_1', {\sf Z}_2',{\sf Z}_3,{\sf Z}_4).
\fe
The kernel is given by
\begin{equation}
K_q({\sf Z}_1,{\sf Z}_2;{\sf Z}_3,{\sf Z}_4) = 
(q-1)\, J_{q}^2\, \mathcal{G}_q({\sf Z}_{31}) \, 
\mathcal{G}_q({\sf Z}_{24}) \, \mathcal{G}_q({\sf Z}_{34})^{q-2}.
\label{eq:kernel4}
\end{equation}	
The series can be formally resummed as
\ie
{\cal F}={1\over 1-K}{\cal F}_0.
\fe
The above formal expression can be made more precise by expanding the right hand side in terms of the eigenfunctions of the kernel \eqref{eq:kernel4}. In the conformal limit, the kernel \eqref{eq:kernel4} commutes with the superconformal Casimir operators, and the eigenfunctions become superconformal partial waves
\ie
\Xi_{h,\bar{h}}(\chi,\overline{\chi})&={h\bar{h}\, \sin\pi h\over 2\cos\pi\bar{h}}\;
\bigg(\varphi_h(\chi)\, \varphi_{\bar{h}}(\overline \chi)-\varphi_{-h}(\chi)\,\varphi_{-\bar{h}}(\overline \chi)\bigg),
\\
\varphi_{ h}( \chi)&=F_h(\chi)-F_{h+1}(\chi),\qquad F_h(\chi)={\Gamma(h)^2\over \Gamma(2h)}\; \chi^h\; {}_2F_1(h,h,2h;\chi),
\fe
where the super-cross-ratios are given by
\ie
\chi = \frac{\Theta_{12}\Theta_{34}}{\Theta_{14}\Theta_{32}} \qquad \mathrm{and} \qquad \overline{\chi} = \frac{\overline{\Theta}_{12}\overline{\Theta}_{34}}{\overline{\Theta}_{14}\overline{\Theta}_{32}}.
\fe
The inner product of the zero-rung ladder diagram $\mathcal{F}_{0}$ with the superconformal partial wave gives
\ie
\langle \Xi_{h,\bar{h}},\mathcal{F}_{0} \rangle = (-1)^{\bar{h}-h}\frac{4\pi^2\Delta_{q}}{\Delta_{q}-1}k(h,\bar{h}),
\fe
where $k(h,\bar{h})$ is the eigenvalue of the kernel $K$. The normalization of the superconformal partial wave is
\ie
\langle \Xi_{h,\bar{h}},\Xi_{h',\bar{h}'} \rangle = (2\pi)^{4}\, h\bar{h}\bigg(\delta_{ll'}\,\delta(s-s')+\delta_{-ll'}\,\delta(s+s')\bigg),
\fe
where the conformal dimensions are parametrized as
\ie
h={\ell\over 2}+is,\quad \bar{h}=-{\ell\over 2}+is.
\fe
Putting all of this together, the four-point function can be explicitly written as
\begin{equation}
\begin{split}
&\mathcal{F}(\chi,\overline{\chi}) = \sum_{h,\bar{h}}\Xi_{h,\bar{h}}\; {1\over 1-k(h,\bar{h})}\; {\la\Xi_{h,\bar{h}}, {\cal F}_0\ra\over \la\Xi_{h,\bar{h}}, \Xi_{h,\bar{h}}\ra} 
\\
&\quad\quad\quad\,=\frac{1}{4\pi}\frac{\Delta_{q}}{\Delta_{q}-1} \; \sum_{\ell \in \mathbb{Z}}\int_{-\infty}^{\infty}\frac{ds}{2\pi}\,(-1)^{\bar{h}-h}\;\frac{k(h,\bar{h})}{1-k(h,\bar{h})} \; {\sin\pi h\over \cos\pi\bar{h}}\;\varphi_h(\chi)\,\varphi_{\bar{h}}(\overline \chi)\,.
\end{split}
\label{eq:4pt_integral_formula}
\end{equation}

The eigenvalue of the kernel $k(h,\bar{h})$ is computed by\footnote{We have fixed some typos in the kernel eigenvalue computation in \cite{Bulycheva:2018qcp}, namely there is an extra factor of $(-1)$ from the Grassmann integration and there is a factor $(-1)^{h-\bar{h}}$ from swapping the chiral and anti-chiral operators in the eigenfunction. We thank Ksenia Bulycheva for helpful correspondence on this issue.}
\begin{equation}
\begin{split}
k(h,\overline {h}) & \ \langle \mathscr{X}^{A_q}({\sf Z}_1)\overline{\mathscr{X}}^{A_q}({\sf Z}_2)\mathcal{O}_{h,\bar{h}}(\infty)\rangle  \\
&= \int d^{2}z_{3}\,d^{2}z_{4}\,d^{2}\overline{\theta}_{3}\,d^{2}\theta_{4}\;K_q({\sf Z}_1,{\sf Z}_2;{\sf Z}_3,{\sf Z}_4) \; \langle \mathscr{X}^{A_q}({\sf Z}_4)\overline{\mathscr{X}}^{A_q}({\sf Z}_3)\mathcal{O}_{h,\bar{h}}(\infty)\rangle.
\end{split}
\label{eq:kevals_setup}
\end{equation}
We use the superconformal algebra $\mathfrak{su}(1,1|1)\oplus\mathfrak{su}(1,1|1)$ to set $Z_{1} = (0,0,0)$ and $Z_{2} = (1,0,0)$ in order to simplify the integral. We then find that \eqref{eq:kevals_setup} evaluates to
\begin{equation}
\begin{split}
 k(h,\overline {h}) &=
 	 \int d^{2}z_{3}\,d^{2}z_{4}\,d^{2}\overline{\theta}_{3}\,d^{2}\theta_{4}\;
 	K_q(1,0;{\sf Z}_3,{\sf Z}_4)\; \langle \mathscr{X}^{A_q}({\sf Z}_4)\overline{\mathscr{X}}^{A_q}({\sf Z}_3)\mathcal{O}_{h,\bar{h}}(\infty)\rangle\\	
&= 
 	(q-1)J_{q}^2b_{q}^{q}\int d^{2}z_{3}\,d^{2}z_{4}\,d^{2}\overline{\theta}_{3}\,d^{2}\theta_{4}\;
 	\frac{1}{|\Theta_{31}|^{2\Delta_{q}}|\Theta_{24}|^{2\Delta_{q}}|\Theta_{34}|^{2(q-2)\Delta_{q}}}
 	\frac{1}{\Theta_{34}^{\Delta_{q}-h}\overline{\Theta}_{34}^{\Delta_{q}-\bar{h}}}\\	
&= 
 	n_{q}(h,\bar{h})\int d^{2}z_{3}\,d^{2}z_{4}\,
 		\frac{z_{34}^{h}\, \overline{z}_{34}^{\bar{h}}}{|z_{3}|^{2\Delta_{q}}|1-z_{4}|^{2\Delta_{q}}|z_{34}|^{4-2\Delta_{q}}}\\	
&= 
	n_{q}(h,\bar{h})\int d^{2}z_{4}\,\frac{z_{4}^{h}\overline{z}_{4}^{\bar{h}}}{|z_{4}|^{2}|1-z_{4}|^{2\Delta_{q}}}\int d^{2}\widetilde{z}_{3}\,\frac{(\widetilde{z}_{3}-1)^{h}(\overline{\widetilde{z}}_{3}-1)^{\bar{h}}}{|\widetilde{z}_{3}|^{2\Delta_{q}}|\widetilde{z}_{3}-1|^{4-2\Delta_{q}}} \qquad \;\; 
	\text{with} \;\; \widetilde{z}_{3} = \frac{z_{3}}{z_{4}}\\	
&= 
	n_{q}(h,\bar{h})(-1)^{h-\bar{h}}\int d^{2}z_{4}\,\frac{z_{4}^{h}\overline{z}_{4}^{\bar{h}}}{|z_{4}|^{2}|1-z_{4}|^{2\Delta_{q}}}\int d^{2}\widetilde{z}_{3}\,\frac{(1-\widetilde{z}_{3})^{h}(1-\overline{\widetilde{z}}_{3})^{\bar{h}}}{|\widetilde{z}_{3}|^{2\Delta_{q}}|1-\widetilde{z}_{3}|^{4-2\Delta_{q}}},
\end{split}
\label{eq:kevals_comp}
\end{equation}
where in the last line we have rotated $\widetilde{z}_{3}$ around $1$ by $\widetilde{z}_{3} \rightarrow 1+(1-\widetilde{z}_{3})e^{i\pi}$, 
$\overline{\widetilde{z}}_{3} \rightarrow 1+(1-\overline{\widetilde{z}}_{3})e^{-i\pi}$, which gives the additional factor $(-1)^{h-\bar{h}}$. We have also defined
\ie
n_{q}(h,\bar{h})=\frac{1-\Delta_{q}}{\pi^2\Delta_{q}}(h+\Delta_{q}-1)(\bar{h}+\Delta_{q}-1).
\fe
Evaluating this product of two integrals by standard techniques gives the kernel eigenvalue
\begin{equation}
k(h,\bar{h}) = (-1)^{h-\bar{h}}\Delta_{q}(\Delta_{q}-1)\frac{\Gamma^2(-\Delta_{q})}{\Gamma^2(\Delta_{q})}\frac{\Gamma(-h+\Delta_{q})\Gamma(\bar{h}+\Delta_{q})}{\Gamma(1-h-\Delta_{q})\Gamma(1+\bar{h}-\Delta_{q})}.
\label{eq:kevals}
\end{equation}
These eigenvalues are related to the boson-boson superoperator channel eigenvalues $k_{BB}$ in the $\mathcal{N}=(1,1)$ 2d SYK model \cite{Murugan:2017eto} by the relation: $k(h,\bar{h}) = (-1)^{h-\bar{h}}k_{BB}(h,\bar{h})$.

The OPE of the superfield ${\mathscr{X}}$ with its conjugate can be studied by expanding the four-point function at the point $\chi=\overline\chi=0$. To compute such an expansion, we need to deform the contour of the integral formula \eqref{eq:4pt_integral_formula} in a way that the integral becomes a sum over residues of the poles of the integrand. The contour is chosen to be in the complex $s$-plane along the negative imaginary axis, which we close toward $u = is \in \mathbb{R}_{\geq 0}$ for the convergence of the integral.  There are physical poles in \eqref{eq:4pt_integral_formula} coming from 
\ie\label{eqn:eigenvalue_eq}
k(h,\bar{h}) = 1.
\fe
The locations of them give the spectrum of superconformal primaries that appear in the $\overline{\mathscr{X}}\times {\mathscr{X}}$ OPE, and the residues give the squared of the OPE coefficients. There are potentially solutions to the equation \eqref{eqn:eigenvalue_eq} with $h$ and $\bar h$ outside the range $h \geq 0$ and $\bar{h} \geq 0$. Such solutions would violate the unitarity of the theory. We will show in Appendix \ref{sec:unitarity} that solutions violating the unitarity bound do not exist (essentially by bounding the kernel eigenvalue). A related issue is that we require the conformal weights to be real, and thus also need  to check that there are no poles with $s$ having a non-vanishing real part. We have checked this numerically for $|\ell| \leq 100$ in a large range of $u$ and we do not find any such poles.

In addition to the physical poles, there are various spurious poles coming from $h=\bar{h}=0$ and from the zeros of $\cos \pi\bar{h}$ at $\bar{h} \in \frac{1}{2}+\mathbb{Z}$. The former spurious pole is removed by infinitesimally deforming the contour away from $s=0$ in the complex $s$-plane as discussed in \cite{Bulycheva:2018qcp}, but the latter spurious poles are more subtle. We will demonstrate in \S\ref{sec:spuriouspoles} that the latter set of poles delicately cancel amongst themselves by adapting the argument given in \cite{Murugan:2017eto}. For now we focus our attention on the physical poles.

We find that the equation \eqref{eqn:eigenvalue_eq} has solutions $(h,\bar{h})= (1,0)$ and $(0,1)$, which correspond to the holomorphic and anti-holomorphic supercurrent supermultiplets. The holomorphic supercurrent supermultiplet is organized as 
\ie
\mathcal{J} = \mathcal{R}+\theta^{+}S+\overline{\theta}^{+}\overline{S}+\theta^{+}\overline{\theta}^{+}T,
\fe
where $\mathcal{R}$, $S$ and $T$ are the $R$-current, supercurrent, and the stress-tensor, respectively. Unlike the $1d$ SYK model, the superconformal primary $\mathcal{J}$ only contributes a single pole to the four-point function. Therefore once we pull the contour to pick up the pole, there is no divergence in the four-point function due to this operator. As discussed in \cite{Murugan:2017eto} this implies that the low energy dynamics is not solely determined by this multiplet alone.  From a holographic perspective, this implies that the dual theory is akin to classical string theory (at large $N$), which does not truncate to supergravity. 

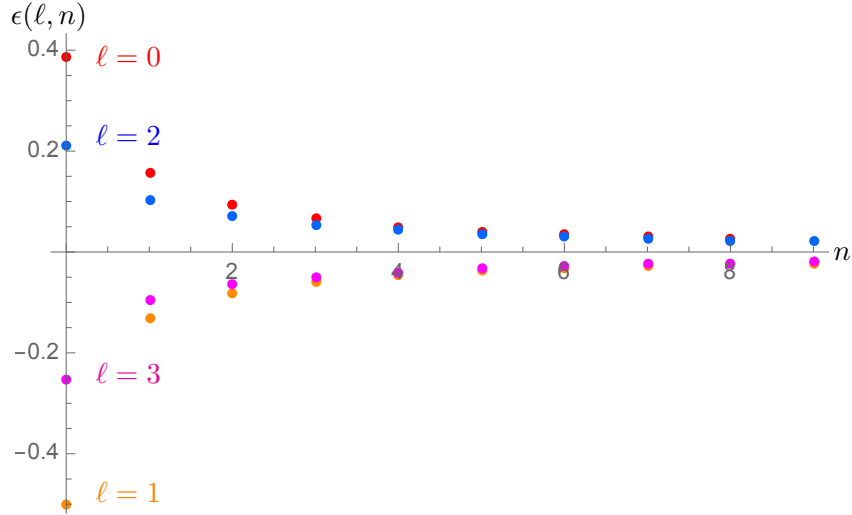
\begin{figure}[h!]
\centering
\includegraphics[width=0.7\linewidth]{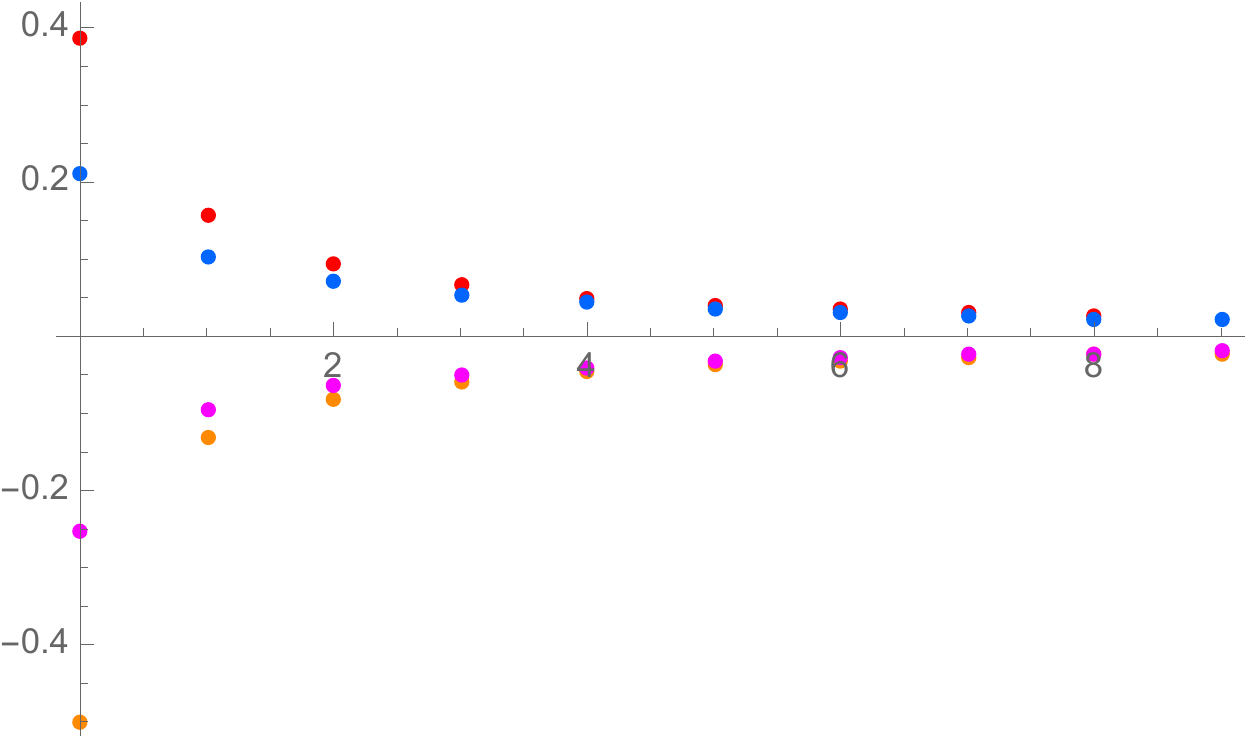}
\begin{picture}(0,0)
\put(0,100){\makebox(0,0){$n$}}
\put(-300,190){\makebox(0,0){$\epsilon(\ell,n)$}}
\put(-270,175){\makebox(0,0){$\color{red}{\ell=0}$}}
\put(-270,145){\makebox(0,0){$\color{blue}{\ell=2}$}}
\put(-270, 55){\makebox(0,0){$\color{magenta}{\ell=3}$}}
\put(-270,10){\makebox(0,0){$\color{orange}{\ell=1}$}}
\end{picture}
\caption{ A plot of the anomalous dimensions  $\epsilon(\ell,n)$ of the composite operators described in \eqref{eq:compops} in the spectrum of the isotropic model.}
\label{fig:spectrumeps}
\end{figure}

The operator spectrum can be organized into `Regge trajectories': sequences of operators with increasing spins and an approximately fixed twist. More precisely, the twist $\tau=\Delta - |\ell|$ ($\Delta = h+\bar{h}$ and $\ell=h-\bar{h}$) of the operators take the form
\ie
\tau = 2\Delta_q + 2n + \epsilon(\ell,n),
\label{eqn:Regge_twists}
\fe
where $n \in \bZ_{\ge 0}$ labels the different Regge trajectories, and $\epsilon(\ell,n)$ is a nonzero function approaching zero in the large $\ell$ or large $n$ limit. The operators with twists \eqref{eqn:Regge_twists} can be identified with the composite operators in the asymptotically free UV theory
\ie \label{eq:compops}
{\mathscr X}^{A_q}(D_+)^{2s+2n}(D_-)^{2n} {\mathscr X}^{A_q}
\fe
and $\epsilon(\ell,n)$ is the anomalous dimension in the IR generated by the RG flow. We plot $\epsilon(\ell,n)$ for low-spin values in Fig.~\ref{fig:spectrumeps}.

We can also compute the central charge explicitly and check that it agrees with the analysis from the chiral algebra as presented in Table~\ref{tab:qfolddata}. In the OPE, the stress tensor contributes the ${\cal O}(\chi^2)$ terms to the four-point function with the coefficient related to the central charge of the theory. More explicitly, in the four-point function $\mathcal{F}(\chi,\bar \chi)$ expanded at $\chi=\overline\chi=0$, there exists a term
\begin{equation}
\label{eqn:c_in_4pt}
\frac{\Delta^2}{2c}\chi^2 \in \frac{1}{N^3}\mathcal{F}.
\end{equation}	
Since the (anti)holomorphic stress tensor is in the same supermultiplet as the (anti)holomorphic $R$-symmetry current, the central charge can be read of from the $\chi$ expansion of the residue at that point at $k(1,0)=1$. The result matches with the central charges displayed in Table \ref{tab:qfolddata}.

Finally, let us note that the knowledge of the Euclidean four-point function computed above, is sufficient to obtain the out-of-time-ordered  Lorentzian  thermal four-point function at inverse temperature $\beta$ that probes the scrambling and chaotic dynamics of the theory. Thermal correlation functions  can be conformally mapped to vacuum correlation functions on $\bR^2$. Using this map, it follows that the chaos limit for the out-of-time-ordered Lorentzian thermal four-point function is equivalent to the Regge limit of an analytically continued vacuum four-point function \cite{Murugan:2017eto}. 
By this relation, the chaos exponent can be easily computed using the eigenvalue of the kernel \eqref{eq:kevals}.  Analytically continuing the spin in the $n^{\rm th}$ Regge trajectory to intersect the principal series at
\ie
h={1\over 2}(1+\ell),\quad \bar{h}={1\over 2}(1-\ell),
\fe
we obtain the Regge intercept $\ell_n$. The chaos exponent $\lambda_L$ is related to the leading Regge intercept by 
\ie
\lambda_L = (\ell_0-1){2\pi\over \beta}.
\fe
For our model, the leading Regge intercept is roughly
\ie
\ell_0\approx 1.55,
\fe
which leads to a sub-maximal chaos exponent. The intuition for why we find sub-maximal chaos exponent is that all the operators on the Regge trajectory containing the stress-tensor lead to growing contributions in the out of  time-ordered correlator (OTOC) so the stress-tensor does not dominate the OTOC, as explained in \cite{Murugan:2017eto}.

%~~~~~~~~~~~~~~~~~~~~~~~~~~~~~~~~~~~~~~~~~~~~~~
\subsubsection{Cancellation of spurious poles}
\label{sec:spuriouspoles}
%~~~~~~~~~~~~~~~~~~~~~~~~~~~~~~~~~~~~~~~~~~~~~~

We have discussed the four-point function above, under the assumption that the only poles of relevance in evaluating \eqref{eq:4pt_integral_formula} are those coming from solving \eqref{eqn:eigenvalue_eq}.  To finish off, we need to demonstrate that the other poles of the integrand, which we refer to as the spurious poles,  cancel amongst themselves. We can adapt for our purposes the discussion in  \cite{Murugan:2017eto} where they argue for a similar cancellation in the $\mathcal{N} = (1,1)$ SYK model. Modulo some differences, the essence of the argument follows along similar lines -- we show that the poles cancel in pairs once we have suitable chosen a contour for the integration in the $s$-plane. We find it however useful to assemble the pieces in a slightly different manner to simplify the argument.

Let us first see where the spurious poles are located. Ignoring the factor $k/(1-k)$, the integrand of the four-point function \eqref{eq:4pt_integral_formula} is
\ie
I_{h,\bar{h}}(\chi,\overline\chi) = (-1)^{\bar{h}-h}{\sin\pi h\over \cos\pi \bar{h}}\varphi_{h}(\chi)\varphi_{\bar{h}}(\overline \chi).
\label{eq:4-ptfnintegrand}
\fe
To identify the poles, it is helpful to rewrite the ${\cal N}=2$ superconformal partial wave $\varphi_{h}(\chi)$ as
\ie
\varphi_h(\chi)=\Gamma(h)^2\, \chi^h \; {}_2\widetilde{F}_1(h,h,2h;\chi)-\Gamma(h+1)^2\, \chi^{2h+2} \; {}_2\widetilde{F}_1(h+1,h+1,2(h+1);\chi),
\label{eq:N=2superconfpartwave_reg}
\fe
where $\widetilde F_1(a,b,c;\chi) \equiv F_1(a,b,c;\chi)/\Gamma(c)$ is the regularized hypergeometric function, which has the benefit of being regular (in particular, unlike the hypergeometric function it has no poles at $c \in {\mathbb Z}_{\leq 0}$). We  then rewrite the integrand as
\ie
I_{h,\bar{h}}(\chi,\overline\chi) = (-1)^{\bar{h}-h}\; {\sin\pi h\over \cos\pi \bar{h}} \; \Gamma(h)\, \Gamma(\bar{h})\; {\varphi_{h}(\chi)\over \Gamma(h)} \; {\varphi_{\bar{h}}(\overline \chi)\over \Gamma(\bar{h})}\,,
\label{eq:integrand_rewrite}
\fe
to make the singularity structure manifest.  For one, the function ${\varphi_{h}(\chi)\over \Gamma(h)}$ has no poles and its zeros are not at $h\in {1\over 2}\bZ$. Therefore, the poles and zeros of the integrand $I_{h,\bar{h}}(\chi,\overline\chi)$ come from $\frac{\sin \pi h}{\cos \pi \bar{h}}\Gamma(h)\Gamma(\bar{h})$, which are summarized as
\ie
&{\rm poles}:\,\bar{h}\in{1\over 2}+\bZ,\, \bZ_{\le 0},
\\
&{\rm zeros}:\, h\in \bZ_{\ge 1}.
\label{eq:poles+zeros}
\fe
Since we close the contour so that $u = is \in \mathbb{R}_{\geq 0}$, we are interested in the following domain in the $(h,\bar{h})$ plane:
\ie
h-\bar{h}\in\bZ,\quad h+\bar{h}\ge 0.
\label{eq:confdimrestriction}
\fe
Therefore, we are only interested in the poles at
\ie
(h,\bar h)=(0,0)\quad {\rm and}\quad (h,\bar h)\in \bigg({1\over 2},{1\over 2}\bigg)+\bZ^2, \quad (h+\bar{h}\ge 0)\,.
\label{eq:polesofinterest}
\fe
As discussed previously, we deform the contour to avoid the pole at $(h,\bar h)=(0,0)$. 

It remains to show that the residues of the poles at $(h,\bar h)\in ({1\over 2},{1\over 2})+\bZ^2$ cancel. Let us examine the properties of the integrand for  $h = m+\frac{1}{2}$ and $\bar{h} = n+\frac{1}{2}$. The residues of the integrand $I_{h,\bar{h}}(\chi,\overline\chi)$ at these locations are
\ie
&{(-1)^{1+m+n}\over\pi}\, \varphi_{m+{1\over 2}}(\chi) \, \varphi_{n+{1\over 2}}(\overline\chi) \,, \qquad m,n\in\bZ.
\label{eq:residues}
\fe
Let us define
\ie
\phi_h(\chi)=\Gamma(h)^2\, \chi^h\ {}_2\widetilde F_1(h,h,2h;\chi),
\label{eq:superconfpartwavepiece}
\fe
which has the property
\ie
\phi_{n+{1\over 2}}(\chi)=\phi_{-n+{1\over 2}}(\chi),
\label{eq:superconfpartwavepiece_property}
\fe
where we have used the following identity for the regularized hypergeometric function
\ie
{}_2 \widetilde F_1(a,b,-n;z)=z^{n+1}\, (a)_{n+1}\, (b)_{n+1}\ {}_2 \widetilde F_1(a+n+1,b+n+1,n+2;z), \qquad n \in \mathbb{Z}_{\geq 0}.
\label{eq:reghypergeo_identity}
\fe
Thus, we find the property of the $\mathcal{N}=2$ superconformal partial wave:
\ie
\varphi_{m+{1\over 2}}(\chi) = \phi_{m+{1\over 2}}(\chi)-\phi_{m+{3\over 2}}(\chi) = \phi_{-m+{1\over 2}}(\chi)-\phi_{-m-{1\over 2}}(\chi)=-\varphi_{-m-{1\over 2}}(\chi).
\label{eq:superconfpartwave_property}
\fe
Furthermore, one can verify the following symmetry of the kernel when $h,\bar{h} \in \frac{1}{2}+\mathbb{Z}$:
\ie
k(-h,\bar{h}) = k(h,-\bar{h}) = k(h,\bar{h}).
\fe

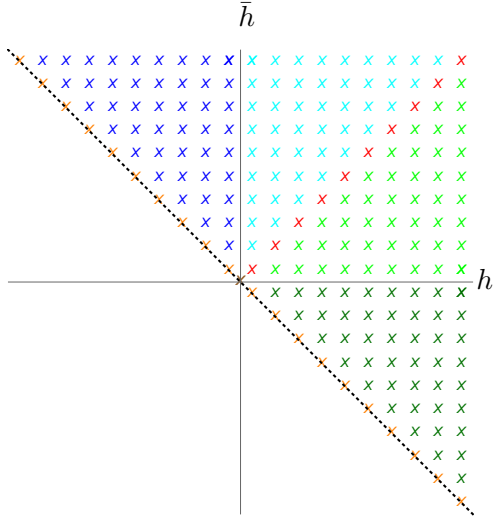
\begin{figure}[h!]
\centering
\includegraphics[width=0.4\linewidth]{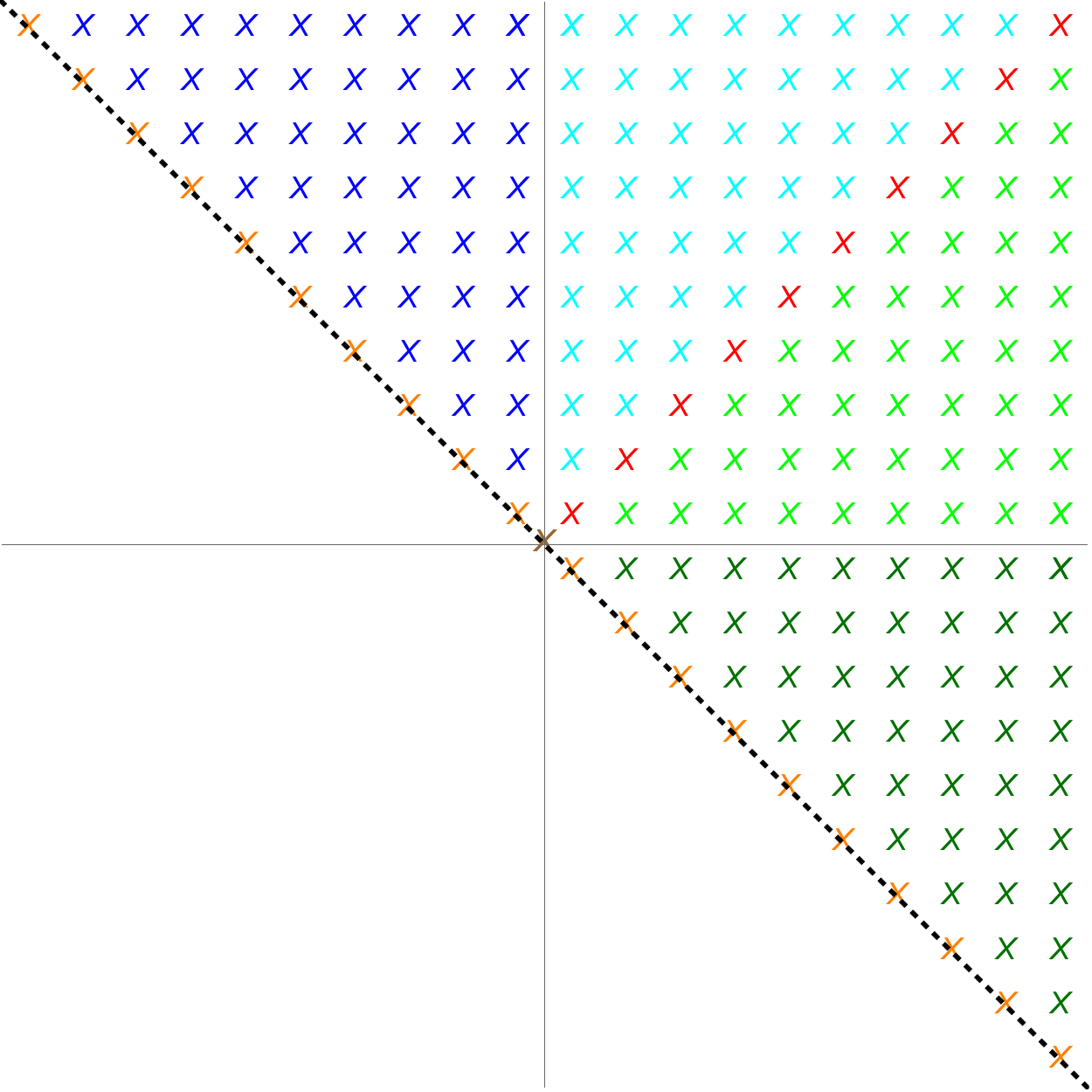}
\begin{picture}(0,0)
\put(0,90){\makebox(0,0){$h$}}
\put(-90,190){\makebox(0,0){$\bar{h}$}}
\end{picture}
\caption{The spurious poles in the $(h,\bar{h})$-plane: the light green poles above the $h$-axis cancel the dark green poles below the $h$-axis, the light blue poles to the right of the $\bar{h}$-axis cancel the dark blue poles to the left of the $\bar{h}$-axis, and the orange poles along the line $h+\bar{h}=0$ cancel the red poles along the line $h=\bar{h}$. We deform the contour away from the brown pole at $h=\bar{h}=0$. }
\label{fig:spuriouspoles}
\end{figure}

Putting together all these pieces we can now see how the residues of the poles at $(h,\bar h)\in ({1\over 2},{1\over 2})+\bZ^2$ cancel. Denoting the residues at these locales by $\mathrm{Res}(h,\bar{h})$ we see that in the half-plane $h+\bar{h} > 0$, the poles cancel by virtue of the relations
\begin{equation}
\begin{split}
\mathrm{Res}(h,\bar{h}) &= -\mathrm{Res}(h,-\bar{h}), \qquad h > \bar{h}
\\ \mathrm{Res}(h,\bar{h}) &= -\mathrm{Res}(-h,\bar{h}), \qquad \bar{h} > h.
\end{split}
\label{eq:polecancel}
\end{equation}
This cancellation can be seen visually in Fig.~\ref{fig:spuriouspoles}. 

We are left with having to address the poles at $h+\bar{h} = 0$ ($\ell \neq 0$) that lie along the integration contour and the poles at $h = \bar{h}$. The trick is to rewrite the contour integral along the real $s$ axis in the four-point function \eqref{eq:4pt_integral_formula} as 
\ie
\int_{-\infty}^{\infty} ds \rightarrow \frac{1}{2}\bigg(\int_{-\infty+i\epsilon}^{\infty+i\epsilon} ds+\int_{-\infty-i\epsilon}^{\infty-i\epsilon} ds\bigg), \qquad \ell \neq 0.
\label{eq:contourdeform}
\fe
The first term in the sum avoids all the poles at $h+\bar{h} = 0$ and the second term in the sum picks up all the poles at $h+\bar{h} = 0$. We reiterate that this contour deformation is only for the $\ell \neq 0$ case while for $\ell = 0$ ($h=\bar{h}=0$) we always deform the contour to avoid the pole at $s=0$. We can now see how the poles at $h+\bar{h} = 0$ and $h = \bar{h}$ cancel:
\ie
\frac{1}{2}\bigg(\mathrm{Res}(h,-h)+\mathrm{Res}(-h,h)\bigg) = -\mathrm{Res}(h,\bar{h}).
\label{eq:polecancel_edge}
\fe
We conclude that all the spurious poles in the four-point function \eqref{eq:4pt_integral_formula} cancel.

%~~~~~~~~~~~~~~~~~~~~~~~~~~~~~~~~~~~~~~~~~~~~~~~
\subsection{Anisotropic model}
\label{sec:aam}
%~~~~~~~~~~~~~~~~~~~~~~~~~~~~~~~~~~~~~~~~~~~~~~

We now turn to the anisotropic models which are of primary interest to us. Fortunately, we can use the results for the isotropic models discussed above, with suitable modifications, to quickly infer the answers in this case. The main novelty we will find is that the anisotropy induces a non-trivial convolution between the position space  and the index contraction structures in  the correlation functions. For the two-point function the change is mild, but the four-point function will turn out to be more involved.  In what follows, we first derive the deformed Schwinger-Dyson equations, and then write down a formula for the deformed four-point function by resumming the ladder diagrams. 

%~~~~~~~~~~~~~~~~~~~~~~~~~~~~~~~~~~~~~~~~~~~~~~~
\subsubsection{Two-point function}
\label{sec:anisotropic_SD}
%~~~~~~~~~~~~~~~~~~~~~~~~~~~~~~~~~~~~~~~~~~~~~~

\begin{figure}[H]
\centering
\subfloat{
\begin{tikzpicture}
\draw[red, thick] (0,0) circle (1.75cm);

\draw[red, thick] (-3,0) node[left]{$a_1$} to (-1.9,0)
(-1.6,0) to (1.6,0)
(1.9,0) to (3,0);
\draw[blue, thick] (-3,0.3) node[left]{$a_3$};
\draw [blue, thick,domain=7.5:172.5] (0,0) plot ({2*cos(\x)}, {2*sin(\x)});
\draw [blue, thick] (-3,{2*sin(172.5)}) to ({2*cos(172.5)},{2*sin(172.5)});
\draw [blue, thick] (3,{2*sin(7.5)}) to ({2*cos(7.5)},{2*sin(7.5)});

\draw[green, thick] (-3,-0.3) node[left]{$a_2$};
\draw [green, thick,domain=-7.5:-172.5] (0,0) plot ({2*cos(\x)}, {2*sin(\x)});
\draw [green, thick] (-3,{2*sin(-172.5)}) to ({2*cos(-172.5)},{2*sin(-172.5)});
\draw [green, thick] (3,{2*sin(-7.5)}) to ({2*cos(-7.5)},{2*sin(-7.5)});

\draw [green, thick,domain=10:170] (0,0) plot ({1.5*cos(\x)}, {1.5*sin(\x)});
\draw [green, thick] ({1.5*cos(10)},{1.5*sin(10)}) to ({1.5*cos(170)},{1.5*sin(170)});

\draw [blue, thick,domain=-10:-170] (0,0) plot ({1.5*cos(\x)}, {1.5*sin(\x)});
\draw [blue, thick] ({1.5*cos(-10)},{1.5*sin(-10)}) to ({1.5*cos(-170)},{1.5*sin(-170)});

\draw[green, thick] (0.3,0.5) node[left]{$b_2$};
\draw[blue, thick] (0.3,-0.5) node[left]{$b_3$};
\draw[red, thick] (0.3,1.73) node[left]{$b_1$};
\end{tikzpicture}
}
\caption{The two-loop supergraph contributing to the self-energy $\Sigma^{a_1a_2a_3}$ drawn in triple-line notation.}
\label{fig:2-loop_2-point_supergraph}
\end{figure}
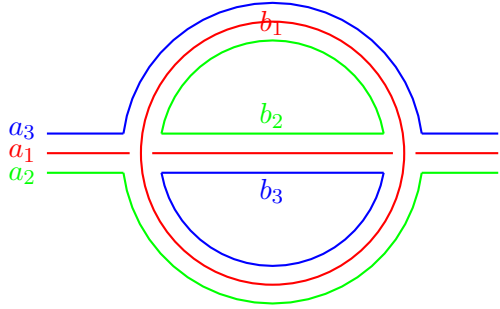

We start by considering the two-point function \eqref{eq:superG}, which in the large $N$ limit can be computed by the same set of melonic diagrams as in the isotropic model. The super-Schwinger-Dyson equation with the anisotropic deformation is
\ie
&D_{+1} \overline D_{-1} {\cal G}^{a_1a_2a_3}({\sf Z}_{13})+J^2\int d^2z_2 d^2 \theta_2 {\cal G}^{a_1a_2a_3}({\sf Z}_{12})\Sigma^{a_1a_2a_3}({\sf Z}_{32})=\overline{\theta}_{13}^{+}\overline{\theta}_{13}^{-}\delta(\Theta_{13})\overline\delta(\Theta_{13}),
\\
&\Sigma^{a_1a_2a_3}({\sf Z}_{12})={1\over N^3}\sum_{b_1,b_2,b_3=1}^N |\A_{a_1b_1,a_2b_2,a_3b_3}|^2  {\cal G}_{a_1b_2b_3}({\sf Z}_{12}) {\cal G}_{b_1a_2b_3}({\sf Z}_{12}) {\cal G}_{b_1b_2a_3}({\sf Z}_{12}).
\fe
In the conformal limit, we drop the first term of the first equation, and assume the conformal ansatz
\ie
{\cal G}^{a_1a_2a_3}({\sf Z}_{12})={b \over \Theta^{\Delta}_{12}\overline\Theta^{\Delta}_{12}}\beta_{a_1a_2a_3}.
\fe
The super-Schwinger-Dyson equation gives
\ie\label{eqn:beta_eq}
\Delta = \frac{1}{4}\,,\quad \beta^{-1}_{a_1a_2a_3}={1\over N^3}\sum_{b_1,b_2,b_3=1}^N|\A_{a_1b_1,a_2b_2,a_3b_3}|^2 \beta_{a_1b_2b_3}\beta_{b_1a_2b_3}\beta_{b_1b_2a_3}.
\fe
The right hand side of the second equation in \eqref{eqn:beta_eq} is coming from the self-energy $\Sigma^{a_1a_2a_3}$. The index structure represented in the triple-line notation is shown in Fig.~\ref{fig:2-loop_2-point_supergraph} where the closed loop correspond to the indices being summed over (viz., $b_1$, $b_2$, and $b_3$).

The wavefunction renormalization in the effective K\"ahler potential \eqref{eqn:effective_kahler}, \eqref{eqn:Z3N_constraint} is given by
\ie\label{eqn:2pt_wave_renor}
Z_{a_1a_2a_3}=b^{-{1\over 2}}\, \beta^{-{1\over 2}}_{a_1a_2a_3}.
\fe
The physical couplings are therefore 
\ie
\widetilde\A_{a_1b_1,a_2b_2,a_3b_3}&=
	g\, \A_{a_1b_1,a_2b_2,a_3b_3}\, b^2\, \sqrt{\beta_{a_1a_2a_3}\, \beta_{a_1b_2b_3}\, \beta_{b_1a_2b_3}\,
	\beta_{b_1b_2a_3}}
\\
&={1\over 2\pi N^{3\over 2}}\, \A_{a_1b_1,a_2b_2,a_3b_3}\,
	 \sqrt{\beta_{a_1a_2a_3}\,\beta_{a_1b_2b_3}\,\beta_{b_1a_2b_3}\,\beta_{b_1b_2a_3}}\,.
\fe
As promised, the physical couplings are independent of the overall coupling $g$. As a consequence we have a low energy conformal manifold, which is a projective space parametrized by the (projective) coordinates $\A_{a_1b_1,a_2b_2,a_3b_3}$. However, there is a caveat to this statement:  we must remove from this space the choices of $\A_{a_1b_1,a_2b_2,a_3b_3}$ for which the superpotential \eqref{eqn:deformed_superpotential} has flat directions.

%~~~~~~~~~~~~~~~~~~~~~~~~~~~~~~~~~~~~~~~~~~~~~~~
\subsubsection{Four-point function}
\label{sec:anisotropic_SD}
%~~~~~~~~~~~~~~~~~~~~~~~~~~~~~~~~~~~~~~~~~~~~~~

\begin{figure}[H]
\centering
\subfloat{
\begin{tikzpicture}

\draw[red, thick] (-3,2)  to (3,2);
\draw[red, thick] (-3,-2) to (3,-2);

\draw[green, thick] 
(-3,1.5) to (-.5,1.5)
(-.5,1.5) to[out=210,in=150] (-.5,-1.5)
(-.5,-1.5) to (-3,-1.5)  node[left]{$a_2$};

\draw[green, thick] 
(3,1.5) node[right]{$b_2$} to (.5,1.5)
(.5,1.5) to[out=330,in=30] (.5,-1.5)
(.5,-1.5) to (3,-1.5);

\draw[red, thick] 
(0,1.5) to[out=330,in=30] (0,-1.5)
(0,1.5) to[out=210,in=150] (0,-1.5);
\draw[color=red, thick] 
(0,2) node[above]{$a_1$}
(0.2,0) node[right]{$c_1$} 
(0,-2) node[below]{$b_1$};

\draw[blue, thick] 
(3,1.75) to (0.7,1.75)
(0.7,1.75) to[out=180,in=30] (-.25,1.5)
(-.25,1.5) to[out=210,in=150] (-.25,-1.5)
(-.25,-1.5) to[out=330,in=180] (.7,-1.75)
(.7,-1.75) to (3,-1.75) node[right]{$b_3$};
{}
\draw[blue, thick] 
(-3,1.75)  node[left]{$a_3$} to (-0.7,1.75)
(-0.7,1.75) to[out=0,in=150] (-.25,1.675)
(.25,1.5) to[out=330,in=30] (.25,-1.5)
(-0.7,-1.75) to[out=0,in=210] (-.25,-1.675)
(-.7,-1.75) to (-3,-1.75);
\end{tikzpicture}
}
\caption{The one-rung ladder supergraph computing the matrix $M$ in \eqref{eq:mmatrix} in the triple-line notation.}
\label{fig:2-loop_2-point_supergraph_1rung}
\end{figure}
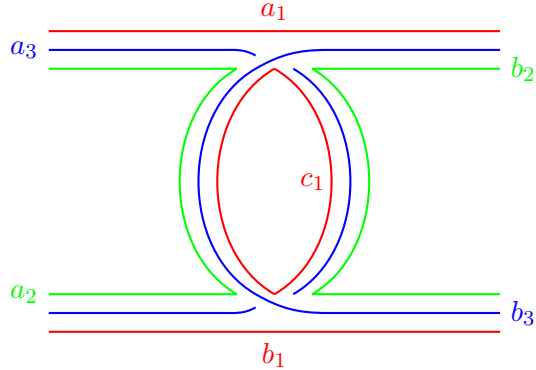

While the story for the two-point function was reasonably similar to the isotropic case, the four-point function analysis is significantly affected by the presence of anisotropy. After turning on the anisotropic deformation, the sum over ladder diagrams in \eqref{eqn:sum_ladders} becomes
\ie\label{eqn:sum_ladders_anisotropic}
{\cal F}({\sf Z}_1, {\sf Z}_2,{\sf Z}_3,{\sf Z}_4) = \sum_{n=0}^\infty r_n(\A) {\cal F}_n({\sf Z}_1, {\sf Z}_2,{\sf Z}_3,{\sf Z}_4) ,
\fe
where each ladder diagram is weighted by the factor $r_n(\A)$, which is a function of the anisotropic deformation parameters. More explicitly, $r_n(\A)$ are obtained by the following iterative formula
\ie
r_0(\A)&=N^{-3}\sum_{a_1,a_2,a_3=1}^N\beta_{a_1a_2a_3}^2,
\\
r_n(\A)&=N^{-3-2n}\sum_{a_1,a_2,a_3,b_1,b_2,b_3=1}^N\delta_{a_1b_1}(M_{a_1b_1}^n)_{a_2a_3}{}^{b_2b_3}\beta_{a_1b_2b_3}\beta_{b_1b_2b_3},\quad n>0,
\fe
where $M_{a_1b_1}^n$ denotes the $n^{\rm th}$ power of the matrix $M_{a_1b_1}$ with the entries given by
\ie
(M_{a_1b_1})_{a_2a_3}{}^{b_2b_3}={1\over N}\beta_{a_1a_2a_3}\beta_{b_1a_2a_3}\sum_{c_1=1}^N\beta_{c_1a_2b_3}\beta_{c_1b_2a_3}\A_{a_1c_1,a_2b_2,a_3b_3}\A^*_{b_1c_1,a_2b_2,a_3b_3}\,.
\label{eq:mmatrix}
\fe
The index structure in the definition of this matrix $M$ can be easily visualized using  the triple-line notation as shown in Fig.~\ref{fig:2-loop_2-point_supergraph_1rung}. As before the convention is that the indices $a_1$, $b_1$, $c_1$ correspond to the red lines, $a_2$, $b_2$ correspond to the green lines, and $a_3$, $b_3$ correspond to the blue lines, respectively.

The series \eqref{eqn:sum_ladders_anisotropic} can be resummed formally as
\ie\label{eqn:F_rn_sum}
{\cal F} =A(\A,K){\cal F}_0, \qquad A(\A,k)=\sum_{n=0}^\infty r_n(\A)k^n.
\fe
$A(\A,k)$ is a homogeneous function of degree $-1$ in the anisotropic parameters, i.e.,
\ie\label{eqn:homogeneous_A}
A(\lambda\A,k)=\lambda^{-1}A(\A,k).
\fe
More explicitly, using the eigenfunctions of the Casimir operators, the four-point function can be written as
\begin{equation}
\begin{split}
&\mathcal{F}(\chi,\overline{\chi})=\frac{1}{4\pi}\frac{\Delta_{q}}{1-\Delta_{q}}\sum_{\ell \in \mathbb{Z}}\int_{-\infty}^{\infty}\frac{ds}{2\pi}\,k(h,\bar{h})A(\A,k(h,\bar{h})){\sin\pi h\over \cos\pi\bar{h}}\varphi_h(\chi)\,\varphi_{\bar{h}}(\overline \chi)\,.
\end{split}
\label{eq:deformed_4pt_integral_formula}
\end{equation}

The spectrum of composite operators in the theory can as before  be solved by the location of the poles of the function
\ie
A(\A,k(h,\bar h)),
\fe
where $k(h,\bar h)$ is the eigenvalue of the kernel $K$. Therefore, in general, the spectrum of operators would be deformed by the anisotropic deformation.  As discussed in \S\ref{sec:spectrum}, the chaos exponent can be computed by analytically continuing the spin of the leading Regge trajectory, and hence would also be deformed by the anisotropic deformation in general.

While we have not solved the resulting eigenvalue problem, there are special choices of the deformation parameter $\alpha$ for which we can obtain some  immediate results. For instance, in the $(\bZ_2^N\rtimes S_N)^3$ invariant anisotropic model \eqref{eqn:Z2xSN_model}, since there are only ${\cal O}(N^5)$ number of anisotropic deformation terms while there are ${\cal O}(N^6)$ number of terms in the isotropic superpotential, we expect that the effect of the anisotropic deformation on the two- and four-point functions would be suppressed by ${1\over N}$. In particular, we have
\ie\label{eqn:rn_ZS}
r_n(\A)=1+{\cal O}(N^{-1}).
\fe
Hence, in the $(\bZ_2^N\rtimes S_N)^3$ invariant anisotropic model, large $N$ scaling would imply that the spectrum and chaos exponent would not be deformed by the anisotropic deformation in the leading large $N$ limit. Another simple situation is when we can treat the anisotropy perturbatively; we next turn to analyze this case.

%~~~~~~~~~~~~~~~~~~~~~~~~~~~~~~~~~~~~~~~~~~~~~~~
\subsubsection{Infinitesimal anisotropic deformation}
\label{sec:infAD}
%~~~~~~~~~~~~~~~~~~~~~~~~~~~~~~~~~~~~~~~~~~~~~~~

As a particular example, which turns out to be quite tractable,  let us consider an infinitesimal anisotropic deformation of the isotropic tensor models. For simplicity, we further make the specialization that the deformation parameter factorizes across the three sub-indices $1$, $2$, and $3$ (or the red, green, and blue colors of the figures), viz.,
\ie
\A_{a_1b_1,a_2b_2,a_3b_3}&=\A_{a_1b_1}\, \A_{a_2b_2}\,\A_{a_3 b_3},
\\
\A_{ab}&=1+\epsilon_{ab},
\fe
for real and symmetric $\epsilon_{ab}$ and
\ie
1\gg \epsilon_{ab}\gg {1\over N}.
\fe
By explicit computation, we find the following results for the coefficients $r_n(\A)$ in \eqref{eqn:F_rn_sum},
\ie\label{eqn:r_coefficients}
r_n(\A)\big|_{{\cal O}(\epsilon)}&=-3(\epsilon)_1,
\\
r_0(\A)\big|_{{\cal O}(\epsilon^2)}&=-{3\over 2}(\epsilon^2)_1+12(\epsilon^2)_2-{9\over 2}(\epsilon^2)_3,
\\
r_n(\A)\big|_{{\cal O}(\epsilon^2)}&=-{3\over 2}(\epsilon^2)_1+10(\epsilon^2)_2-{5\over 2}(\epsilon^2)_3\quad{\rm for}\quad n\ge1,
\\
r_0(\A)\big|_{{\cal O}(\epsilon^3)}&=-36(\epsilon^3)_1-{9\over 2}(\epsilon^3)_2+60(\epsilon^3)_3-{35\over 2}(\epsilon^3)_4-24(\epsilon^3)_5+12(\epsilon^3)_6,
\\
r_1(\A)\big|_{{\cal O}(\epsilon^3)}&=-24(\epsilon^3)_1-{5\over 2}(\epsilon^3)_2+32(\epsilon^3)_3-{15\over 2}(\epsilon^3)_4-18(\epsilon^3)_5+10(\epsilon^3)_6,
\\
r_n(\A)\big|_{{\cal O}(\epsilon^3)}&=-28(\epsilon^3)_1-{5\over 2}(\epsilon^3)_2+40(\epsilon^3)_3-{23\over 2}(\epsilon^3)_4-18(\epsilon^3)_5+10(\epsilon^3)_6\quad{\rm for}\quad n\ge2.
\fe
where the various structures constructed from the $\epsilon_{ab}$ are
\ie
&(\epsilon)_1={1\over N^2}\sum_{a,b=1}^N\epsilon_{ab},&&(\epsilon^2)_1={1\over N^2}\sum_{a,b=1}^N\epsilon_{ab}\epsilon_{ba},
\\
&(\epsilon^2)_2={1\over N^3}\sum_{a,b,c=1}^N\epsilon_{ab}\epsilon_{bc},&&(\epsilon^2)_3={1\over N^4}\left(\sum_{a,b=1}^N\epsilon_{ab}\right)^2,
\\
&(\epsilon^3)_1={1\over N^4}\sum_{a,b,c,d=1}^N\epsilon_{ab}\epsilon_{bc}\epsilon_{cd},&&(\epsilon^3)_2={1\over N^4}\Big(\sum_{a,b=1}^N\epsilon_{ab}\epsilon_{ba}\Big)\Big(\sum_{c,d=1}^N\epsilon_{cd}\Big),
\\
&(\epsilon^3)_3={1\over N^5}\Big(\sum_{a,b,c=1}^N\epsilon_{ab}\epsilon_{bc}\Big)\Big(\sum_{c,d=1}^N\epsilon_{cd}\Big),&& (\epsilon^3)_4={1\over N^6}\Big(\sum_{a,b=1}^N\epsilon_{ab}\Big)^3,
\\
&(\epsilon^3)_5={1\over N^4}\sum_{a,b,c,d=1}^N\epsilon_{ad}\epsilon_{bd}\epsilon_{cd},&&(\epsilon^3)_6={1\over N^3}\sum_{a,b,c=1}^N\epsilon_{ac}\epsilon_{bc}\epsilon_{bc}.
\fe

The result suggests that the function $A(\A,k)$ in \eqref{eqn:F_rn_sum} takes the form
\ie\label{eqn:A_structure}
A(\A,k)={r_0(\A)\over 1-k}+\sum_{n,m=1}^\infty   \sum_i c_{m,n,i}(\epsilon^{m+n})_i k^n.
\fe
By the homogeneity of the function $A(\A,k)$ \eqref{eqn:homogeneous_A}, the coefficients $c_{m,i}$ must satisfy
\ie\label{eqn:sum_c=0}
\sum_ic_{m,n,i}=0.
\fe
From \eqref{eqn:r_coefficients}, we find the coefficients
\ie
&c_{1,1,1}=0\,,&& c_{1,1,2}=-2\,,&& c_{1,1,3}=2\,,
\\
&c_{2,1,1}=12\,,&& c_{2,1,2}=2\,,&& c_{2,1,3}=-28\,,&& c_{2,1,4}=10\,,&& c_{2,1,5}=6\,,&& c_{2,1,6}=-2\,,
\\
&c_{1,2,1}=8\,,&& c_{1,2,2}=2\,,&& c_{1,2,3}=-20\,,&& c_{1,2,4}=6\,,&& c_{1,2,5}=6\,,&& c_{1,2,6}=-2\,,
\fe
which indeed do satisfy \eqref{eqn:sum_c=0}. 

The location of the poles of $A(\A,k(h,\bar{h}))$ can be determined  by solving 
\ie
0=&{1\over A(\A,k(h,\bar{h}))}
\\
=&{1\over r_0(\A)}\left[1-k(h,\bar{h})\right]-{1\over r_0(\A)^2}\left[1-k(h,\bar{h})\right]^2k(h,\bar{h})\sum_{i=1}^3c_{1,1,i}(\epsilon^2)_i
\\
&-{1\over r_0(\A)^2}\left[1-k(h,\bar{h})\right]^2k(h,\bar{h})\sum_{i=1}^6\left[k(h,\bar{h})c_{1,2,i}+c_{2,1,i}\right](\epsilon^3)_i+{\cal O}(\epsilon^4),
\fe
where we have used the infinitesimal nature of $\epsilon$ to expand out the relation.  We see that the solutions to $k(h,\bar{h})=1$ remain the solution to the deformed equation at any finite order in the $\epsilon$-expansion. There are extra solutions located an order $\epsilon$ distance away from the poles of $k(h,\bar{h})$, at which, however, the $\epsilon$-expansion also breaks down. This is of course not surprising, since new solutions in perturbation theory can only arise at the scale set by the perturbation parameter. Therefore, to deform the spectrum of operators, we would have to turn on finite anisotropic deformations. We hope to return to this question in the future.

However, insofar as establishing that we have a reliable large $N$ fixed point without a non-trivial moduli space we have succeeded.  Note also that for the $(\bZ_2^N\rtimes S_N)^3$ invariant anisotropic model, the infinitesimal parameter $\epsilon_{ab}$ would be proportional to the Kronecker delta,
\ie
\epsilon_{ab}=\epsilon\,\delta_{ab}.
\fe
We verify our earlier observation that the fixed point spectrum is unchanged for  \eqref{eqn:rn_ZS} does indeed hold owing to $(\epsilon^n)_i={\cal O}(N^{-n})$.

%~~~~~~~~~~~~~~~~~~~~~~~~~~~~~~~~~~~~~~~~~~~~~~~
\section{Moduli space}
\label{sec:moduli}
%~~~~~~~~~~~~~~~~~~~~~~~~~~~~~~~~~~~~~~~~~~~~~~

Our analysis thus far has been confined to examining the large $N$ diagrammatic structure. Exploiting the melonic dominance, we have been able to argue that the models flow to an IR fixed point with the low energy dynamics essentially determined by the superpotential. We would now like to examine these theories more closely at finite $N$, which we are able to do thanks to the $(2,2)$ supersymmetry. The main question we would like to address is the reliability of our large $N$ analysis. As will become clear in the following, the theories we are considering typically have a number of flat directions, which as 
indicated above in \S\ref{sec:rg}, could potentially pose problems. We will start with the simple models analyzed above, and attempt to modify them while retaining the melonic large $N$ structure, in several steps. We will find that we can remove these flat directions by considering the anisotropic model with certain choices of the anisotropic deformation parameters.

%~~~~~~~~~~~~~~~~~~~~~~~~~~~~~~~~~~~~~~~~~~~~~~~
\subsection{Flat directions in the isotropic models}
\label{sec:flatlg}
%~~~~~~~~~~~~~~~~~~~~~~~~~~~~~~~~~~~~~~~~~~~~~~

Recall that the bosonic potential of our models is given by the  gradient squared of the holomorphic superpotential $W(\mathscr{O} )$, viz.,
\begin{equation}
V(\mathscr{O} ) = \bigg| \frac{\partial W}{\partial \mathscr{O} } \bigg|^2.
\label{eq:bospot}
\end{equation}	
We can quickly infer that the models constructed above have flat directions. 

\begin{itemize}
\item{\it Colored tensors:} We have $q$-independent tensor fields, $\{\mathscr{B}_a\}$, which appear exactly once in the monomial, and therefore
we have (schematically)
\begin{equation}
\frac{\partial W}{\partial \mathscr{B}_a} = \prod_{b\neq a} \, \mathscr{B}_a \,.
\label{eq:ctWgrad}
\end{equation}	
We can clearly minimize the potential by setting any two of the chiral superfields to zero, say $\mathscr{B}_{k}=\mathscr{B}_{l}=0$  for $k\neq l$. There are then 
$q-2$ flat directions parameterized by the other bottom components of the other superfields $\{\mathscr{B}_a\}_{a \neq k,l}$. Thus, the moduli space of vacua contains ${{q} \choose {2}}$ subspaces isomorphic to $\mathbb{C}^{q-2}$, and hence the potential has a large space of flat directions. 
One can consider variants of this simple model, but in each case we find them plagued with flat directions.
\item {\it Uncolored tensors \& matrix-vectors:} The gradient of the potential now transforms as a tensor under the group $G$. The bosonic potential in these cases reads:
\begin{equation}
V(\mathscr{X}) = \sum_{d,e,f =1}^N \, \big| \mathscr{X}^{ade} \, \mathscr{X}^{fbe}\, \mathscr{X}^{fdc}\big|^2 \,, \qquad 
V(\mathscr{Y}) = \sum_{I=1}^M\, \big| \mathscr{Y}^I\, \mathscr{Y}^K\, \mathscr{Y}^I \big|^2\,,
\label{eq:ucWgrad}
\end{equation}	
respectively in the two cases of interest.\footnote{ The notation $|\cdot|^2$ has an implicit contraction of the dangling index, e.g. $\sum_{I=1}^{M}\big| \mathscr{Y}^I\, \mathscr{Y}^K\, \mathscr{Y}^I \big|^2 = \sum_{I,J,K=1}^{M}(\mathscr{Y}^I\, \mathscr{Y}^K\, \mathscr{Y}^I)(\mathscr{Y}^{\ast J}\, \mathscr{Y}^{\ast K}\, \mathscr{Y}^{\ast J})$.} The potential is minimized at the origin, but the presence of flat directions can be inferred by noting that
\begin{equation}
\mathscr{Y}^1 = x \left(\begin{matrix} \sigma^{1} & 0 \\ 0 & \mathbf{0}_{N-2}\end{matrix}\right), 
\quad 
\mathscr{Y}^2  = x\left(\begin{matrix} \sigma^{2} & 0 \\ 0 & \mathbf{0}_{N-2}\end{matrix}\right),  \quad \mathscr{Y}^I = 0\,, \;\text{for}\  I \neq 1,2\,, 
\label{eq:mvmoduli}
\end{equation}	
with $x\in {\mathbb R}$ and $\mathbf{0}_{n}$ being the $n \times n$ zero matrix, also gives a vanishing potential. We additionally still have the freedom to conjugate 
$Y^I$ by a $U(N)$ matrix. We have not attempted to chart out the full moduli space of solutions though it is clear that it has a large dimension. 
Note that given a solution for the matrix-vector model we can immediately find an embedding for our tensors $\mathscr{X}^{abc}$, making it clear that we always have flat directions.\footnote{ For representations of $O(N,\mathbb{R}) \times O(M,\mathbb{R})$ or $O(N,\mathbb{R})^3$, we have checked that the origin is the unique vacuum up to $N=M=5$. However, these representations are real and we need complex representations for chiral superfields owing to holomorphy. We could consider representations of $O(N,\mathbb{C}) \times O(M,\mathbb{C})$ where fields transform in the adjoint of $O(N,\mathbb{C})$ and in the fundamental representation of $O(M,\mathbb{C})$. However, one can check that for $N=M=2$, $\mathscr{X}_{1} = i\, \mathscr{X}_{2}$ gives zeros of the bosonic potential and then this solution generalizes to all $N$ and $M$ in the same way as \eqref{eq:mvmoduli}.}

\end{itemize}
Thus in the simplest examples we have examined so far we always have some number of moduli, rendering the low-energy CFT non-compact and possibly unstable. We will now see that the anisotropic model lifts all the non-trivial moduli. 

%~~~~~~~~~~~~~~~~~~~~~~~~~~~~~~~~~~~~~~~~~~~~~~~
\subsection{Absence of moduli in the anisotropic tensor models}
\label{sec:trivialmodulianiso}
%~~~~~~~~~~~~~~~~~~~~~~~~~~~~~~~~~~~~~~~~~~~~~~

The anisotropic deformation was introduced to remove all moduli so that the theory has a unique classical vacuum given by the origin in field space. This crucially leads to the absences of higher order terms in the effective K\"ahler potential and thereby gives a stable fixed point in the IR as previously discussed in $\S$\ref{sec:Rkahler}. We will now prove that there indeed exist choices of the anisotropic deformation parameters $\A_{a_1b_1,a_2b_2,a_3b_3}$ such that the moduli space is trivial. The proof actually holds for any $q \geq 4$ tetrahedral superpotential, but we will restrict ourselves to $q=4$ to simplify the proof.

Consider the anisotropic superpotential $W_{4}$ defined in \eqref{eqn:deformed_superpotential} whose critical points define the classical moduli space $\mathcal{M}$. The critical points of $W_{4}$ are determined by the common zeros of the partial derivatives of $W_{4}$ with respect to the superfields $\mathscr{X}^{c_1c_2c_3}$:
\ie\label{eqn:anisomodulispace}
\mathcal{M} = \bigg\{\mathscr{X}^{a_1a_2a_3} \in \mathbb{C}^{N^3}\,\bigg|\,\frac{\partial W_{4}}{\partial \mathscr{X}^{c_1c_2c_3}}(\mathscr{X}^{a_1a_2a_3}) = 0\,\ \forall\;1 \leq c_1,c_2,c_3 \leq N\bigg\}.
\fe
The partial derivatives of $W_{4}$ are explicitly given by
\ie\label{eqn:partderivW4}
\mathfrak{f}_{c_1c_2c_3} = \frac{W_{4}}{\partial \mathscr{X}^{c_1c_2c_3}} = \sum_{b_{1},b_{2},b_{3}}\;\A_{c_1b_{1},c_2 b_{2},c_3 b_{3}}\,
\mathscr{X}^{c_1b_{2}b_{3}}\, \mathscr{X}^{b_{1}c_2b_{3}}\, \mathscr{X}^{b_{1}b_{2}c_3},
\fe
where we have used the symmetry \eqref{eqn:Z2xZ2} to simplify this expression. Then we state the claim that $\mathcal{M} = \{\mathbf{0}\}$ as follows:\\

\noindent \textbf{\underline{Theorem 1}}: There exist nonzero $\A_{a_{1}b_{1},a_{2}b_{2},a_{3}b_{3}}$ such that the set of equations
\ie\label{eqn:nomoduli_claim}
\{\mathfrak{f}_{c_1c_2c_3} = 0\,|\,1 \leq  c_1,c_2,c_3 \leq N\}
\fe
has no non-trivial solution. \\

We shall provide a sketch of the proof below relegating technical details to Appendix~\ref{sec:nomoduliproof}. 
The idea is to convert this set of non-linear equations to a linear algebra problem. This can be done by working in the space of monomials built from the components of our tensor field. We have $N^3$ equations given by the zero sets of the polynomials $\mathfrak{f}_{c_1c_2c_3}$, each of which is a linear combination of degree $3$ monomials in the $N^3$ variables $\mathscr{X}^{a_1a_2a_3}$. We however have many more monomials than equations, which is sub-optimal.

What we need to do is the following. We should find a related system of equations where the number of monomials equals the number of equations. Then working in the space of all monomials, we have a system of linear equations, easily visualized as an operator $\mathfrak{C}$ acting on the vector of  monomials of degree $d$. 

Consider the set $\mathcal{S}$ of all monomials of degree $d = 2N^3+1$ in the $N^3$ variables $\mathscr{X}^{a_1a_2a_3}$. The cardinality of this set is huge
\begin{equation}\label{eq:setS}
|\mathcal{S}| \equiv {\sf S}= {{3\,N^3}\choose{N^3-1}}
\end{equation}	
We can partition $\mathcal{S}$ into $N^3$ subsets $\mathcal{S}_{c_1c_2c_3} 
\subset \mathcal{S}$ and then construct $N^3$ sets of polynomials
\ie\label{eqn:polysdegd}
\mathcal{P}_{c_1c_2c_3} = \bigg\{\frac{\mathscr{X}^{\gamma}}{(\mathscr{X}^{c_1c_2c_3})^{3}} \; \mathfrak{f}_{c_1c_2c_3}\,\bigg|\,\mathscr{X}^{\gamma} \in \mathcal{S}_{c_1c_2c_3}\bigg\},
\fe
where $\mathscr{X}^{\gamma}$ denotes a monomial of degree $d$. Note that there is no summation over the indices $c_i$ in the above.
The union of these sets 
\ie\label{eqn:setofpolys}
\mathcal{P} = \bigcup_{c_1,c_2,c_3=1}^{N} \, \mathcal{P}_{c_1 c_2 c_3}
\fe
consists of $\sf{S}$ polynomials which are now linear combinations of the $\sf{S}$ monomials in the set $\mathcal{S}$. 

Now let $\mathbf{M}_{d}$ be the vector consisting of all elements of $\mathcal{S}$ and let $\mathfrak{C}$ be the $\sf{S} \times \sf{S}$ matrix of coefficients of the polynomials in $\mathcal{P}$. Then the common zero sets of all the polynomials in $\mathcal{P}$ defines a linear system of equations, which has the simple form
\ie\label{Pzeroset}
\mathfrak{C}\, \mathbf{M}_{d} = \mathbf{0}_{\sf{S}}\,,
\fe
where $\mathbf{0}_{\sf{S}}$ is the zero vector in $\mathbb{C}^{\sf{S}}$. Therefore, if the polynomials in $\mathcal{P}$ have a non-trivial common zero, then linear operator $\mathfrak{C}$ should have vanishing determinant, i.e., $\det(\mathfrak{C}) = 0$. The crucial observation is that if the system of equations $\{\mathfrak{f}_{c_1 c_2 c_3} = 0\}$ as in \eqref{eqn:nomoduli_claim} has a nontrivial solution $\mathscr{X}_{0}^{i_1i_2i_3} \neq \mathbf{0}_{N^{3}}$, then $\mathscr{X}_{0}^{i_1i_2i_3}$ is also a non-trivial common zero of the set of polynomials in $\mathcal{P}$, since we constructed $\mathcal{P}$ from the $\mathfrak{f}_{c_1 c_2 c_3}$. Thus the crux of the proof is in establishing that the matrix $\mathfrak{C}$ has a non-vanishing determinant.

In Appendix~\ref{sec:nomoduliproof}, we explicitly construct the coefficient matrix $\mathfrak{C}$ and show that its determinant does not vanish identically. We specifically establish that there exists a choice of coefficients $\A_{a_1b_1,a_2b_2,a_3b_3}$ such that $\det(\mathfrak{C}) \neq 0$, which furnishes a proof of the Theorem. 

We actually need more than the statement of the existence of some choice of coefficients $\A_{a_1b_1,a_2b_2,a_3b_3}$ for which the moduli space is trivial. Recall that our diagrammatics requires all the $\A_{a_1b_1,a_2b_2,a_3b_3}$ to be positive and $\mathcal{O}(1)$ after using the scaling symmetry \eqref{eqn:melonic_cond}. Furthermore, for the infinitesimal anisotropic deformation examined in \S\ref{sec:infAD} we needed the $\A_{a_1b_1,a_2b_2,a_3b_3}$ to lie within a distance $\epsilon$ of the isotropic coefficients $\mathbf{1}_{N^{6}} = (1,\ldots,1)$ for arbitrarily small $\epsilon$. We prove in Appendix~\ref{sec:nomoduliproof} that such a choice of $\A_{a_1b_1,a_2b_2,a_3b_3}$ exists, and furnish many of the necessary details. Furthermore, we also show there that the $(\bZ_2^N\rtimes S_N)^3$-invariant anisotropic deformation \eqref{eqn:Z2xSN_model} admits a choice of $\{\alpha_{1},\ldots,\alpha_{8}\}$ for which there are no non-trivial moduli.

%~~~~~~~~~~~~~~~~~~~~~~~~~~~~~~~~~~~~~~~~~~~~~~~
\section{Analysis on gauged models}
\label{sec:gauge}
%~~~~~~~~~~~~~~~~~~~~~~~~~~~~~~~~~~~~~~~~~~~~~~

The situation with the colored tensor models appears hopeless vis a moduli, since the flat directions arise from the multiplicity of fields, and the fact that our superpotential is constrained to be a particular monomial combination for melonic dominance. The uncolored tensors and matrix-vectors, on the other hand, are slightly better in this regard. For one, the explicit flat directions we have exhibited in \eqref{eq:mvmoduli} transform non-trivially under the symmetry group $G$. This suggests that we could perhaps lift the moduli by gauging the model to project onto $G$-singlets, which would provide a different route from the anisotropic deformation discussed hitherto. We will see that gauging the model by $G$, or by a subgroup $H$ after using partial anisotropy to break $G$ down to $H$ is insufficient to lift the moduli. While the result is negative, there are some interesting special cases we encounter along the way, and therefore we have chosen to provide some details of these gauged models. 

%~~~~~~~~~~~~~~~~~~~~~~~~~~~~~~~~~~~~~~~~~~~~~~~
\subsection{Gauged models}
\label{sec:gaugelg}
%~~~~~~~~~~~~~~~~~~~~~~~~~~~~~~~~~~~~~~~~~~~~~~

To gauge the Landau-Ginzburg models we have been discussing, we need to include gauge multiplets and upgrade the K\"ahler term to gauge covariant interactions. To write Lagrangians, we need both the vector multiplet as well as twisted chiral multiplets (in which the field strength resides) \cite{Witten:1993yc}. Aspects of non-abelian $(2,2)$ models are discussed in \cite{Hori:2006dk,Hori:2011pd}, and we refer the reader to these references for further details. We will first write down the gauged model, and then argue that this does not spoil our requirement of having melonic dominance.

We will primarily consider the matrix-vector model, where the symmetry group $G = {\rm SU}(N) \times {\rm O}(M)$. Let us first consider gauging the entire group and see where this leads us. We introduce vector multiplets $\mathscr{V}_u$ and $\mathscr{V}_o$ associated with the  ${\rm SU}(N)$ and ${\rm O}(M)$ transformations, respectively. The explicit form of the vector superfield is given in Appendix~\ref{sec:conventions}. The vector and chiral superfields transform under the two gauge groups as follows:
\ie
{\rm SU}(N)&: && \quad
\mathscr{Y} \rightarrow e^{i\Lambda} \mathscr{Y}  e^{-i\Lambda^{\dagger}}, &&\quad \overline{\mathscr{Y} }\rightarrow e^{i\Lambda}\overline{\mathscr{Y} }e^{-i\Lambda^{\dagger}}, &&\quad e^{2g_u \,\mathscr{V}_u} \rightarrow e^{i\Lambda^{\dagger}}e^{2g_u \,\mathscr{V}_u}e^{-i\Lambda}
\\ 
{\rm O}(M)&: &&\quad \mathscr{Y}  \rightarrow e^{i\Omega}\mathscr{Y} , &&\quad \overline{\mathscr{Y} } \rightarrow  \overline{\mathscr{Y} }e^{-i\Omega^{\dagger}}, &&\quad e^{2g_o\, \mathscr{V}_o} \rightarrow e^{i\Omega^{\dagger}}e^{2g_o\, \mathscr{V}_o}e^{-i\Omega},
\label{eqn:gaugetrans}
\fe
where $\Lambda \in {\rm SU}(N)$ and $\Omega \in {\rm O}(M)$ are both adjoint valued superfields of the respective gauge groups. 
We define the gauge covariant superderivatives $\mathcal{D}_{\alpha}$ and $\overline{\mathcal{D}}_{\dot{\alpha}}$ in the standard way from the superderivatives $D_{\alpha}$ and $\overline{D}_{\dot{\alpha}}$ and we define the twisted chiral superfield
\begin{equation}
\Sigma = \frac{1}{2\sqrt{2}}\{\overline{\mathcal{D}}_{+},\mathcal{D}_{-}\},
\label{eq:twch}
\end{equation}
which contains the field strength. 

The action for the gauged theory then takes the form:\footnote{ We retain $\rm{Tr}$ to refer to trace over the $SU(N)$ indices. On occasions where we need to trace over the generators of the $O(M)$ Lie algebra we indicate this explicitly with $\mathrm{Tr}_\mathfrak{o}$.}
\begin{equation}
\begin{split}
S &= - \frac{1}{2g_{u}^2}\int d^{2}x\,d^{4}\theta\,\mathrm{Tr}\Big(\overline{\Sigma}_{u}\Sigma_{u}\Big)
	- \frac{1}{2g_{o}^2}\int d^{2}x\,d^{4}\theta\,\mathrm{Tr}_{\mathfrak{o}}\Big(\overline{\Sigma}_{o}\Sigma_{o}\Big) \\
&\qquad \quad 
+\;	
	2\int\, d^2x\, d^4 \theta\, \sum_{I=1}^M \, \mathrm{Tr} \left( \overline{\mathscr{Y}}^I e^{4(\mathscr{V}_{u}+\mathscr{V}_{o})}\, \mathscr{Y}^I  \right)- \int d^2 x \, d^2\theta \; \frac{g}{4}\, 
	\mathrm{Tr} \left(\mathscr{Y}^I \, \mathscr{Y}^J \, \mathscr{Y}^I \,\mathscr{Y}^J\right) + \textrm{h.c}.
\end{split}
\label{eq:gaugeLag}
\end{equation}
Integrating out the auxiliary fields $D_u$  and $D_o$ from the vector multiplets gives us the $D$-term constraints:
\begin{equation}
\begin{split}
D_o^{IJ} &= -g_o^2\, \mathrm{Tr}\left(\overline{Y}^{[J}\, Y^{I]}\right)\,,\quad 
D_u = -g_u^2 \, [Y^{I},\overline{Y}^{I}].
\end{split}
\label{eqn:Dterm}
\end{equation}
We still have the $F$-term constraint from the superpotential:
\begin{equation}
F = - g \sum_{I=1}^M \, Y^I\, Y^J \, Y^I \,.
\label{eq:Fterm}
\end{equation}	
Let us for completeness record the bosonic potential obtained after integrating out the auxiliary fields:
\begin{align}
V(X,\sigma) &= \frac{1}{2\,g_o^2}\, \mathrm{Tr}_{\mathfrak{o}} \bigg( [\sigma_o,\overline{\sigma}_o]^2\bigg)
 +\mathrm{Tr} \bigg(\overline{Y}^{I}\,\{\overline{\sigma}_o,\sigma_o\}^{IJ}\, Y^{J}\bigg)
+\frac{g_o^2}{2} \, \mathrm{Tr}_{\mathfrak{o}} \bigg(\left[\mathrm{Tr} \left( Y^{[I}\overline{Y}^{J]} \right) \right]^2
\bigg) 
\nonumber \\	
& \quad +\frac{1}{2\,g_u^2}\, \mathrm{Tr}\bigg([\sigma_u,\overline{\sigma}_u]^2\bigg)+\mathrm{Tr}
\bigg(\big[\overline{Y}^{I},\overline{\sigma}_u\big]\big[\sigma_u,Y^{I}\big]\bigg)+
\mathrm{Tr} \bigg( \big[\overline{Y}^{I},\sigma_u\big]\big[\overline{\sigma}_u,Y^{I}\big] \bigg)
\nonumber \\	
&\quad +\frac{g_u^2}{2}\, \mathrm{Tr} \bigg([Y^{I},\overline{Y}^{I}]^2 \bigg)
+|g|^2\, \mathrm{Tr}\bigg (
Y^{J}Y^{I}Y^{J}\overline{Y}^{K}\overline{Y}^{I}\overline{Y}^{K}\bigg).
\label{eqn:bosonpot}
\end{align}
Since we have introduced ${\rm SU}(N)\times {\rm O}(M)$ gauge dynamics, we have now the gauge couplings $g_o$ and $g_u$, apart from the matter coupling $g$. All of them have the same canonical scaling dimension. We will fine tune the system to choose $g_o \sim g_u\sim g$, 
so that the tetrahedral coupling is much larger than the associated 't Hooft couplings
\begin{equation}
g\, N\sqrt{M} \gg \{g_u\,  \sqrt{N} , g_o\, M\}.
\label{eq:hierarchy}
\end{equation}	
We expect this choice of scalings to suppress all diagrams involving gauge fields so that we retain the standard melonic analysis. However, we have not constructed a proof since the model has more immediate issues due to potential moduli. It would be interesting to prove that the melonic analysis indeed holds.

Let us summarize the $R$-charges and gauge symmetry representations for the chiral and gauge multiplets in Table~\ref{tab:Rgreps}. 
\begin{table}[H]
\begin{center}
  \begin{tabular}{|c|c|c|c|c|c|c|}
  \hline
   & ${\rm U}(1)_R$ & ${\rm U}(1)_L$ & ${\rm U}(1)_V$ & ${\rm U}(1)_A$ & ${\rm SU}(N)$ &${\rm O}(M)$
  \\
    \hline
  $Y^I$ & $\frac{1}{4}$ & $\frac{1}{4}$ & $\frac{1}{2}$ & $0$ & {\bf adj+1} & {\bf vec}
  \\
    \hline
  $\Sigma_u$ & $1$ & $-1$ & $0$ & $2$ & {\bf adj} & {\bf 1}
  \\
    \hline
      $\Sigma_o$ & $1$ & $-1$ & $0$ & $2$ & {\bf 1} & {\bf adj}
  \\
    \hline
  \end{tabular}
\end{center}
\caption{R-charges and representation content for the chiral and twisted chiral multiplets in the matrix-vector model}
\label{tab:Rgreps}
\end{table}
By the anomaly matching of the ${\rm U}(1)_R$ symmetry, the IR central charge is given by
\begin{equation}
\begin{split}
 c&=3\left[{1\over 2}(\text{number of chiral superfields})-(\text{number of vector superfields})\right]
 \\
 &=\begin{cases}
  {3\over 2}(MN^2-2N^2+2)&{\rm for}~ {\rm SU}(N)~\text{gauge theory},
\\
 {3\over 2}(MN^2-2N^2-M^2+M+2)&{\rm for}~ {\rm SU}(N)\times {\rm O}(M)~\text{gauge theory}.
 \end{cases}
\end{split}
\label{eq:caniso}
\end{equation}	
%

%~~~~~~~~~~~~~~~~~~~~~~~~~~~~~~~~~~~~~~~~~~~~~~~
\subsection{Moduli}
\label{sec:gaugelg}
%~~~~~~~~~~~~~~~~~~~~~~~~~~~~~~~~~~~~~~~~~~~~~~

We examine the moduli space of this theory to see whether gauging lifts the moduli. One can check that all the terms in \eqref{eqn:bosonpot} are positive, and hence each term must vanish independently for the potential to be minimized.
We can thus examine the vacuum structure by setting each of these terms to vanish. The first thing we learn is that 
\begin{equation}\
[\sigma_o,\overline{\sigma}_o] = [\sigma_u,\overline{\sigma}_u]  = 0,
\label{eq:Dsigma}
\end{equation}	
and hence $\sigma_{u,o}$ are diagonalizable. Since $\sigma_o$ is anti-symmetric, we must have $\sigma_o = 0$. However, $\sigma_u$ is not necessarily zero implying that we have a non-trivial  Coulomb branch. This is of course expected (see eg., \cite{Hori:2006dk}), and we will analyze this sector in greater detail in \S\ref{sec:cbranch}.

The constraints on the matter fields are 
\begin{equation}
\begin{split}
\mathrm{Tr}\bigg([Y^{I},\overline{Y}^{I}]^2\bigg) &= 0 \implies \sum_{I=1}^{M}\; [Y^{I},\overline{Y}^{I}] = 0, \\
\mathrm{Tr}_{\mathfrak{o}}\bigg(\bigg[ \mathrm{Tr}\left(Y^{[I}\overline{Y}^{J]}\right) \bigg]^2\bigg) &= 0 \implies 
\mathrm{Tr}\big(Y^{I}\overline{Y}^{J}\big) \in \mathbb{R}, \\
\mathrm{Tr}\bigg( Y^{J}Y^{I}Y^{J}\overline{Y}^{K}\overline{Y}^{I}\overline{Y}^{K}\bigg) &= 0.
\end{split}
\label{eq:Fexplicit}
\end{equation}
Let us first solve these constraints for some simple cases:
\begin{itemize}
\item For $M=1$ and arbitrary $N$, the first condition implies that $Y$ is diagonalizable. Denote by $\lambda_{i}$ ($i=1,\ldots,N$) the eigenvalues of $Y$. Then the third condition becomes $\sum_{i=1}^{N}|\lambda_{i}|^{6} = 0$, and hence $\lambda_{i}=0$ for all $i$ so $Y = 0$. 
\item For $N=1$, arbitrary $M$, the third condition gives $\sum_{I=1}^{M}|Y^{I}|^2\sum_{J,K}(Y^{J}Y^{K\ast})^2=0$. The second condition gives $Y^{J}Y^{K\ast} \in \mathbb{R}$ and thus the third condition implies either $\sum_{I=1}^{M}|Y^{I}|^2 = 0$ or $\sum_{J,K}(Y^{J}Y^{K\ast})^2=0$. In the first case, we have $|Y^{I}|^2=0$ for all $I$ so $Y=0$. In the second case, we have $Y^{J}Y^{K\ast}=0$ for all $J$ and $K$ and thus again $|Y^{I}|^2 = 0$ for all $I$ so $Y=0$.
\item Things however unravel when $M=N=2$. Our previous solution \eqref{eq:mvmoduli} continues to solve \eqref{eq:Fexplicit} leading to a non-compact moduli space of vacua. Since we can embed this solution for larger values of $M$ and $N$ we need a new strategy.
\end{itemize}

The upshot of this discussion is that while we can gauge the models to retain melonic dominance, the moduli are not entirely lifted. We will next try to deform the superpotential to attempt to find a theory without flat directions.

%~~~~~~~~~~~~~~~~~~~~~~~~~~~~~~~~~~~~~~~~~~~~~~~
\subsection{Anisotropic gauged matrix-vector models}
\label{sec:gmv}
%~~~~~~~~~~~~~~~~~~~~~~~~~~~~~~~~~~~~~~~~~~~~~~
 
We have seen that the gauging of $G = {\rm SU}(N) \times {\rm O}(M)$ was insufficient to lift the flat directions of the model. We need to do more, and the only option left is to deform the $F$-term constraints so as to lift \eqref{eq:mvmoduli}  (while hopefully not introducing other flat directions). We had two sets of issues: a non-compact Coulomb branch coming from the ${\rm SU}(N)$ sector and non-trivial solutions to the $F$-term equations.   It will turn out that the Coulomb branch can be tamed without much ado. We will discuss this in 
\S\ref{sec:cbranch}. The Higgs branch is however tricker to tame.

As we have seen in the previous section, for the ungauged models, the moduli can be lifted by turning on the anisotropic deformations, which generically break all the continuous flavor symmetries to discrete symmetries. We consider a special class of the anisotropic deformations that breaks only the ${\rm O}(M)$ part of the flavor symmetry,
%Let us start by considering the following anisotropic deformation of our superpotential 
%
\begin{equation}
W_{4\,\alpha}(Y) = \frac{1}{4} g\, \sum_{I,J=1}^M \, \alpha_{IJ} \, {\rm Tr} \left(Y^I \, Y^J \, Y^I\, Y^J \right) \,.
\label{eq:anisoW}
\end{equation}	
The ${\rm SU}(N)$ symmetry is preserved by this superpotential, and can be gauged. As we argued in Section \ref{sec:anisotropic}, melonic dominance is preserved as long as the deformation parameters $\alpha_{IJ}$ are of order unity. The superpotential \eqref{eq:anisoW} is not the most generic single-trace superpotential for our fields, which would have been determined by a four-tensor $\xi_{IJKL}$ of ${\rm O}(M)$ with cyclic symmetry. More generally, we could have allowed double-trace interactions as well (or multi-traces if we consider $q$-body interactions). Our choice is predicated by requiring that we still retain solvability in the large $N$ limit.

For the gauged anisotropic model we still need to handle the $F$-term constraint from the superpotential and the $D$-term constraint from the gauge couplings. 
For the deformed matrix-vector model these read:
\begin{equation}
\sum_{I =1}^M\, \alpha_{IJ} \, Y^I \, Y^J \, Y^I = 0 \,, \qquad \sum_{I=1}^M\, [\overline{Y}^I , Y^I ]  =0 \,.
\label{eq:FDcons}
\end{equation}	
The full moduli space is given by the solution to the above two equations, quotiented by the ${\rm SU}(N)$ gauge symmetry.\footnote{ Usually, one solves the $F$-term constraints and quotients by the complexified gauge group, ignoring in the process the $D$-term constraint. This is usually justified by arguing that there exists a complex gauge transformation that allows one to trivialize the D-term constraint. More formally, as explained in \cite{Witten:1993yc} the actual moduli space is a symplectic quotient, which is birationally equivalent to the holomorphic quotient obtained by relaxing the D-term constraints and quotienting by the complexified gauge group.} If we only consider anisotropy without gauging, then we still find flat directions given by 
\begin{equation}
Y^{I} = \left(\begin{matrix} 0 & a_{I} \\ \mathbf{0}_{N-1} & 0 \\  \end{matrix}\right), \qquad a_{I} \in \mathbb{C},\;I = 1,\ldots,M.
\label{eq:FDgaugeaniso}
\end{equation}
Therefore, we need both anisotropy and gauging to have any hope of removing the flat directions.

We will now proceed to discuss the moduli space of the gauged matrix-vector theory with the deformed superpotential \eqref{eq:anisoW} in some detail. We proceed by first exhibiting that the theory with gauge group ${\rm SU}(N)$ has a non-compact Coulomb branch, which however can be lifted if we consider the group ${\rm PSU}(N) \cong {\rm SU}(N)/\mathbb{Z}_N$. We will find that for specific odd/even parity choices of $N,M$ and suitable choices of bare theta angles, we can end up with a compact Coulomb branch. The real issue for us is whether we have a compact Higgs branch. This turns out to be the case for some specific choices such as  $M=1$, $N$ arbitrary, or $M=N=2$. For $M=N=3$, we numerically found nontrivial solutions to the equations \eqref{eq:FDcons} for generic $\A_{IJ}$. More precisely, we numerically minimized a modified potential 
\ie
\!\!\!\!\!\!\!V(Y,\overline Y) = & |g|^2\sum_{I,J,K=1}^M\,\alpha_{IJ} \,\alpha_{KJ}^* \,\tr( Y^I \, Y^J \, Y^I \, \overline Y^K \,\overline Y^J \,\overline Y^K )+{g_u^2\over 2}\sum_{I,J=1}^M\,\tr([\overline{Y}^I , Y^I ] [\overline{Y}^J , Y^J ] )
\\
&+{m^4\over 4g}-{m^2\over 2}\sum_{I=1}^M\tr(Y^I\,\overline Y^I)+{\xi\over 4}\left[\sum_{I=1}^M\tr(Y^I\,\overline Y^I)\right]^2,
\fe
which is obtained from the bosonic potential of our model by adding a double-well type potential. This potential is bounded from below  $V(Y,\overline Y)\ge 0$ if the parameters $m,\xi $ are real and positive. If we could find a minimum of the potential at $(Y^I,\overline Y^I)=(Y^I_*,\overline Y^I_*)$ such that $V(Y_*,\overline Y_*)=0$, then $Y^I_*,\,\overline Y^I_*$ would be a nontrivial solution to the equations \eqref{eq:FDcons}. Moreover, solutions for smaller values of $M,\,N$ can always be embedded into the solutions for larger values of  $M,\,N$.  We have found this to be case for small values of $M$ and $N$ (our checks were carried out for $M\leq 5$ and $N\leq5 $).

Based on this we conjecture that one will find a non-compact Higgs branch for generic $\alpha_{IJ}$. However, this does not rule out the possible existence of some special choices of $\alpha_{IJ}$ such that the Higgs branch is compact. We will give a broad discussion of the Higgs and Coulomb branches below, and supplement this analysis with a computation of the elliptic genera in the sequel. 

%~~~~~~~~~~~~~~~~~~~~~~~~~~~~~~~~~~~~~~~~~~~~~~~
\subsection{Coulomb branch}
\label{sec:cbranch}
%~~~~~~~~~~~~~~~~~~~~~~~~~~~~~~~~~~~~~~~~~~~~~~

The classical Coulomb branch of the theory is non-compact and $N-1$ dimensional. At a generic point on this moduli space, ${\rm SU}(N)$ gauge symmetry 
is broken down to the maximal abelian subgroup ${\rm U}(1)^{N-1}$. The quantum Coulomb branch could however be compact as there is a twisted superpotential generated from loop effects \cite{Witten:1993yc}. Following \cite{Hori:2006dk} we parameterize the twisted superpotential 
at large $\sigma$ (see Appendix~\ref{sec:conventions} for multiplet structure):
\begin{equation}
\widetilde W_{\rm eff} = (M+1)\sum_{i\neq j =1}^{N} \, (\Sigma_i-\Sigma_j)\big[\log(\Sigma_i-\Sigma_j) - 1\big].
\label{eq:tw1}
\end{equation}	
Without loss of generality, we assume $\Sigma_1<\Sigma_2<\cdots<\Sigma_N$, and use the traceless condition 
$\sum_{i=1}^N\Sigma_i=0$. We find that \eqref{eq:tw1} simplifies to 
\begin{equation}
\widetilde W_{\rm eff} =2 i (M+1)\pi \sum_{i=1}^{N-1}(N-i)\Sigma_i.
\label{eq:tw2}
\end{equation}	
The twisted $F$-flatness condition is
\begin{equation}
\theta_i \in 2\pi P,
\label{eqn:vacuum_equation_suN}
\end{equation}	
where $P$ is the weight lattice ${\rm SU}(N)$, and $\theta_i$ are the effective theta angles of the unbroken ${\rm U}(1)^{N-1}$ symmetries
\begin{equation}
\theta_i ={\rm Im}\, \frac{\partial \widetilde W_{\rm eff}}{\partial \Sigma_i}  = 2(M+1)\pi(N-i).
\label{eqn:thetas}
\end{equation}	
Given a weight vector $w_i\in P$, we have a character
\begin{equation}
\chi_{w}(z)=z_1^{w_1}\cdots z_{N-1}^{w_{N-1}},
\end{equation}	
which is a (single-valued) function on the maximal torus of ${\rm SU}(N)$
\begin{equation}
T=\{z_1,\cdots, z_N\in\bC\, |\, |z_i|=1,\,z_1\cdots z_N=1 \}.
\end{equation}	
The character of the weight vector given by the theta angles \eqref{eqn:thetas} is a well-defined monomial; hence, the twisted $F$-flatness condition \eqref{eqn:vacuum_equation_suN} is satisfied. We conclude that the Coulomb branch is noncompact for ${\rm SU}(N)$ gauge theory with $M$ adjoint matters.

%\footnote{
%For $N=2$ we can immediately see that the theta angle is $\theta = 2(M+1)\pi$. If the gauge group were $SU(2)$, the periodicity condition 
%$\theta \sim \theta + 2\pi$, implies that we can simply set $\theta = 0$. Then the theory has a flat direction. We can similarly analyze larger $N$. 
%
%However, the Coulomb branch can be lifted, if we consider $PSU(N)\cong  SU(N)/\mathbb{Z}_N$ gauge theory. Lets again start with $N=2$ where 
% $SO(3)\cong  SU(2)/\mathbb{Z}_2$. The effective twisted superpotential computed in \cite{Hori:2011pd} reads:
% %
%\begin{equation}
%\widetilde W_{\rm eff} = \left[(M+1)\pi+\theta_{\rm bare}\right] i \Sigma_1,
%\label{eq:twN2}
%\end{equation}	
%%
%where $\theta_{\rm bare}\in \pi_1(SO(3))\cong\bZ_2$ is the bare theta angle. Now the Coulomb branch can be rendered compact by a suitable choice of bare theta angle. Picking the tree-level $\bZ_2$ theta angle 
%$\theta_{\rm bare}=0$ (or $\theta_{\rm bare}=\pi$), the Coulomb branch is compact for $M$ even (odd).}

The Coulomb branch can be lifted, if we consider ${\rm PSU}(N)\cong  {\rm SU}(N)/\mathbb{Z}_N$ gauge theory.\footnote{We thank Kentaro Hori for a discussion on this point.} In this case, the twisted $F$-flatness condition is modified to
\begin{equation}
\theta_i \in 2\pi P / \bZ_N \cong 2\pi Q,
\label{eqn:vacuum_equation_psuN}
\end{equation}	
where $Q$ is the root lattice of ${\rm SU}(N)$.\footnote{The center $\bZ_N$ of ${\rm SU}(N)$ is isomorphic to $P/Q$.}
The maximal torus of ${\rm PSU}(N)$ is
\begin{equation}
T=\bigg\{z_1,\cdots, z_N\in\bC\, \bigg|\  |z_i|=1,\, \prod_{i=1}^N \, z_i =1 \bigg\} \Big/ \left[(z_1,\cdots,z_N)\sim(z_1\omega,\cdots,z_N\omega)\right], 
\label{eq:maxT}
\end{equation}	
where $\omega$ is a $N^{\rm th}$ root of unity, $\omega^N=1$. The character of a vector $w\in Q$ can be obtained to be 
\begin{equation}
\chi(z)=z_1^{w_1}\cdots z_{N-1}^{w_{N-1}} \,, \qquad 
\omega^{\sum_{i=1}^{N-1} \, w_i}=1 \; \ \Longrightarrow\;\;    \sum_{i=1}^{N-1} \, w_i \in N \mathbb{Z}.
\end{equation}	
Going back to \eqref{eqn:thetas}, we learn therefore that the condition to the lift the Coulomb branch is to require:
\begin{equation}
\sum_{i=1}^{N-1}\theta_i=\pi\, (M+1) \, (N-1)N \; \not\in\;  2\pi N\mathbb{Z}\,.
\end{equation}
Therefore, for $M,N\in 2\bZ$, the Coulomb branch is lifted. For $M\not\in 2\bZ$ or $N\not\in 2\bZ$, to lift the Coulomb branch, we need to turn on bare theta angle $\theta_i^{\rm bare}$, which takes values in $\pi_1({\rm PSU}(N))\cong \bZ_n$. The twisted F-flatness condition becomes
\begin{equation}
\theta_i^{\rm bare}+\theta_i \in  2\pi Q.
\label{eqn:vacuum_equation_psuN}
\end{equation}	
Since the twisted F-flatness condition is satisfied for the theories with zero bare theta angle, any choice of nontrivial bare theta angle would break the twisted F-flatness condition, and the Coulomb branch is lifted.

Thus, as presaged, for the gauge group ${\rm PSU}(N)$ with suitable choice of bare theta angles we have a compact Coulomb branch. The next question we need to address is the compactness of the Higgs branch.

%~~~~~~~~~~~~~~~~~~~~~~~~~~~~~~~~~~~~~~~~~~~~~~~
\subsection{Higgs branch}
\label{sec:hbranch}
%~~~~~~~~~~~~~~~~~~~~~~~~~~~~~~~~~~~~~~~~~~~~~~

A useful way to study the Higgs branch is to directly analyze the Higgs branch chiral ring of the theory \cite{Vafa:1988uu,Lerche:1989uy}. The way we approach the question is to ask what are all the gauge invariant monomials that we can build out of our chiral superfields, and quotient them by the Higgs branch chiral ring relations resulting from the $F$-term equations. To wit, the Higgs branch chiral ring is given by the polynomial ring (quotiented by an ideal)
\begin{equation}
\mathcal{R}_H = \left(\mathbb{C}[Y^I] \bigg/\left( \sum_J\, \alpha_{IJ}\, Y^J Y^I Y^J\right)\right)  \bigg/ \bigg(Y^I \sim M^{-1}\, Y^I\, M \,, \quad M \in {\rm SU}(N)_{\mathbb{C}}\bigg) . 
\label{eq:Hring}
\end{equation}	

For small values of $N$ and $M$ one can explicitly analyze the problem and see that the resulting Higgs branch chiral ring is finite-dimensional. For instance, we explicitly analyze the $N=M=2$ example in Appendix~\ref{sec:higss2} demonstrating that as long as $\alpha_{IJ}$ are all unequal, one indeed recovers a finite-dimensional ring of chiral operators. To encode the information about the Higgs branch chiral ring generators we can compute a suitable generating function (a Poincar\'e polynomial) for the ring. We define:
\begin{equation}
\mathcal{P}_\mathcal{R}(y) = {\rm Tr}_{\mathcal{R}_H} \left( y^\frac{R}{2}\right)  = \sum_{m=1}^\infty \mathfrak{a}_m\, y^\frac{m}{4}\,, \\
\label{eq:chiyR}
\end{equation}	
where $y$ can be viewed as a fugacity for the vector ${\rm U}(1)_V$ R-charge.  If $\mathfrak{a}_m =0$ for $m> n$ with $n\in \mathbb{Z}_+$ then we can conclude that the Higgs branch chiral ring is finite-dimensional, implying a compact Higgs branch. If the expansion however is an infinite sum, then we have an infinite-dimensional chiral ring.

For this analysis, we find it easier to work in the simplified case where each $Y^{I}$ is traceless: ${\rm Tr}(Y^I) = 0$ for every $1 \leq I \leq M$. Some basic results for $\mathcal{P}_\mathcal{R}(y)$ for the traceless model are tabulated below in 
Table~\ref{tab:Hpoin}. We have always chosen the $R$-charge to be that given by the IR fixed point value for a quartic superpotential $R = \frac{1}{2}$  to facilitate comparison, 
We see that the unconstrained count for the free theory, which of course has an infinite-dimensional chiral ring, gets somewhat reduced by the $F$-term constraints, but generically this reduction does not seem strong enough to cull down the ring to a finite one even with the most generic superpotential.
Curiously, however, the generic superpotential ends up culling a lot of states relative to that in the free theory. For example for $M=N=3$ we see a rather rapid growth of the free count, but a very slow growth in the generating function for the generic $W$ case.\footnote{ The situation is even more remarkable at $\mathcal{O}(y^\frac{9}{4})$ which is not presented in Table~\ref{tab:Hpoin}. At this order we find the free theory having $1785$ states, of which only $17$ survive for a generic choice of the superpotential! The constraints however are seemingly insufficient to truncate the spectrum of chiral operators.}
\begin{table}[h]
\centering
	\begin{tabular}{| c | c| c |}
	\hline
	$(M,N)$ &
		$\alpha_{IJ}$ &
			$\mathcal{P}_\mathcal{R}(y)$ \\
\hline\hline
		& \shadeR{$0 $} &  
			\shadeR{$1+3\, y^\frac{1}{2}+6\, y+10\, y^\frac{3}{2}+15\, y^2+\cdots$} \\
	$(2,2)$ 
		& $1  $ & 
			$1+3\, y^\frac{1}{2}+3\, y+3\, y^\frac{3}{2}+3\, y^2+\cdots$ \\
		& \shadeB{$\begin{pmatrix}1&2\\2&1\end{pmatrix}$} & 
			\shadeB{$1+3\, y^\frac{1}{2}+2\, y$} \\
\hline \hline
		& \shadeR{$0$} 	 & 
			\shadeR{$1+6\, y^\frac{1}{2}+11\, y^\frac{3}{4}+30\, y+75\, y^\frac{5}{4}+186\, y^\frac{3}{2}+381\, y^\frac{7}{4}+
			885\, y^2 + \cdots$}\\
		& $1$ & 
			$1+6\, y^\frac{1}{2}+8\, y^\frac{3}{4}+24\, y+51 \, y^\frac{5}{4}+84\, y^\frac{3}{2}+115\, y^\frac{7}{4}+\cdots$ \\
	$(3,3)$ & 
		\shadeB{$\begin{pmatrix}1&2&3\\2&1&2\\3&2&1\end{pmatrix}$} &	
				\shadeB{$1+6\, y^\frac{1}{2}+8\, y^\frac{3}{4}+21\, y+51 \, y^\frac{5}{4}+66\, y^\frac{3}{2}+74\, y^\frac{7}{4}+
				65\, y^2 + \cdots$}\\

		& generic single trace $W$ &
			$1+6\, y^\frac{1}{2}+8\, y^\frac{3}{4}+21\, y+51 \, y^\frac{5}{4}+64\, y^\frac{3}{2}+71\, y^\frac{7}{4}+\cdots$\\	
		& 	generic $W$	 &
			$1+6\, y^\frac{1}{2}+8\, y^\frac{3}{4}+21\, y+51 \, y^\frac{5}{4}+64\, y^\frac{3}{2}+71\, y^\frac{7}{4}
			+64\, y^2 +\cdots$\\
\hline\hline
		& \shadeR{$0$} &
			\shadeR{$1+10\, y^\frac{1}{2}+24\, y^\frac{3}{4}+90\, y+\cdots$} \\
	$(4,3)$	& 	
		\shadeB{$\begin{pmatrix}1&2&3&4\\2&1&2&3\\3&2&1&2\\4&3&2&1\end{pmatrix}$} &
			\shadeB{$ 1+10\, y^\frac{1}{2}+20\, y^\frac{3}{4}+74\, y+\cdots$} \\
\hline\hline			
	 &	\shadeR{$0$} &
			\shadeR{$1+6\, y^\frac{1}{2}+11\, y^\frac{3}{4}+45\, y+\cdots$} \\
	$(3,4)$& 
		\shadeB{$\begin{pmatrix}1&2&3\\2&1&2\\3&2&1\end{pmatrix}$} &
			\shadeB{$1+6\, y^\frac{1}{2}+8\, y^\frac{3}{4}+36\, y+\cdots$} \\		
\hline			
	\end{tabular}
\caption{Higgs branch chiral ring Poincar\`e polynomial for various choice of $\alpha_{IJ}$. We have highlighted the results for the free theory and the anisotropic model for ease of visualization.}
\label{tab:Hpoin}
\end{table}

It is possible to give a succinct formula for the counting of gauge invariant monomials using Polya counting at least for the theory without a superpotential, i.e. $\alpha_{IJ} =0$. Define the single particle generating function which simply enumerates the alphabets in our theory weighted by their $R$-charge (taken to be the non-trivial IR charge)
\begin{equation}
z_s(\{y_i\}) = \sum_{I=1}^M \, y_I^\frac{1}{4}  \,,
\label{eq:zsinglep}
\end{equation}	
where we have fine-grained the $R$-fugacity to account for contributions from each of the fields $Y^I$ independently. 
The number of ${\rm SU}(N)$ singlets is then obtained from the multiparticle generating formula:
\begin{equation}
Z(\{y_i\}) = \int dU \, \exp\left( \sum_{\ell =1}^\infty \,\frac{1}{\ell}\, z(y_1^\ell, y_2^\ell, \cdots, y_M^\ell) \, \chi_{adj}(U^\ell) \right), 
\label{eq:Zmulti}
\end{equation}	
where the integral is over the Haar measure for ${\rm SU}(N)$. The character in the adjoint representation for the holonomy matrix $U$ can be simplified to $\chi_{adj}(U) = {\rm Tr}(U) \, {\rm Tr}(U^{-1}) -1$, where the trace is taken in the fundamental representation. To obtain the number of gauge invariant operators made out of $k$-alphabets of the $Y^I$ we simply need the coefficient of $y^\frac{k}{4}$ in the expansion of 
\begin{equation}
\mathcal{P}_\mathcal{R}(y)\bigg|_{\alpha =0} = \int dU \, \exp\left( \sum_{\ell =1}^\infty \,\frac{M}{\ell}\, y^\frac{\ell}{4}   \, \chi_{adj}(U^\ell) 
 \right) .
\label{eq:Zygen}
\end{equation}	

This result valid for the free theory $\alpha =0$ should be quotiented by the relations arising due to the $F$-term.  We have thus far not managed to come up with a closed form expression for our anisotropic model \eqref{eq:anisoW}. The best we can do is provide an upper bound on the number of monomials which will be culled in the quotient  \eqref{eq:Hring}.  We start with the $F$-term equations \eqref{eq:FDcons} which are linear combinations of cubic monomials transforming in the adjoint of ${\rm SU}(N)$. We can take these objects and construct invariants by contracting with ${\rm SU}(N)$ invariant tensors $\mathcal{T} = \mathcal{T}_{a_1 a_2\, \ldots a_m} $.\footnote{ Such tensors can be expanded in a basis of the tensor products of the dual algebra $\mathfrak{su}(N)^*$ generators $\omega^a$, viz.,   $\mathcal{T} = \mathcal{T}_{a_1 a_2\, \ldots a_m} \omega^{a_1} \otimes \omega^{a_2} \otimes \cdots \otimes \omega^{a_m}$ and are required to satisfy $\sum_{p=1}^{m} \, f^c_{a_p b}\, \mathcal{T}_{a_1 a_2\ldots a_{p-1} c a_{p+1} \ldots a_m} =0$ (we assumed for simplicity 
$\omega_a(t^b) = \delta_a^b$). One can alternately view them as being built from the  $N$ independent Casimirs of $SU(N)$ along with the structure constants.} 
For example when $N=3$ we have three building blocks: $\delta_{ab}, d_{abc}, f_{abc}$ corresponding to ${\rm Tr} \left(t_a t_b\right)$, $ {\rm Tr} \left(\{t_a, t_b\} t_c\right)$, and 
${\rm Tr} \left([t_a, t_b] t_c\right)$, respectively for ${\rm SU}(3)$.

Viewing the constraint as a generic adjoint tensor we could write down a constraint single partition sum, assuming that there are no redundancies. That is to say we build monomials of the form:
\begin{equation}
\mathcal{T} \cdot \left( \left( \sum_{I =1}^N\, \alpha_{IJ}\, Y^I\, Y^J \, Y^I \right) Y^{K_1}  Y^{K_2} \cdots Y^{K_{m-3}} \right),
\label{eq:}
\end{equation}	
which will give us ${\rm SU}(N)$ singlets. The trouble with performing the counting is that even for generic $\alpha_{IJ}$ we find some redundant relations. If we pretend that such redundancies do not exist then the counting is feasible. This will give us an over-count, but one that can be useful to understand the structure of the chiral operators in a theory with a superpotential.

We take the constraint at every order to mean the following: one solves for $[(Y^I)^3\times \textrm{monomial from} \; (m-3) \; Y^K]$ in terms of the other monomials. Then all we have to do is subtract out the contribution from monomials that can be built this way. This can be seen to be obtained by considering the expectation value of $\left(\sum_{K=1}^M\, y_K^\frac{3}{4}\right) \, \chi_{adj}(U)$. Thus the 
number of invariants after removing the monomials of the aforementioned form is:
\begin{equation}
\begin{split}
\mathcal{P}_\mathcal{R}(y) &= \sum_{m=1}^\infty \mathfrak{a}_m\, y^\frac{m}{4}\, \\
\mathfrak{a}_m& \geq \text{Coefficient of}\; \, y^m \;\, \text{in} 
	\bigg[ 
	 \int dU \left[ 1 - M\, y^\frac{3}{4} \, \chi_{adj}(U)\right]
	 \exp\left( \sum_{\ell =1}^\infty \,\frac{M}{\ell}\, y^\frac{\ell}{4}   \,\chi_{adj}(U^\ell) \right) \bigg]. \\
\end{split}
\label{eq:amdef}
\end{equation}	
If we have $\mathfrak{a}_m \geq 0$ for all $m$ we can conclude that the chiral ring is infinite-dimensional, for we have potentially removed more operators than suggested by the $F$-term constraints. However, if $\mathfrak{a}_m \leq 0$ we should exercise care as we could potentially be removing too many operators from the free ring, without accounting for the interdependencies in \eqref{eq:FDcons}. From our numerical experiments we find that we are indeed over-counting the constraints, so the bound in Eq.~\eqref{eq:amdef} is not very helpful. As noted earlier the constraints appear to almost do the job, in that the number of states in the chiral ring is quite small compared to those in the free theory, but nevertheless not powerful enough to force a finite-dimensional chiral ring.

%~~~~~~~~~~~~~~~~~~~~~~~~~~~~~~~~~~~~~~~~~~~~~~~
\section{Elliptic genera}
\label{sec:egen}
%~~~~~~~~~~~~~~~~~~~~~~~~~~~~~~~~~~~~~~~~~~~~~~
  
We now turn to the computation of elliptic genera for the Landau-Ginzburg tensor models and some of the rank-one gauged models. This provides checks and a bit more insight on some of our conclusions on the general structure of the Coulomb and Higgs branches.

The computation of the elliptic genus for $\mathcal{N}=(2,2)$ theories was first described in \cite{Witten:1993jg} for Landau-Ginzburg models. Recently, \cite{Benini:2013nda,Benini:2013xpa} used localization techniques to give general expressions for the elliptic genus of ${\cal N}=(2,2)$ gauged linear sigma models. We will mostly use their results and check the elliptic genus for rank-$1$ theories.  

The elliptic genus is a Hilbert space trace over the Ramond-Ramond sector of the IR SCFT, explicitly defined by
\begin{equation}
\mathcal{Z}(q, y, \{x_a\} ) = {\rm Tr}_{\text{RR}} \left( (-1)^F \, q^{\Delta_L} \, \bar{q}^{\Delta_R} \, y^{J_L}
\right),
\label{eq:egenus}
\end{equation}	
where $J_L$ is the ${\rm U}(1)_L$ charge for the left-moving $R$-symmetry. The parameters $q,\bar{q}$ encode the inverse temperature and the rotation chemical potential in terms of the complex structure parameter of the two-torus on which we place the theory and $y$ is the $R$-symmetry fugacity. We will write these in terms of the associated chemical potentials,
\begin{equation}
q = e^{2\pi i\, \tau} \,, \qquad y = e^{2\pi i\, z} \,.
\label{eq:chempots}
\end{equation}	
The chemical potentials $z$ can be viewed as the holonomy of a background $R$-gauge field. Other indices are related to various limits of the elliptic genus. The $\chi_y$-genus is obtained by taking the limit $q\to 0$ (equivalently $\tau \to i\,\infty$). 
Further setting the $R$-symmetry holonomy to zero,  
 $z=0$, we recover the Witten index $W_\text{RR}$ (which can be interpreted as giving us the Euler number of the target space):
\begin{equation}
\begin{split}
\chi_y(z) &= \lim_{q\to0}\mathcal{Z}(q, z)\,,\quad W_\text{RR}  =\chi_y(0)\,.
\end{split}
\label{eq:Wchi}
\end{equation}	

\subsection{Landau-Ginzburg tensor models}

The elliptic genus for Landau-Ginzburg models can be computed by a path integral with a certain twisted boundary condition \cite{Witten:1993jg}. The path integral is invariant under continuous variations of the superpotential, and hence it can be evaluated by the one-loop determinant of the chiral superfields. The elliptic genera of the tensor models introduced in Section \ref{sec:models} are
\begin{equation}
\begin{split}
\{\mathcal{Z}_{\mathscr{B}}(q,y),\mathcal{Z}_{\mathscr X}(q,y), \mathcal{Z}_{\mathscr Y}(q,y)\}
&=\{\mathcal{Z}_{0}(q,y)^{4N^3},
\mathcal{Z}_{0}(q,y)^{N^3},
\mathcal{Z}_{0}(q,y)^{N^2M}\},
\\
\mathcal{Z}_{0}(q,y)&=\frac{\theta_1(\tau|-{3\over 4}z) }{\theta_1(\tau|\frac{1}{4}z) }.
\end{split}
\end{equation}	
The $\chi_y$-genera and the Witten indices are
\begin{equation}
\begin{split}
\{\mathcal{\chi}_{y,\mathscr{B}}(y),\mathcal{\chi}_{y,\mathscr{X}}(y),\mathcal{\chi}_{y,\mathscr{Y}}(y)\}
&=\{\mathcal{\chi}_{y,0}(y)^{4N^3},
\mathcal{\chi}_{y,0}(y)^{N^3},
\mathcal{\chi}_{y,0}(y)^{N^2M}\},
\\
\mathcal{\chi}_{y,0}(y)&=y^{-\frac{1}{4}}+ 1+ y^\frac{1}{4},
\\
\{W_{RR,\mathscr{B}},W_{RR,\mathscr{X}},W_{RR,\mathscr{Y}}\}&=\{3^{4N^3},3^{N^3},3^{N^2M}\}.
\end{split}
\label{eq:}
\end{equation}
One can verify the central charge values in \eqref{eq:c4} via the modular property of the elliptic genus, viz., 
\begin{equation}
\mathcal{Z}\left(- \frac{1}{\tau}, \frac{z}{\tau}\right)=e^{c \, \frac{i\, \pi z^2}{3\tau} }\, \mathcal{Z}(\tau,z) \,.
\label{eq:}
\end{equation}	

\subsection{Gauged tensor models}

The elliptic genus of gauged linear sigma models can be computed by supersymmetric localization.  The result for the elliptic genus obtained in \cite{Benini:2013nda,Benini:2013xpa} uses as building blocks the contribution of the one-loop determinants from the chiral and gauge multiplets,
%The result is expressed as a sum over residues of the one-loop determinant of fields as a meromorphic form on the moduli space of flat connections on the torus \cite{Benini:2013nda,Benini:2013xpa}. The one-loop determinants of the chiral and gauge multiplets are
\ie
\mathcal{Z}_{\rm chiral}=\prod_{\rho\in{\cal R}}{\theta_1(q,y^{J_L-1}x^\rho)\over \theta_1(q,y^{J_L}x^\rho)},\quad \mathcal{Z}_{\rm gauge}=\left({i\eta(q)^3\over \theta_1(q,y^{-1})}\right)^r\prod_{\A:\,{\rm roots}}{\theta_1(q,x^\A)\over \theta_1(q,y^{-1}x^\A)},
\fe
where $r$ is the rank of the gauge group, $\rho$ is the weight of the representation ${\cal R}$, and $x$ is the gauge fugacity
\ie
 x^\rho = e^{2\pi i\, \rho(u)}.
\fe
The parameter $u$ is gauge holonomy valued in the maximal torus and is to be integrated over to pick out the contribution from the gauge invariant sector. Explicit expressions can be found in the references cited. The non-trivial aspect of the computation is to perform the integral over the gauge holonomies by a residue calculus. The prescription involves picking up the contributions from a subset of residues in the $\{u_a\}$-plane. We will implement their strategy below to extract the final answer. 

\subsubsection{${\rm U}(1)$ theory}

A special case worth considering is the isotropic model with $N=1$, $M=2$ and gauged ${\rm SO(2)}$ symmetry. In this case the superpotential is 
$({\mathscr Y}^1+{\mathscr Y}^2)^2$ which can be rotated to a single monomial ${\mathscr X}^2\,{\mathscr Y}^2$.  The superpotential and twisted superpotential after the field redefinitions are:
\begin{equation}
\int d\theta^+d\theta^-{\mathscr X}^2\,{\mathscr Y}^2+\int d\theta^+d\overline\theta^-\Sigma + c.c.
\label{eq:xysup}
\end{equation}
The superfield ${\mathscr X}$ has charge $+1$ and ${\mathscr Y}$ has charge $-1$ under the ${\rm SO}(2)\cong {\rm U}(1)$ gauge symmetry. 
This model clearly has flat directions which amusingly can be lifted by gauging.

By \eqref{eq:caniso}, the central charge of the resulting theory vanishes. The elliptic genus evaluates to unity. 
\begin{equation}
\begin{split}
\mathcal{Z}(\tau,z)&={i\eta(q)^3\over \theta_1(q,y^{-1})}\oint_{u=\frac{1}{4}z} du {\theta_1(q,y^{-{3\over 4}}x)\over \theta_1(q,y^{\frac{1}{4}}x)}\times {\theta_1(q,y^{-{3\over 4}}x^{-1})\over \theta_1(q,y^{\frac{1}{4}}x^{-1})}
\\
&=2\pi i\times{i\eta(q)^3\over \theta_1(q,y^{-1})}{\theta_1(q,y^{-\frac{1}{2}})\over \theta_1(q,y^{\frac{1}{2}})}\times {\theta_1(q,y^{-1})\over \theta_1'(q,1)}
\\
&=1,
\end{split}
\label{eq:}
\end{equation}

\subsubsection{${\rm SU}(2)$ and ${\rm SO}(3)$ gauge theories}

We will consider the ${\rm SU}(2)$ and the ${\rm PSU}(2)\cong{\rm SO}(3)$ theories, and focus on traceless chiral multiplets, i.e., fix $\Tr(Y^I)=0$, since the trace part is uncharged under the gauge group and would only contribute an overall factor:
\begin{equation}
\mathcal{Z}_{\rm trace}(q,y) =\left( \frac{\theta_1(\tau|-{3\over 4}z) }{\theta_1(\tau|\frac{1}{4}z) } \right)^M.
\label{eq:}
\end{equation}	

\paragraph{Case I ($N=2$, $M=1$):} To begin with, let us start with a single ${\rm SU}(2)$ adjoint chiral multiplet, i.e., with $N=2$ and $M=1$. The superpotential is $\Tr{Y^4}$ and the central charge $\hat{c} =-1$ from \eqref{eq:caniso}. The elliptic genus and associated indices evaluate to: 
\begin{equation}
\begin{split}
\mathcal{Z}(q,y)  &= \left({\theta_1(q,y^{-{1\over 4}})\over \theta_1(q,y^{{5\over 4}})}-{\theta_1(q,y^{-1})\theta_1(q,y^{{7\over 4}})\over\theta_1(q,y^{5\over 4})\theta_1(q,y^{2})}\right)\,,\\
\chi_y(y) &= \frac{y^{1\over 4}-\sqrt{y}+y^{3\over 4}}{1+y},  \\
W_\text{RR} &= \frac{1}{2}\,.
\end{split}
\label{eq:}
\end{equation}	
The singularity in the $\chi_y$-genus at $y=-1$ owes to the non-compact Coulomb branch of the theory, cf., \S\ref{sec:cbranch}. When the Higgs and Coulomb branches are compact we expect to see a finite polynomial for the $\chi_y$-genus reflecting the finiteness of the chiral ring \cite{Benini:2013nda}. 

\paragraph{Case II ($N=2$, arbitrary $M$):} We can extend this result to the case of interest where we have $M$ chiral multiplets transforming in the adjoint of ${\rm SU}(2)$.  The central charge is $\hat{c} = 2\, M -3$ from \eqref{eq:caniso} and the elliptic genus is given by 
\begin{equation}
\begin{split}
\mathcal{Z}(q,y) &=
	\frac{1}{2} \, \frac{i\eta(q)^3}{\theta_1(q,y^{-1})} \; \sum_{k,l=0}^1\oint_{\subalign{\\&u=\frac{1}{8} z+\frac{1}{2}(k+l\tau)\\ &u=-\frac{1}{2}z+\frac{1}{2}(k+l\tau)}} \, du \frac{\theta_1(q,x^2)}{ \theta_1(q,x^2y^{-1})} \, \frac{\theta_1(q,x^{-2})}{ \theta_1(q,x^{-2}y^{-1})} 
	\\
&\quad\times\left[{\theta_1(q,y^{-{3\over 4}})\over \theta_1(q,y^{{1\over 4}})}{\theta_1(q,y^{-{3\over 4}}x^2)\over \theta_1(q,y^{{1\over 4}}x^2)} {\theta_1(q,y^{-{3\over 4}}x^{-2})\over \theta_1(q,y^{{1\over 4}}x^{-2})}\right]^M
	\\
&=
	 \frac{i\eta(q)^3}{\theta_1(q,y^{-1})} \, \left(1+y^{2(M-1)}\right) 
	 \left(\frac{\theta_1(q,y^{-\frac{3}{4}})}{\theta_1(q,y^{\frac{1}{4}})}\right)^{M}
\\
&\quad \times  \oint_{\subalign{\\u&=\frac{1}{ 8}z\\u&=-\frac{1}{2}z}} du \, \frac{\theta_1(q,x^2)}{\theta_1(q,x^2y^{-1})}
	 	\frac{\theta_1(q,x^{-2})}{ \theta_1(q,x^{-2}y^{-1})} \; 
	\left( \frac{\theta_1(q,y^{-\frac{3}{4}}x^2)}{\theta_1(q,y^{\frac{1}{4}}x^2)}  \ 
	\frac{\theta_1(q,y^{-\frac{3}{4}}x^{-2})}{\theta_1(q,y^{\frac{1}{4}}x^{-2})}\right)^M \,.
\end{split}
\label{eq:}
\end{equation}
The integral over $u$ is localized about the zeros of some of the $\theta_1$ in the denominator. They can be computed using residues which requires us to deal with the $M^{\rm th}$ order zero arising from $\theta_1(q,y^{\frac{1}{4}}x^{-2})$.

The non-compact Coulomb branch is again manifested by the singular behaviour of the $\chi_y$-genus. One can more simply examine the 
Witten indices to confirm this. They are given by
\begin{equation}
W_\text{RR}\bigg|_{M=2} = -\frac{5}{2} \,,  \qquad W_\text{RR}\bigg|_{M=3} = \frac{61}{2} \,,  \cdots 
\label{eq:}
\end{equation}	

\paragraph{Case III (${\rm PSU}(2)$ arbitrary $M$):} The situation with ${\rm SU}(2)$ gauge theory is that we have a non-compact Coulomb branch giving rise to some of the singular features seen above. Our discussion in \S\ref{sec:cbranch} and \S\ref{sec:hbranch} leads us to believe that the moduli space is compact when we consider gauging ${\rm PSU}(2) \cong {\rm SO}(3)$ in a theory with an anisotropic superpotential.

The computation for the elliptic genus can then be adapted from the results of \cite{Kim:2017zis}. It is given by 
\begin{equation}
\begin{split}
\mathcal{Z}(\tau, z) & = 
	\mathcal{Z}_+(\tau,z)  + e^{i\theta_{\rm bare}} \mathcal{Z}_-(\tau,z) \,,  \\
\mathcal{Z}_+(\tau,z)  & = 
	\frac{1}{2} \, \frac{i\eta(q)^3}{\theta_1(\tau|-z)}
	\oint_{u=\frac{1}{4}z,u=-z} du  \, \frac{\theta_1(\tau|u)\theta_1(\tau|-u)}{\theta_1(\tau|u-z)\theta_1(\tau|-u-z)}
\\ &
	\qquad \qquad \times\left(
	\frac{\theta_1(\tau|-\frac{3}{4} \, z)\,\theta_1(\tau|u-\frac{3}{4} \, z)\theta_1(\tau|-u-\frac{3}{4} \, z)}{\theta_1(\tau|\frac{1}{4}z)\, 
	\theta_1(\tau|u+\frac{1}{4}z) \, \theta_1(\tau|\frac{1}{4}z-u)}\right)^M \,, \\
\mathcal{Z}_-(\tau,z)  & = 
	\frac{1}{4} \,	\frac{\theta_2(\tau|0)\theta_3(\tau|0)\theta_4(\tau|0)}{\theta_2(\tau|-z)\theta_3(\tau|-z)\theta_4(\tau|-z)}
	\left(
	\frac{\theta_1(\tau|-\frac{3}{4}\, z)\,\theta_2(\tau|-\frac{3}{4}\, z)\theta_3(\tau|-\frac{3}{4}\, z)}{\theta_2(\tau|\frac{1}{4}z)
	\,\theta_3(\tau|\frac{1}{4}z)\,\theta_4(\tau|\frac{1}{4}z)}\right)^M\,.
\end{split}
\label{eq:so3eg}
\end{equation}
Given this we can obtain the $\chi_y$ genus as:
\begin{equation}
\begin{split}
\chi_y(z)  &= 
	\chi_+(z) + \chi_-(z) \,, \\
\chi_+(z)&=
	-\frac{i}{4\,\sin\pi z} \,
	\oint_{u=\{\frac{1}{4}z,-z\}} du \,
	\frac{\sin^2\pi u}{\sin\pi(u-z)\sin\pi(u+z)}
	\left(\frac{\sin\left(\frac{3\pi}{4} z\right)	\sin\pi(u-\frac{3}{4}\, z)\, \sin\pi(u+\frac{3}{4}\, z) }{\sin\left( \frac{\pi}{4}z\right)
	\, \sin\pi(u+\frac{1}{4}\, z)\sin\pi(u-\frac{1}{4}\, z)}\right)^M,\\
\chi_-(z)&=
	\frac{1}{4\, \cos\pi z}\left(- \frac{\cos \left(\frac{3\pi}{4}z\right)}{ z \,\cos \left(\frac{\pi}{4}z\right)}\right)^M .
\end{split}
\label{eq:chiyo3}
\end{equation}
This is in perfect agreement with our analysis of Higgs and Coulomb branches. For specific values of $M$ and choices of the bare theta angle, $\theta_\text{bare}$, we have
\begin{align}
&\chi_y=0&&{\rm for}~M=1,~\theta_{\text{bare}}=\pi,
\nonumber \\
&\chi_y=-1&&{\rm for}~ M=2,~\theta_{\text{bare}}=0,
\nonumber \\
&\chi_y=y^{-{3\over 4}}+6y^{-\frac{1}{4}}+1+6y^{\frac{1}{4}}+y^{{3\over 4}}&&{\rm for}~ M=3,~\theta_{\text{bare}}=\pi.
\label{eq:chiyspeco3}
\end{align}
The Witten indices are also likewise commensurate; we obtain integral answers for the expected cases where the moduli space is compact. To wit, 
\ie
&\theta_{\text{bare}} = 0&:&\quad \frac{1}{2},\,-1,\,{31\over 2},\,-202,\,\cdots,
\\
&\theta_{\text{bare}} = \pi&:&\quad 0,\,-{3\over 2},\,15,\,-{405\over 2},\,\cdots.
\label{eq:windo3}
\fe

%~~~~~~~~~~~~~~~~~~~~~~~~~~~~~~~~~~~~~~~~~~~~~~
\section{Discussion}
\label{sec:discussion}
%~~~~~~~~~~~~~~~~~~~~~~~~~~~~~~~~~~~~~~~~~~~~~~

The main thrust of our analysis was to construct large central charge CFTs in two spacetime dimensions, exploiting the large $N$ solvability of melonic theories. The advantage of the melonic dominance, as is by now well known, is that the Schwinger-Dyson equations for the theory truncate within sub-sectors of few-body interactions, leading to a closed set of equations for the physical correlation functions. To mitigate issues arising from tensor valued bosonic field theories having Hamiltonians that are not bounded from below, we focused on theories with $(2,2)$ supersymmetry, where we moreover can exploit non-renormalization theorems. 

Somewhat surprisingly, the simplest class of models with tensor valued fields with superpotentials constrained to have tetrahedral index contractions (or generalizations thereof), to guarantee melonic dominance, suffer generically from a classical moduli space of vacua. This makes it hard to unambiguously argue, without more detailed insight into the dynamics, that the theories do flow to an IR superconformal field theory. While we were able to establish that such constructions work for low $\mathcal{O}(1)$ central charge, large $c$ theories required us to consider generalizations in the form of anisotropic superpotentials. We demonstrated that with suitable choice of such anisotropy one can indeed construct LG theories with an isolated vacuum which is picked out by the low energy theory. Not only could we argue for the presence of a superconformal fixed point using standard arguments relying on supersymmetry, but we could also establish the same by solving the truncated set of Schwinger-Dyson equations to verify the claim. Furthermore, the latter analysis enabled us to determine a part of the low energy spectrum, especially the states appearing in the OPE in the singlet channel between a chiral operator and its anti-chiral partner. 

In the current discussion we only examined the complete details of the anisotropic models in two special cases: perturbation theory around the isotropic point and for a specific choice of anisotropy which preserves a large discrete symmetry. From the viewpoint of establishing a low energy fixed point, this was sufficient. However, the anisotropic models themselves warrant further investigation, especially to determine how the spectrum shifts as we vary the anisotropy. For instance, in the class of examples we studied, one finds no real hierarchy in the spectrum between the stress tensor and other composite operators. This, in particular, ends up implying that these theories have a sub-maximal chaos exponent. It would be interesting to know how much better one can do, to inch closer towards the bound  of \cite{Maldacena:2015waa}, by tuning the anisotropy.

There is one advantage of the anisotropic models in that the large continuous global symmetry of the isotropic models is completely broken to a discrete subgroup. For one this should imply a lifting of degeneracy in the non-singlet spectrum, perhaps even mitigating the rapid growth of states noted in \cite{Bulycheva:2017ilt,Choudhury:2017tax} who found that the isotropic models in quantum mechanics have a Hagedorn growth of states with a vanishing Hagedorn temperature.  Furthermore, the explicit breaking of continuous global symmetries in the microscopic description should also help with taming the instability of the IR fixed point noted in \cite{Murugan:2017eto}. Theories with large global symmetry have a plethora of marginal operators which could end up moving one away from the naive fixed point.

It is also worth comparing the anisotropic tensor models with disordered models of the SYK type. Apart from the fact that the anisotropic tensor models are genuine quantum systems, in turning on the anisotropic deformation we are inching closer towards the disordered models. The key difference is that the tensor structure of the fields ensures the melonic dominance directly without having to average the couplings over some ensemble.  Should we choose, we could nevertheless average the anisotropy parameters; we expect the averaged observables to coincide with those of the isotropic model. 

Apart from breaking the symmetry by making the couplings anisotropic, we have also examined whether we can gauge the global symmetry of the isotropic tensor model, and obtain a reliable low energy superconformal fixed point. This turns out to be insufficient, with the resulting gauged models generically having a non-compact Higgs branch. As noted in the text, there appear to be particular choices for which the theory has compact Higgs and Coulomb branches, but thus far we have only been able to establish such for low values of the central charge. 

Speaking of gauging, it is also worth remarking that one could have contemplated upgrading the models to gauged linear sigma models and examined the resulting phase structure for potential connections to non-linear sigma models with Calabi-Yau target \cite{Witten:1993yc}. For instance, focusing on the anisotropic uncolored tensor model \eqref{eqn:deformed_superpotential}, consider introducing an additional chiral superfield $\mathscr{U}$, upgrading the superpotential to $\mathscr{U} \, W_4(\mathscr{X})$ and gauge an abelian symmetry where $\mathscr{X}$ has charge $+1$ and $\mathscr{U}$ has charge $-4$ (by homogeneity). This theory suffers from a chiral anomaly for the ${\rm U}(1)_A$ symmetry, with the anomaly being proportional to $N^3 -4$. In the UV one can realize the theory as a non-linear sigma model with target, the hypersurface $W_4(\mathscr{X}) =0$ in $\mathbb{CP}^{N^3}$. However, owing to the anomaly we are not guaranteed that the low energy theory is described by the LG model we favored. The theory could instead flow to a gapped phase, see \cite{Witten:1993yc,Deligne:1999qp} for details. 

Let us finally turn to potential applications to holography and various generalizations of our construction that are worth pursuing.

\paragraph{Bulk dual:} The first question we need to address for holographic considerations is  what is the bulk dual of our LG models. 
In our LG (isotropic or anisotropic) models, the number of operators below a fixed finite dimension $\Delta_*$ grows with $N$ in the large $N$ limit. A subset of them are the chiral ring operators given by polynomials of the chiral superfields ${\mathscr X}^{a_1a_2a_3}$ subject to the $F$-term constraints.  Under the AdS/CFT correspondence, these operators correspond to massive bulk fields with finite masses.  The existence of such large number of light fields prevents the bulk theory to be a gravity theory. In particular, it violates the sparseness condition \cite{Hartman:2014oaa} and hence, cannot have a Hawking-Page phase transition at finite temperature.

To kill these light modes, we could consider the $(\bZ_2^N\rtimes S_N)^3$ invariant anisotropic tensor model and gauge (orbifold) the $S_N^3$ subgroup. The gauging clearly kills a lot of light operators, but the question is whether it kills enough light operators for the theory to admit a bulk gravity dual, in particular, to satisfy the sparseness condition. This question can be partially answered by studying the elliptic genus of the orbifold theory. The elliptic genus depends on the superpotential only via the ${\rm U}(1)_V$ charge assignment of the chiral superfields such that the superpotential has ${\rm U}(1)_V$ charge 2. Therefore, we can simply consider the elliptic genus of the $S_N^3$ orbifold of a free chiral superfield ${\cal X}$. From the results in 
\cite{Haehl:2014yla,Belin:2014fna,Benjamin:2015hsa,Belin:2015hwa}, the density of states of the $S_N^3$ orbifold grows at least exponentially in the dimension of the operator, which is still in tension with the sparseness condition. We will leave a more detailed analysis of the elliptic genus and obtaining the bulk dual of these LG models for the future.

\paragraph{Generalizations:}

The anisotropic deformation can be generalized to theories in other dimensions. In three and four dimensions, the quartic superpotentials are either marginally irrelevant or irrelevant.  In order to have nontrivial IR dynamics, one can instead consider non-supersymmetric tensor models with the anisotropic tetrahedral interaction, which is relevant in three dimensions and  marginally irrelevant in four dimensions. The $\bZ_2^{3N}$ symmetry restricts the RG flow such that only the anisotropic versions of the tetrahedral, pillow, and double-sum terms can be generated.  In \cite{Giombi:2017dtl}, by studying the $\epsilon$-expansion in $d=4-\epsilon$ dimensions, the authors found that the ${\rm O}(N)^{q-1}$ invariant fixed points only exist with complex pillow and double-sum coupling constants. It would be interesting to examine if finite anisotropy helps in obtaining real fixed points.

The anisotropic deformations of the quantum mechanical melonic tensor model in $d=1$ deserves further scrutiny. For instance, they could prove helpful for numerical explorations of the tensor models (cf., \cite{Pakrouski:2018jcc} for the state of the art), especially to determine whether the spectrum displays random matrix characteristics (which one suspects they do). Likewise as noted earlier, they help lift the $\mathcal{O}(N^2)$ light modes in the isotropic theory, which are associated with time-dependent ${\rm O}(N)^{q-1}$ rotations \cite{Choudhury:2017tax}, by breaking the global symmetry explicitly. The low energy dynamics is likely to be governed by the reparametrization mode alone. One might therefore wonder if they might admit a holographic dual in terms of the  Jackiw-Teitelboim dilaton gravity theory as SYK  at low energies.  

The anisotropic deformations can be generalized to other large $N$ models. For example, consider the ${\rm O}(N)$ vector model, where one can turn on an anisotropic deformation in the potential $V(\phi)$ that breaks the ${\rm O}(N)$ global symmetry down to the discrete $\bZ_2^N$ subgroup,
\ie
V(\phi)={1\over 4}g\sum_{a,b=1}^N \A_{ab}(\phi^a)^2(\phi^b)^2.
\fe
Similar to the anisotropic tensor models, when all the anisotropic coefficients $\A_{ab}$ are of the same order in the large $N$ limit, the model is solvable in the same way as the (isotropic) critical ${\rm O}(N)$ vector model. In general, the anisotropic deformation parameters $\A_{ab}$ flow under the RG. The IR fixed points could be studied by performing an $\epsilon$-expansion in $d=4-\epsilon$ dimensions. For a special class of anisotropic vector models that preserve the $\bZ_2^N\rtimes S_N$ symmetry, it has been argued that there exist four fixed points: the free ${\rm O}(N)$ fixed point, the critical ${\rm O}(N)$ fixed point, the $N$ decoupled Ising models fixed point, and an anisotropic fixed point \cite{Fei:2015oha}. For $N>4$, the anisotropic fixed point is IR stable, and all the other fixed points can flow to it by turning on relevant deformations. However, this anisotropic fixed point lies outside the purview of large $N$ techniques, for at this fixed point the anisotropy parameters are not all of the same order in the large $N$ limit.

%~~~~~~~~~~~~~~~~~~~~~~~~~~~~~~~~~~~~~~~~~~~~~~~
\acknowledgments
%~~~~~~~~~~~~~~~~~~~~~~~~~~~~~~~~~~~~~~~~~~~~~~

It is a pleasure to thank  Kentaro Hori, Veronika Hubeny, Igor Klebanov, Emil Martinec, Shiraz Minwalla, Sun Woo Park, Cheng Peng, Douglas Stanford, Cumrun Vafa, Johannes Walcher,  Edward Witten, and Xi Yin for illuminating discussions. We would like to thank KITP, UCSB for hospitality during the workshop ``Chaos and Order: From strongly correlated systems to black holes'', where the research was supported in part by the National Science Foundation under Grant No. NSF PHY17-48958 to the KITP.
The authors are supported  by U.S.\ Department of Energy grant DE-SC0009999 and by funds from the University of California. 

\appendix
 
%~~~~~~~~~~~~~~~~~~~~~~~~~~~~~~~~~~~~~~~~~~~~~~~
\section{Supersymmetry conventions}
\label{sec:conventions}
%~~~~~~~~~~~~~~~~~~~~~~~~~~~~~~~~~~~~~~~~~~~~~~

We provide our conventions for $\mathcal{N}=(2,2)$ supersymmetry in $d=2$ dimensions following \cite{Witten:1993yc}, which we rewrite in complex coordinates in Euclidean spacetime. 
We work in superspace with coordinates ${\sf Z} = (z,\overline{z},\theta^{\pm},\overline{\theta}^{\pm})$. We contract spinor indices using the tensor $\epsilon_{\alpha\beta}$, where we adopt the convention $\epsilon^{12} = \epsilon_{12} = 1$, i.e. $\psi^{\alpha} = \epsilon^{\alpha\beta}\psi_{\beta}$ and $\psi_{\alpha} = \psi^{\beta}\epsilon_{\beta\alpha}$.

The supersymmetry generators are
\begin{equation}
  Q_{-} = \frac{\partial}{\partial \theta^{-}}+\overline{\theta}^{-}\partial_{\overline{z}} \qquad \mathrm{and} \qquad Q_{+} = \frac{\partial}{\partial \theta^{+}}+\overline{\theta}^{+}\partial_{z},
\end{equation} 
 and
\begin{equation}
  \overline{Q}_{-} = -\frac{\partial}{\partial \overline{\theta}^{-}}-\theta^{-}\partial_{\overline{z}} \qquad \mathrm{and} \qquad \overline{Q}_{+} = -\frac{\partial}{\partial \overline{\theta}^{+}}-\theta^{+}\partial_{z}.
\end{equation}
The superderivatives are 
\begin{equation}
 D_{-} = \frac{\partial}{\partial \theta^{-}}-\overline{\theta}^{-}\partial_{\overline{z}} \qquad \mathrm{and} \qquad D_{+} = \frac{\partial}{\partial \theta^{+}}-\overline{\theta}^{+}\partial_{z}, 
 \end{equation}
and
\begin{equation}
 \overline{D}_{-} = -\frac{\partial}{\partial \overline{\theta}^{-}}+\theta^{-}\partial_{\overline{z}} \qquad \mathrm{and} \qquad \overline{D}_{+} = -\frac{\partial}{\partial \overline{\theta}^{+}}+\theta^{+}\partial_{z}.
\end{equation}
The chiral variables which we work with are defined to be: 
\begin{equation}
  y^+ = z  - \theta^{+}\overline{\theta}^{+},\quad y^- = \overline z  - \theta^{+}\overline{\theta}^{+}, \qquad \mathrm{and} \qquad \overline{y}^+ = z + \theta^{-}\overline{\theta}^{-},\quad \overline{y}^- = \overline{z} + \theta^{-}\overline{\theta}^{-}. 
\label{eq:ydef}
 \end{equation}
The Grassmann integration is defined to be 
\begin{equation}
\begin{split}
\int d^{4}\theta &= \int d\theta^{+}\,d\theta^{-}\,d\overline{\theta}^{-}\,d\overline{\theta}^{+}, \\ 
\int d^{2}\theta &= \int d\theta^{+}\,d\theta^{-}, \qquad \int d^{2}\overline{\theta} = \int d\overline{\theta}^{-}\,d\overline{\theta}^{+}. 
\end{split}
\end{equation}
Finally, the Grassmann delta function  $\delta^{(4)}(\theta) = \theta^{-}\theta^{+}\overline{\theta}^{+}\overline{\theta}^{-} = \theta^2\overline{\theta}^2 \; \Longrightarrow \; \int d^{4}\theta\,\delta^{(4)}(\theta) = 1$.

While we do not explicitly need the information below, let us also record the superfield expansion of the vector multiplet:
\begin{equation}
\begin{split}
2V_{a} = \theta^{-}\overline{\theta}^{-}[(v_{a})_{0}-(v_{a})_{1}]+\theta^{+}\overline{\theta}^{+}[(v_{a})_{0}+(v_{a})_{1}]-\sqrt{2}\theta^{-}\overline{\theta}^{+}\sigma-\sqrt{2}\theta^{+}\overline{\theta}^{-}\overline{\sigma}  
\\	+\sqrt{2}i\overline{\theta}^{+}\overline{\theta}^{-}[\theta^{+}(\lambda_{a})_{+}+\theta^{-}(\lambda_{a})_{-}]+\sqrt{2}i\theta^{-}\theta^{+}[\overline{\theta}^{+}(\overline{\lambda}_{a})_{+}+\overline{\theta}^{-}(\overline{\lambda}_{a})_{-}]
\\	+\theta^{-}\theta^{+}\overline{\theta}^{+}\overline{\theta}^{-}D_{a}.
\end{split}
\label{eqn:vectorsuperfield}
\end{equation}
%

%~~~~~~~~~~~~~~~~~~~~~~~~~~~~~~~~~~~~~~~~~~~~~~~
\section{Unitarity of spectrum}
\label{sec:unitarity}
%~~~~~~~~~~~~~~~~~~~~~~~~~~~~~~~~~~~~~~~~~~~~~~

In this appendix, we prove the assertion made in \S\ref{sec:spectrum} that the spectrum of the isotropic tensor model is unitary. That is, we show that the kernel eigenvalue equation \eqref{eqn:eigenvalue_eq} only has solutions for $h \geq 0$ and $\overline{h} \geq 0$. Our choice of contour deformation was made to ensure that the four-point function is convergent, but one can alternatively understand this choice in terms of unitarity because we must have $u=is \in \mathbb{R}_{\geq 0}$ to have any hope of finding $h = \frac{\ell}{2}+u \geq 0$ and $\overline{h} = -\frac{\ell}{2}+u \geq 0$. However, this choice of contour deformation is still not enough to ensure unitarity since $\ell$ can be arbitrarily large and potentially lead to negative $h$ or $\overline{h}$. We will now see that this problem does not occur.

The spectrum is clearly unitary for $\ell = 0$ because in this case $h=\overline{h}=u \in \mathbb{R}_{\geq 0}$. For $\ell \neq 0$, it suffices to focus on the choice $\ell > 0$ because  
$\ell \rightarrow -\ell$ is equivalent to $h \rightarrow \overline{h}$. One can verify that the kernel eigenvalue \eqref{eq:kevals} is invariant under such a transformation, namely $k(h,\overline{h}) = k(\overline{h},h)$, and hence any statement about the kernel for $\ell > 0$ also holds for $\ell < 0$. Therefore, we restrict ourselves to the case $\ell > 0$. In this case, we have $h > 0$ so we need to check that there are no solutions to \eqref{eqn:eigenvalue_eq} when $\overline{h} < 0$, or equivalently when $0 \leq u < \frac{\ell}{2}$. To demonstrate this, we will prove the following:
\begin{equation}
k(h,h-\ell) < 1 \qquad \mathrm{for\;\;} \frac{\ell}{2} \leq h < \ell.
\label{eq:unitarityclaim}
\end{equation}
We will restrict ourselves to $q=4$ for simplicity. We split the proof into two cases: $\ell$ even and $\ell$ odd.

\noindent
\underline{{\bf 1.} $\ell$ even:} For $\ell$ even, it is easier to prove the stronger bound $k(h,h-\ell) < 0$. Using the Gamma function identity $\Gamma(x)\Gamma(1-x) = \frac{\pi}{\sin \pi x}$, we can rewrite \eqref{eq:kevals} as
\begin{equation}
k(h,h-\ell) = -\frac{6\pi^2}{\Gamma(\frac{1}{4})^{4}}\frac{\Gamma(h+\frac{1}{4})\Gamma(\frac{1}{4}+\ell-h)}{\Gamma(h+\frac{3}{4})\Gamma(\frac{3}{4}+\ell-h)} < 0.
\label{eq:kevaleven}
\end{equation}
The inequality follows from the fact that in the range $\frac{\ell}{2} \leq h < \ell$, the arguments of all of the Gamma functions are positive, and hence all the Gamma functions are positive.

\noindent
\underline{{\bf 2}. $\ell$ odd:} The $\ell$ odd case is more involved than the $\ell$ even case. For $\ell=1$, we have not been able to rigorously prove \eqref{eq:unitarityclaim} owing to the fact that the bound is saturated at $h=1$. Nevertheless, we have numerically verified that the kernel eigenvalue for $\ell = 1$ satisfies the inequality $\eqref{eq:unitarityclaim}$ when $h \neq 1$. 
For $\ell \geq 3$, we first repeat the step in the $\ell$ even case to rewrite the kernel eigenvalue in a form where all the Gamma functions have a positive argument: 
\begin{equation}
k(h,h-\ell) = \frac{6\pi^2}{\Gamma(\frac{1}{4})^{4}}\frac{\Gamma(h+\frac{1}{4})\Gamma(\frac{1}{4}+\ell-h)}{\Gamma(h+\frac{3}{4})\Gamma(\frac{3}{4}+\ell-h)}.
\label{eq:kevalodd}
\end{equation}
Next, we use Wendel's inequality
\begin{equation}
\frac{\Gamma(x)}{\Gamma(x+t)} \leq \frac{(x+t)^{1-t}}{x}, \qquad 0 < t < 1, \; 0 < x
\end{equation}
to obtain
\begin{equation}
\frac{\Gamma(h+\frac{1}{4})}{\Gamma(h+\frac{3}{4})} \leq \frac{(h+\frac{3}{4})^{\frac{1}{2}}}{(h+\frac{1}{4})} \qquad \mathrm{and} \qquad \frac{\Gamma(\frac{1}{4}+\ell-h)}{\Gamma(\frac{3}{4}+\ell-h)} \leq \frac{(\frac{3}{4}+\ell-h)^{\frac{1}{2}}}{(\frac{1}{4}+\ell-h)}.
\end{equation}
Inserting these inequalities into \eqref{eq:kevalodd}, we find
\begin{equation}
k(h,h-\ell) \leq \frac{6\pi^2}{\Gamma(\frac{1}{4})^{4}}\frac{(h+\frac{3}{4})^{\frac{1}{2}}}{(h+\frac{1}{4})}\frac{(\frac{3}{4}+\ell-h)^{\frac{1}{2}}}{(\frac{1}{4}+\ell-h)}.
\end{equation}
One can check that the righthand side of this inequality is a monotonically increasing function of $h$ in the range $\frac{\ell}{2} \leq h \leq \ell$ so it attains a maximum at $h=\ell$, which gives
\begin{equation}
k(h,h-\ell) < \frac{12\sqrt{3}\pi^2}{\Gamma(\frac{1}{4})^{4}}\frac{(\ell+\frac{3}{4})^{\frac{1}{2}}}{(\ell+\frac{1}{4})}.
\end{equation}
Finally, one can check that the righthand of this inequality is a monotonically decreasing function of $\ell$ for $\ell \geq 3$ so it attains a maximum at $\ell = 3$, and hence
\begin{equation}
k(h,h-\ell) < \frac{12\sqrt{3}\pi^2}{\Gamma(\frac{1}{4})^{4}}\frac{(\frac{15}{4})^{\frac{1}{2}}}{(\frac{13}{4})} = \frac{72\sqrt{5}\pi^2}{13\Gamma(\frac{1}{4})^{4}} \approx 0.7 < 1.
\end{equation}
This completes the proof of unitarity of the spectrum.

% 
%~~~~~~~~~~~~~~~~~~~~~~~~~~~~~~~~~~~~~~~~~~~~~~~
\section{Trivial moduli space for anisotropic tensor model}
\label{sec:nomoduliproof}
%~~~~~~~~~~~~~~~~~~~~~~~~~~~~~~~~~~~~~~~~~~~~~~

In this appendix, we provide the details of the proof of the Theorem in $\S$\ref{sec:trivialmodulianiso}. The mathematical results we need come from the theory of resultants \cite{Cox:2005ab}. We begin by reviewing the general theory of resultants\footnote{We actually will not need the resultant, but a closely related, simpler polynomial which has the resultant as its divisor.} following which we then discuss our specific case of interest. 

%~~~~~~~~~~~~~~~~~~~~~~~~~~~~~~~~~~~~~~~~~~~~~~
\subsection{Resultants}
%~~~~~~~~~~~~~~~~~~~~~~~~~~~~~~~~~~~~~~~~~~~~~~

Consider the homogeneous polynomials $F_{0},\ldots,F_{n} \in \mathbb{C}[x_{0},\ldots,x_{n}]$ of degree $d_{0},\ldots,d_{n}$, respectively. Define
\begin{equation}\label{eqn:dvar}
d = \sum_{i=0}^{n}(d_{i}-1)+1 = \sum_{i=0}^{n}d_{i} - n.
\end{equation}
It is important to note that $d \geq \mathrm{max}(\{d_{i}\})$ so we can multiply each $F_{0},\ldots,F_{n}$ by suitable powers of $x_{0},\ldots,x_{n}$ to obtain polynomials of degree $d$. We write monomials of degree $d$ as $x^{\rho} \equiv x_{0}^{\rho_{1}}\ldots x_{n}^{\rho_{n}}$. The choice of $d$ is due to the following important Lemma.\\

\noindent\underline{\textbf{Lemma:}} Each monomial of degree $d$ is divisible by $x_{i}^{d_{i}}$ for some $0 \leq i \leq n$.

\noindent\textit{Proof:} Let $x^{\rho}$ be a monomial of degree $d$. Suppose, to reach a contradiction, that $x_{i}^{\rho_{i}}$ is not divisible by $x_{i}^{d_{i}}$ for all $0 \leq i \leq n$. Then $\rho_{i} \leq d_{i}-1$ for all $0 \leq i \leq n$, and hence $\sum_{i=0}^{n}\rho_{i} \leq \sum_{i=0}^{n}(d_{i}-1) < d$, which is a contradiction since $x^{\rho}$ has degree $d$. $\square$\\
	
Denote the set of degree $d$ monomials by $\mathcal{S}$. Now consider the partition of $\mathcal{S}$ into $n+1$ mutually disjoint sets determined by the degrees of the polynomials $F_{0},\ldots,F_{n}$:
\begin{equation}\label{eqn:degdpolypartition}
\mathcal{S}_{i} = \{x^{\rho}\,|\,x_{0}^{d_{0}}, \ldots, x_{i-1}^{d_{i-1}} \; \mathrm{\;don't\;divide\;}x^{\rho},\;\mathrm{but\;}x_{i}^{d_{i}}\mathrm{\;does}\}.
\end{equation}
By the Lemma, each degree $d$ monomial lies in one of the $\mathcal{S}_{i}$. Observe that the total number of degree $d$ monomials in $n+1$ variables is given by the multichoose of length $d$ on $n+1$ symbols, viz., 
\begin{equation}\label{eqn:degdpolycount}
|\mathcal{S}|  \equiv {\sf S} = {{n+d}\choose{d}} = {{n+d}\choose{n}}.
\end{equation}
From each of the polynomials $F_{0},\ldots,F_{n}$, we construct the sets of polynomials
\begin{equation}\label{eqn:subsetsofdegdpolys}
\mathcal{P}_{i} = \bigg\{\frac{x^{\rho}}{x_{i}^{d_{i}}}F_{i}\,\bigg|\,x^{\rho} \in \mathcal{S}_{i}\bigg\}.
\end{equation}
Each set $\mathcal{P}_{i}$ contains polynomials of degree $d$ since $x_{i}^{d_{i}}$ divides $x^{\rho} \in \mathcal{S}_{i}$ by definition and $(x^{\rho}/x_{i}^{d_{i}}) \cdot F_{i}$ has degree $d$. The crucial point is that the union of these sets of polynomials
\begin{equation}\label{eqn:largerpolyset}
\mathcal{P} = \bigcup_{i=0}^{n}\, \mathcal{P}_{i}
\end{equation}
consists of ${\sf S}$  polynomials, each of which is a function of the ${\sf S}$  degree $d$ monomials.

We have now linearized the problem of finding whether the set of equations $F_{i} = 0$  for $i=0, \ldots, n$ has no non-trivial solution, as  explained in $\S$\ref{sec:trivialmodulianiso}. Recall that any solution to the set of equations $F_{i} = 0$ is also a solution to the set of equations $\{g = 0\,|\,g \in \mathcal{P}\}$ by the construction of $\mathcal{P}$. The latter set of equations form a system of linear equations in the degree $d$ monomials. That is, if we define $\mathbf{M}_{d}$ to be a vector of all degree $d$ monomials and further define $\mathfrak{C}$ to be the ${\sf S} \times {\sf S}$ matrix of coefficients of the polynomials of $\mathcal{P}$, then the set of equations $\{g = 0\,|\,g \in \mathcal{P}\}$ can be rewritten as the linear equation 
\begin{equation}\label{eqn:linearizingsolns}
\mathfrak{C}\;\mathbf{M}_d = \mathbf{0}_{{\sf S}}\,.
\end{equation}
This linear equation has a non-trivial solution if and only if $\mathfrak{D}_{n} \equiv \det(\mathfrak{C}) = 0$. It follows that if $F_{0} = \ldots = F_{n} = 0$ has a non-trivial solution, then $\mathfrak{D}_{n} = 0$. 

We can construct $\mathfrak{C}$ explicitly as follows. We write the polynomials explicitly as
\begin{equation}\label{eqn:explicitpolys}
F_{i} = \sum_{j=1}^{m_{i}}c_{ij}\, x^{\gamma_{ij}},
\end{equation}
where $m_{i}$ is the number of monomials appearing in $F_{i}$. The exponents of the monomials appearing in $F_{i}$ define a set
\begin{equation}\label{eqn:supportf_i}
\mathcal{A}_{i} = \{\gamma_{i1},\ldots,\gamma_{im_{i}}\} \subset \mathbb{Z}^{n+1},
\end{equation}
which is called the support of $F_{i}$. 

Now let $ \mathcal{R}_{i} \subset \mathbb{Z}^{n+1}$ be the set of exponents $\rho$ appearing in the set $\mathcal{S}_{i}$. We label matrix elements of the linear operator $\mathfrak{C}$ we seek to construct by the monomial exponents. This can be done as follows: for $\rho \in  \mathcal{R}_{i}$ we take $\mathfrak{C}_{\rho\kappa}$ to be the matrix element for the coefficient of the monomial $x^{\kappa}$ appearing in $(x^{\rho}/x_{i}^{d_{i}})\, F_{i}$. Then the matrix $\mathfrak{C}$ is given by 
\begin{equation}\label{eqn:coeffmatrix}
\mathfrak{C}_{\rho\kappa} =
 \begin{cases} c_{ik} & \;
 \mathrm{if\;}\;	\kappa+d_{i}\,\hat{e}_{i}-\rho = \gamma_{ik} \in \mathcal{A}_{i}, \\ 
 0\;	  
 & \mathrm{if\;}\; \kappa+d_{i}\,\hat{e}_{i}-\rho \not \in \mathcal{A}_{i}, \\ \end{cases} \qquad \rho \in  \mathcal{R}_{i}.
\end{equation}
where $\hat{e}_{i}$ is the $i^{\rm th}$ vector in the standard basis on $\mathbb{R}^{n+1}$.

%~~~~~~~~~~~~~~~~~~~~~~~~~~~~~~~~~~~~~~~~~~~~~~
\subsection{Proof of the Theorem}
%~~~~~~~~~~~~~~~~~~~~~~~~~~~~~~~~~~~~~~~~~~~~~~

We are now ready to apply this construction and prove the Theorem quoted in \S\ref{sec:trivialmodulianiso}. 
In our particular case,
\begin{equation}\label{eqn:valscaseofinterest}
n = N^3-1, \qquad d = 2N^3+1, \qquad {\sf S} = {{3\,N^3}\choose{N^3-1}}.
\end{equation}
The set $\mathcal{P}$ of ${\sf S}$ polynomials in ${\sf S}$ degree $d$ monomials is given by
\begin{equation}\label{eqn:polysetcaseofinterest}
\mathcal{P} = \bigcup_{c_1,c_2,c_3=1}^{N}\bigg\{\frac{\mathscr{X}^{\rho}}{(\mathscr{X}^{c_1c_2 c_3})^{3}} \; \mathfrak{f}_{c_1 c_2 c_3}\,\bigg|\ 
\mathscr{X}^{\rho} \in \mathcal{S}_{c_1 c_2 c_3}\bigg\}.
\end{equation}
As noted above, the statement of Theorem 1 reduces to a problem about the determinant of the coefficient matrix of $\mathcal{P}$, viz., 
$\{f_{c_1 c_2 c_3} = 0\,\big|\,1 \leq c_1, c_2, c_3 \leq N\}$  has no non-trivial solution if 
\begin{equation}\label{eqn:restatementofthm}
\mathfrak{D}_{N^3-1}\left(\{\alpha_{a_{1}b_{1},a_{2}b_{2},a_{3}b_{3}}\}\right) = \det(\mathfrak{C}) \neq 0\,.
\end{equation}
Thus, all we need to show is that the polynomial $\mathfrak{D}_{N^3-1}\left(\{\alpha_{a_{1}b_{1},a_{2}b_{2},a_{3}b_{3}}\}\right)$ is not identically zero, which we will now proceed to demonstrate.

\noindent \underline{\textit{Proof:}} Observe that the diagonal elements of the matrix $C$ are given by
\begin{equation}\label{eqn:diagonaleltscoeffmatrix}
C_{\rho\rho} = \begin{cases} c_{ik} \;& \mathrm{if\;}\; \gamma_{ik} = d_{i}\hat{e}_{i} \mathrm{\;for\;some\;}\;1 \leq k \leq m_{i} 
\\ 0 \;& \mathrm{otherwise}. \\ \end{cases} 
\qquad \rho \in  \mathcal{R}_{i}.
\end{equation}
In our case, each polynomial $\mathfrak{f}_{c_1 c_2 c_3}$ has precisely such a term given by $(\mathscr{X}^{c_1 c_2 c_3})^3$ with coefficient 
$\alpha_{c_1 c_1,c_2 c_2,c_3 c_3}$, and hence all diagonal elements of $\mathfrak{C}$ are non-zero. Let 
\begin{equation}\label{eqn:alldegdmonomials}
\mathcal{T} = \bigcup_{i=0}^{n} \mathcal{R}_{i} = \bigg\{\rho = (\rho_{0},\ldots,\rho_{n}) \in \mathbb{Z}^{n+1}\,\bigg|\,\sum_{k=0}^{n}\rho_{k} = d\bigg\}.
\end{equation}
Then the determinant of $\mathfrak{C}$ is given explicitly by
\begin{equation}\label{eqn:detcoeffmatrix}
\mathfrak{D}_{N^3-1}(\{\alpha_{a_{1}b_{1},a_{2}b_{2},a_{3}b_{3}}\}) =  \sum_{\sigma \in S(\mathcal{T})}\mathrm{sgn}(\sigma)
\prod_{\rho \in \mathcal{T}} \; \mathfrak{C}_{\rho\sigma(\rho)}.
\end{equation}

Denote each term in the above sum by $G(\sigma)$. The diagonal entries of $\mathfrak{C}$ contribute a term
\begin{equation}\label{eqn:detdiagonalcontribution}
G(\mathrm{id}) = \prod_{c_1,c_2,c_3=1}^{N}\; \left(\alpha_{c_1 c_1, c_2 c_2, c_3 c_3}\right)^{{\sf S}_{c_1 c_2 c_3}}.
\end{equation}
Any other term $G(\sigma)$ in the determinant \eqref{eqn:detcoeffmatrix} has $\sigma \neq \mathrm{id}$ so there exists some 
$\widetilde{\rho} \in \mathcal{T}$ such that $\widetilde{\rho} \neq \sigma(\widetilde{\rho})$. We have $\widetilde{\rho} \in  \mathcal{R}_{i_1i_2 i_3}$ for some $1 \leq i_1,i_2,i_3 \leq N$ by the definition of $\mathcal{T}$. Therefore, using \eqref{eqn:coeffmatrix},
\begin{equation}\label{eqn:detoffdiagonalcontribution}
\begin{split}
G(\sigma) &= \mathrm{sgn}(\sigma)\prod_{\rho \in \mathcal{T}}\; \mathfrak{C}_{\rho\sigma(\rho)} \\
&= \mathrm{sgn}(\sigma)\; \mathfrak{C}_{\widetilde{\rho}\sigma(\widetilde{\rho})}\;
\prod_{\substack{\rho \in \mathcal{T} \\ \rho \neq \widetilde{\rho}}} \; \mathfrak{C}_{\rho\sigma(\rho)} 
\\	&= \begin{cases} 
	\mathrm{sgn}(\sigma)\alpha_{i_1b_{1},i_2 b_{2},i_3b_{3}}\;
	\prod_{\substack{\rho \in \mathcal{T}\\ \rho \neq \widetilde{\rho}}} \;
		\mathfrak{C}_{\rho\sigma(\rho)} & \; \mathrm{if\;} \;  \sigma(\tilde{\rho})+3\,\hat{e}_{i_1 i_2 i_3}-\tilde{\rho} = \gamma_{i_1b_{1},i_2b_{2},i_3b_{3}} \in \mathcal{A}_{i_1i_2i_3} \\ 0 & \mathrm{otherwise}, \end{cases}
\end{split}
\end{equation}
where we have used $\gamma_{i_1b_{1},i_2b_{2},i_3b_{3}}$ to denote the exponent of the monomial $\mathscr{X}^{i_1b_{2}b_{3}}\mathscr{X}^{b_{1} i_2 b_{3}}\mathscr{X}^{b_{1}b_{2}i_3}$ in $\mathfrak{f}_{i_1i_2 i_3}$.

Since $\widetilde{\rho} \neq \sigma(\widetilde{\rho})$, it follows that $\gamma_{i_1b_{1},i_2b_{2},i_3b_{3}} \neq 3\hat{e}_{i_1i_2 i_3}$ which in turn implies that $\alpha_{i_1b_{1}, i_2b_{2},i_3b_{3}} \neq \alpha_{i_1i_1,i_2i_2,i_3i_3}$. Thus, $G(\sigma) \neq -G(\mathrm{id})$ for any $\sigma \in S(\mathcal{T})$, and hence $\mathfrak{D}_{N^3-1}(\{\alpha_{a_{1}b_{1},a_{2}b_{2},a_{3}b_{3}}\}) \neq 0$ identically. 

Therefore, as required, there exists a choice of $\{\alpha_{a_{1}b_{1},a_{2}b_{2},a_{3}b_{3}}^{(0)}\} \subset \mathbb{C}^{N^{6}}$ such that the determinant  $\mathfrak{D}_{N^3-1}(\{\alpha_{a_{1}b_{1},a_{2}b_{2},a_{3}b_{3}}^{(0)}\}) \neq 0$. $\square$ \\

Now, for the diagrammatics we need to be able to find a choice $\{\alpha_{a_{1}b_{1},a_{2}b_{2},a_{3}b_{3}}\}$ such that each $\alpha_{a_{1}b_{1},a_{2}b_{2},a_{3}b_{3}}$ is positive and $\mathcal{O}(1)$. We would also like this to be valid when we consider infinitesimal deformations of the isotropic model, where we need each $\alpha_{a_{1}b_{1},a_{2}b_{2},a_{3}b_{3}}$ to lie within a distance $\epsilon$ of the point $\mathbf{1} = (1,\ldots,1) \in \mathbb{C}^{N^{6}}$. 

Define the positive wedge of the ball of radius $\epsilon$ centered at $\mathbf{1}$:
\begin{equation}\label{eqn:positivewedgeepsilonball}
B_{\epsilon}^{+}(\mathbf{1}) = \{\mathbf{y} \in \mathbb{C}^{N^{6}}\,|\,y_{i} > 0\,\mathrm{for\;every\;}1 \leq i \leq N^{6},\;|\mathbf{y}-\mathbf{1}| < \epsilon\},
\end{equation}
where the metric $| \cdot |$ on $\mathbb{C}^{N^{6}}$ is the standard induced metric from $\mathbb{R}^{2N^{6}}$. To see that we can always make such a choice, notice that the zero set of $\mathfrak{D}_{N^3-1}$ is a codimension-$1$ variety in $\mathbb{C}^{N^{6}}$, which we denote by $V(\mathfrak{D}_{N^3-1})$. On the other hand, $B_{\epsilon}^{+}(\mathbf{1})$ has codimension-$0$ in $\mathbb{C}^{N^{6}}$. Therefore, $V(D_{N^3-1})$ cannot contain $B_{\epsilon}^{+}(\mathbf{1})$. This immediately can be used to infer that $B_{\epsilon}^{+}(\mathbf{1}) - \left(B_{\epsilon}^{+}(\mathbf{1}) \cap V(\mathfrak{D}_{N^3-1}) \right) \neq \emptyset$, and hence we can make the desired choice of $\{\alpha_{a_{1}b_{1},a_{2}b_{2},a_{3}b_{3}}\}$.

%~~~~~~~~~~~~~~~~~~~~~~~~~~~~~~~~~~~~~~~~~~~~~~
\subsection{Moduli for partial anisotropy}
%~~~~~~~~~~~~~~~~~~~~~~~~~~~~~~~~~~~~~~~~~~~~~~

Our discussion thus far has established that we are able to make a choice of the deformation parameters which lifts all the moduli. Let us look at some special cases to see what forms of anisotropy are needed for us to succeed in our quest. The key observation relies on noting that the proof has two crucial elements. 

Firstly, the proof hinged on the fact that each partial derivative $\mathfrak{f}_{c_1 c_2 c_3} = \partial W_{4}/\partial \mathscr{X}^{c_1 c_2 c_3}$ contains the term $\left(\mathscr{X}^{c_1 c_2 c_3}\right)^{3}$ so that every diagonal element of the matrix is nonzero. This clearly will not hold for general quartic superpotentials because it requires the superpotential to contain $(\mathscr{X}^{c_1 c_2 c_3})^{4}$ for every $1 \leq c_i \leq N$. 

Secondly, the proof required that $\alpha_{i_1b_{1},i_2b_{2},i_3b_{3}} \neq \alpha_{i_1 i_1, i_2 i_2, i_3 i_3}$ whenever $(b_{1},b_{2},b_{3}) \neq (i_1,i_2,i_3)$. This will be true for the most general coefficients $\alpha_{a_{1}b_{1},a_{2}b_{2},a_{3}b_{3}}$, but will not necessarily hold for special choices. Let us examine some of these special cases: 
\begin{enumerate}
\item % #1
For our computation of the anisotropic four-point function in $\S$\ref{sec:aam}, to make the computation easier, we assumed that the coefficients factorized, viz.,  $\A_{a_1b_1,a_2b_2,a_3b_3}= \prod_{k=1}^3 \A_{a_k b_k}$. In this case, we can still make the choice 
$\alpha_{ii} \neq \alpha_{ib_k}$ for $i\neq b_k$. Thus, $\prod_{l=1}^3\,\alpha_{i_l b_l} \neq  \prod_{k=1}^3\, \alpha_{i_k i_k}$ for $(b_{1},b_{2},b_{3}) \neq (i_1, i_2, i_3)$ and so the proof continues to hold.
\item % #2
Suppose we make the choice $\A_{a_1b_1,a_2b_2,a_3b_3}=\A_{a_1b_1}\A_{a_2b_2}$, which we can think of as partially (`two-thirds') anisotropic because only the first and second indices of the tensor contraction have anisotropy. In this case, we can find $\alpha_{i_1b_{1}}\alpha_{i_2b_{2}} = \alpha_{i_! i_1}\alpha_{i_2 i_2}$ for $(b_{1},b_{2},b_{3}) \neq (i_1,i_2,i_3)$ by taking $b_{3} \neq i_3$. The proof no longer holds. Specifically, our linear algebra problem leaves it undetermined whether  or not there are moduli. In fact, we have found numerically that there exists a non-trivial moduli space in this case for small $N$.
\item % #3
By the exact same reasoning as the previous case, the proof also does not hold for the `one-third' anisotropic case $\A_{a_1b_1,a_2b_2,a_3b_3}=\A_{a_1b_1}$. In particular, this applies to the anisotropic matrix-vector model at $N=M$, for which there exist moduli (in the ungauged case) as discussed in \S\ref{sec:gmv}.
\item % #4
For the $(\bZ_2^N\rtimes S_N)^3$-invariant anisotropic deformation \eqref{eqn:Z2xSN_model}, we can see that there are no non-trivial solutions for generic $\{\alpha_{1},\ldots,\alpha_{8}\}$ by focusing on the first and last term $\alpha_{1}+\alpha_{8}\delta_{a_1b_1}\delta_{a_2b_2}\delta_{a_3b_3}$. The diagonal contribution to the determinant $G(\mathrm{id})$ will contain a factor $\alpha_{8}^{\sum_{c_{1},c_{2},c_{3}=1}^{N}{\sf S}_{c_1 c_2 c_3}}$, but every other contribution to the determinant $G(\sigma \neq \mathrm{id})$ will have a strictly smaller power of $\alpha_{8}$, and hence the other contributions cannot cancel the diagonal contribution for sufficiently generic $\alpha_{1}$ and $\alpha_{8}$. This argument is unchanged if we turn on sufficiently generic $\alpha_{2},\ldots,\alpha_{7}$.

\end{enumerate}

%~~~~~~~~~~~~~~~~~~~~~~~~~~~~~~~~~~~~~~~~~~~~~~~
\section{Higgs branch analysis for $N=M=2$}
\label{sec:higss2}
%~~~~~~~~~~~~~~~~~~~~~~~~~~~~~~~~~~~~~~~~~~~~~~

In this appendix, we explicitly analyze the Higgs branch for the $N=M=2$ matrix-vector model. There are two chiral superfields $\mathscr{Y}^1$ and $\mathscr{Y}^2$. For simplicity, we specialize the deformation parameter $\alpha_{IJ}$ to be 
\begin{equation}
\alpha_{11} = \alpha_{22} = 1\,, \qquad \alpha_{12} = \alpha_{21} = a \,.
\end{equation}	
We will examine the problem as a function of $a$ in some detail below. There are three special values of $a$, $a_*  = \{0,1,\infty\}$, while all other values can be treated uniformly.

\subsection{Higgs branch chiral ring}
Let us start by analyzing the Higgs branch chiral ring \eqref{eq:Hring} of the theory. The scalars $Y^1$ and $Y^2$ can be expanded in terms of the Pauli matrices as
\ie
Y^1=y^1_i \sigma^i,\quad Y^2=y^2_i \sigma^i.
\fe
The $F$-term constraints are
\ie
0&=8y^1_i (y^1\cdot y^1)+8a\left[2y^2_i(y^1 \cdot y^2)-y^1_i(y^2 \cdot y^2)\right],
\\
0&=8y^2_i (y^2\cdot y^2)+8a\left[2y^1_i(y^1 \cdot y^2)-y^2_i(y^1 \cdot y^1)\right],
\fe
which are equivalent to the following gauge invariant constraints
\ie\label{eqn:hcr_relations}
0&=8A^2+16aC^2-8aAB,
\\
0&=8AC+8aBC,
\\
0&=8BC+8aAC,
\\
0&=8B^2+16aC^2-8aAB,
\fe
where $A=(y_1\cdot y_1)$, $B=(y_2\cdot y_2)$, and $C=(y_1\cdot y_2)$. The chiral ring is $\bC[A,B,C]$ quotient by the relations \eqref{eqn:hcr_relations}. It is not hard to see that  for generic $a$ the nontrivial classes are represented by 
\ie
1,\quad A,\quad B,\quad C,\quad A^2,\quad B^2,
\fe
which agrees with Table \ref{tab:Hpoin}.

\subsection{Moduli space}

\paragraph{1. Generic $a$:} Let us start with $a$ being generic $a\neq a_*$. We solve the  $F$-term equation to obtain the following parametrization:
\begin{equation}
Y^1 = y_1\begin{pmatrix}
u_1 
\\
u_2
\end{pmatrix}
\begin{pmatrix}
-u_2 & u_1
\end{pmatrix},\quad
Y^2 =y_2 \begin{pmatrix}
u_1 
\\
u_2
\end{pmatrix}
\begin{pmatrix}
-u_2 & u_1
\end{pmatrix}.
\label{eqn:F-term_solution}
\end{equation}	
This can however be further simplified using a complexified gauge transformation ${\rm SU}(2)_\bC$. We can bring the solution \eqref{eqn:F-term_solution} into the form
\begin{equation}
Y^1 = y_1  \, {\sf S}_2 ,\quad
Y^2 = y_2  \, {\sf S}_2 ,\quad  \, {\sf S}_2 =\begin{pmatrix}
0 & 1
\\
0 & 0
\end{pmatrix}=\begin{pmatrix}
1
\\
0
\end{pmatrix}
\begin{pmatrix}
0 & 1
\end{pmatrix}
\end{equation}	
Under the complexified gauge transformation generated by the Cartan sub-algebra, we further have
\begin{equation}
y_1\to e^{2\eta}\, y_1,\quad y_2\to  e^{2\eta}\, y_2\,,
\end{equation}	
where we used $e^{\eta \sigma_3} \, {\sf S}_2 e^{-\eta \sigma_3}=e^{2\eta} \, {\sf S}_2$.

We now can parameterize the holomorphic quotient as a disjoint union of
\begin{equation}
\begin{split}
&(\bC^2 - \{(0,0)\}) / \bC^* = {\mathbb C}{\mathbb P}^1,
\\
 &\{(0,0)\} / \bC^* = \{(0,0)\}.
\end{split}
\label{eq:holc1}
\end{equation}
with the first line arising from $y_1, y_2 \neq 0$, and the second line from when they both vanish. The holomorphic quotient has to be filtered through the D-term constraint $|y_1|^2+|y_2|^2=0$, which then picks out a unique vacuum $Y^1 =  Y^2 =0$. 

\paragraph{2. Isotropic superpotential $a=1$:} Not much to discuss here as we already know from \eqref{eq:mvmoduli} the non-trivial flat directions of the ungauged model. One can convince oneself that they survive the gauging, unfortunately. 

\paragraph{3. Two matrix model ($a=0$):} The $F$-term equation can be now be solved for $Y^1$ and $Y^2$ independently. The general solution is 
\begin{equation}
Y^1 = \begin{pmatrix}
x_1 
\\
x_2
\end{pmatrix}
\begin{pmatrix}
-x_2 & x_1
\end{pmatrix},\quad
Y^2 = \begin{pmatrix}
y_1 
\\
y_2
\end{pmatrix}
\begin{pmatrix}
-y_2 &y_1
\end{pmatrix}.
\end{equation}	
This can be further simplified when $x_1 y_2-x_2y_1\neq 0$, using the ${\rm SU}(2)_{\bC}$ transformation:
\begin{equation}
Y^1 = \begin{pmatrix}
1
\\
0
\end{pmatrix}
\begin{pmatrix}
0 & 1
\end{pmatrix},\quad
Y^2 = \begin{pmatrix}
0 
\\
x_1 y_2-x_2y_1
\end{pmatrix}
\begin{pmatrix}
x_2y_1-x_1 y_2 &\ 0
\end{pmatrix}\,,
\end{equation}	
or
\begin{equation}
Y^1 = \begin{pmatrix}
0
\\
1
\end{pmatrix}
\begin{pmatrix}
1 & 0
\end{pmatrix},\quad
Y^2 = \begin{pmatrix}
x_1 y_2-x_2y_1
\\
0
\end{pmatrix}
\begin{pmatrix}
0 &\ x_2y_1-x_1 y_2
\end{pmatrix}\,.
\end{equation}	
When $x_1 y_2-x_2y_1= 0$, we are back to the situation for generic $a$ discussed above.  The holomorphic quotient is a disjoint union of 
\begin{equation}
{\mathbb C}{\mathbb P}^1,\quad \bC^*=\bC - \{0\}, \quad\{(0,0)\}.
\end{equation}	
The D-term equation selects out $\bC^*$ and $\{(0,0)\}$, and they combine into $\bC$, which is parametrized by $x_1 y_2-x_2y_1$.

\paragraph{Two matrix model ($a=\infty$): } The solutions to the $F$-term equation now fall into 
\begin{enumerate}
\item[i.] $Y^1=0$, arbitrary $Y^2$. 
\item[ii.] $Y^2=0$, arbitrary $Y^1$. 
\item[iii.]  $\det Y^1=\det Y^2=0$ and $Y^1Y^2=0$. 
\item[iv.] $\det Y^1=\det Y^2=0$ and $Y^2Y^1=0$.
\end{enumerate}

Let us analyze the last class in more detail  (the first two clearly have moduli, and the third is similar).  We can write the solution as:
\begin{equation}
 Y^1=u_1 v_1,\quad  Y^2=u_2 v_2,\quad v_2\cdot u_1=0,
\label{eq:}
\end{equation}	
where $u_1$, $u_2$ are column vectors and $v_1$, $v_2$ are row vectors. They are subject to the transformation
\begin{equation}
\begin{split}
&u_{1,2}\to {\sf M} \, u_{1,2},\quad v_{1,2}\to v_{1,2}\, {\sf M}^{-1},
\\
&u_1 \to \kappa \, u_1,\quad v_1\to \kappa^{-1}\, v_1,
\\
&u_2 \to \gamma  \, u_2,\quad v_2\to \gamma^{-1} \, v_2,
\end{split}
\label{eq:}
\end{equation}
where ${\sf M}\in  {\rm SU}(2)_{\bC}$, and $\kappa,\gamma \in\bC^*$. The equation $v_2\cdot u_1=0$ is solved by
\begin{equation}
v_2 = \lambda \, u_1^T \, \epsilon,\quad\lambda\in\bC^*.
\label{eq:}
\end{equation}	
Using the combination of the scaling symmetry with $\kappa=\gamma$, we can fix
\ie
\lambda=1.
\fe
The ${\rm SU}(2)_{\bC}$ and the remaining scaling symmetry combine to ${\rm GL}(2,\bC)$ and completely fix $u_2$ and $v_1$. The holomorphic quotient is $\bC^2$ parametrized by the vector $u_1$. The D-term equation gives
\ie
u_1=(u_{11},u_{12}),\quad |u_{11}|^2=| u_{12}|^2,\quad u_{11}u_{12}^*=0,
\fe
which pick out the solution $Y^1=Y^2=0$.

%%%%%%%%%%%%%%%%%%%%%%%%%%%%%%%%%%%%%%%%%%%%%%%%%%%%%%%

% \bibliography{LGmelons-refs} 
% \bibliographystyle{JHEP}

\providecommand{\href}[2]{#2}\begingroup\raggedright\endgroup

%%%%%%%%%%%%%%%%%%%%%%%%%%%%%%%%%%%%%%%%%%%%%%%%%%%%%%%
\end{document}